\documentclass{article}

\usepackage[pagebackref=true,breaklinks=true,letterpaper=true,colorlinks,bookmarks=true]{hyperref}\usepackage[numbers]{natbib}
\usepackage[preprint]{neurips_2025}

\usepackage[utf8]{inputenc} %
\usepackage[T1]{fontenc}    %
\usepackage{url}            %
\usepackage{booktabs}       %
\usepackage{subcaption}   %
\usepackage{amsfonts}       %
\usepackage{amssymb}
\usepackage{nicefrac}       %
\usepackage{microtype}      %
\usepackage{xcolor}         %

\usepackage{graphicx}
\usepackage{wrapfig}    %

\usepackage{pgfplots,xcolor,siunitx}
\pgfplotsset{compat=1.18}
\definecolor{steelblue}{RGB}{70,120,180} %
\usepackage{tikz}
\usepackage{pgfplots}
\pgfplotsset{compat=1.18}

\usepackage{xspace}
\newcommand{\modelname}{\emph{FLUX.1 Kontext}\xspace}

\newcommand{\figref}[1]{Fig.~\ref{#1}}
\newcommand{\secref}[1]{Section~\ref{#1}}

\makeatletter
\DeclareRobustCommand\onedot{\futurelet\@let@token\@onedot}
\def\@onedot{\ifx\@let@token.\else.\null\fi\xspace}
\def\eg{e.g\onedot} 
\def\ie{i.e\onedot}

\makeatother

\definecolor{darkgreen}{rgb}{0,0.7,0}
\definecolor{darkblue}{RGB}{31,119,180}
\definecolor{darkred}{RGB}{214,39,40}

\usepackage{appendix}

\newlength\savewidth

\renewcommand{\paragraph}[1]{\vspace{1.25mm}\noindent\textbf{#1}}

\usepackage{array}
\newcolumntype{x}[1]{>{\centering\arraybackslash}p{#1pt}}
\newcolumntype{y}[1]{>{\raggedright\arraybackslash}p{#1pt}}
\newcolumntype{z}[1]{>{\raggedleft\arraybackslash}p{#1pt}}

\newcommand{\app}{\raise.17ex\hbox{$\scriptstyle\sim$}}

\definecolor{deemph}{gray}{0.6}

\definecolor{baselinecolor}{gray}{.9}

\usepackage{pifont}%

\usepackage{colortbl}
\usepackage{microtype}
\usepackage{graphicx}
\usepackage{csquotes}
\usepackage{subcaption}
\usepackage{booktabs} %
\usepackage{amsmath}
\usepackage{duckuments}
\usepackage{mathtools}
\usepackage{makecell}
\usepackage{multirow}
\usepackage{comment}
\usepackage{tabularx}
\usepackage{float}
\usepackage{stfloats}
\usepackage{amsfonts}
\usepackage{placeins}
\usepackage[export]{adjustbox} %

\usepackage{fancyvrb}

\RecustomVerbatimCommand{\VerbatimInput}{VerbatimInput}%
{fontsize=\footnotesize,
 frame=lines,  %
 framesep=2em, %
 rulecolor=\color{Gray},
 label=\fbox{\color{Black}Selected Parti Prompts},
 labelposition=topline,
}

\newcommand{\theultimategullteaser}{%
\begin{figure}[p] %
  \centering
  \vspace*{0pt} %
  \begin{center} %
  \begin{subfigure}[t]{0.505\linewidth}
    \centering
    \includegraphics[width=\linewidth]{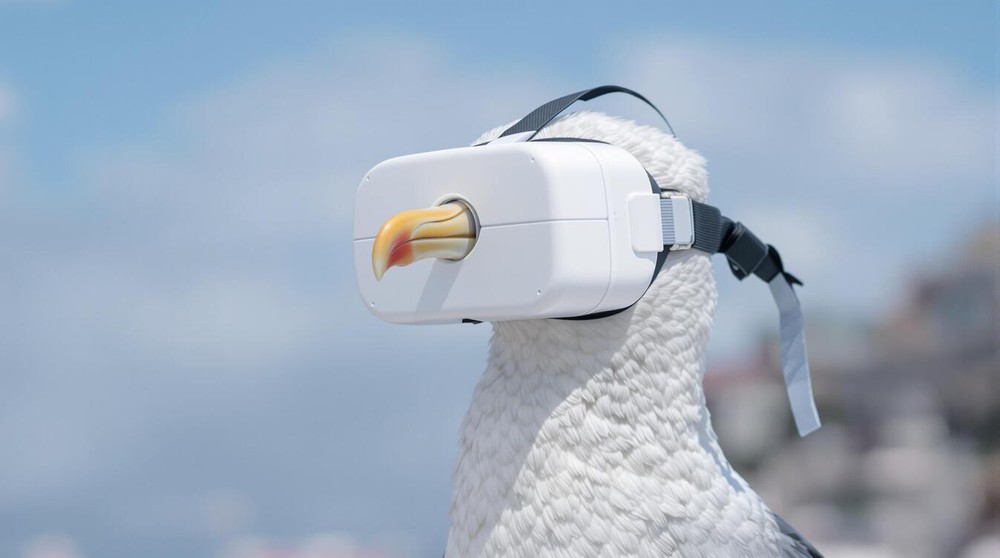}%
    \caption{\textbf{Context image} generated with FLUX.1.}
    \label{fig:gull:a}   %
  \end{subfigure}
  \end{center}
  \vspace{3pt}
  \begin{subfigure}[t]{0.48\linewidth}
    \centering
    \includegraphics[width=\linewidth]{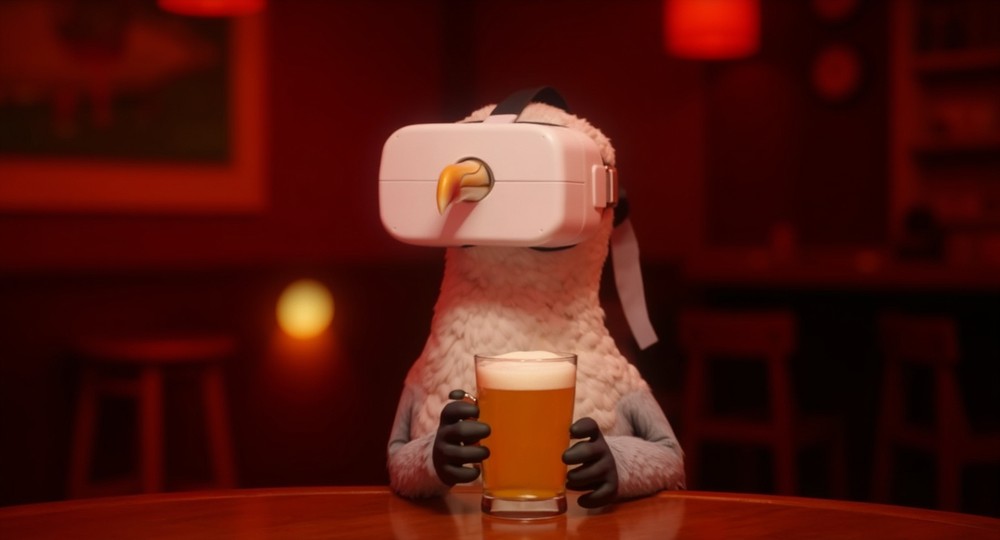}%
    \caption{Image context from \Cref{fig:gull:a}:
    \emph{``The bird is now sitting in a bar and enjoying a beer.''}
    }
    \label{fig:gull:b}
  \end{subfigure}\hfill
  \begin{subfigure}[t]{0.48\linewidth}
    \centering
    \includegraphics[width=\linewidth]{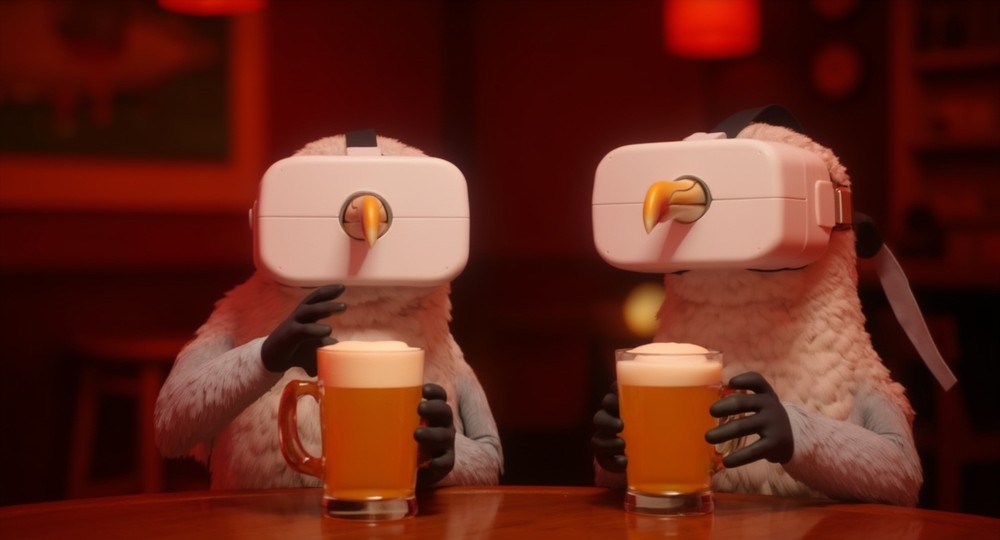}%
    \caption{Image context from \Cref{fig:gull:b}: \\
    \emph{``There are now two of these birds.''}
    }
    \label{fig:gull:c}
  \end{subfigure}
  \vspace{1pt}
  \begin{subfigure}[t]{0.48\linewidth}
    \centering
    \includegraphics[width=\linewidth]{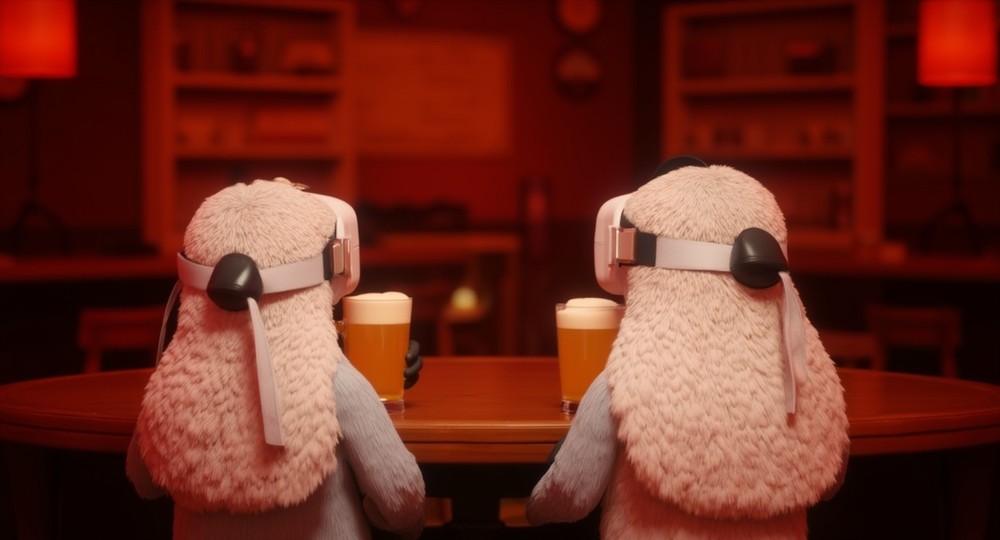}%
    \caption{From \Cref{fig:gull:c}:
    \emph{``Watch them from behind.''}
    }
    \label{fig:gull:d}
  \end{subfigure}\hfill
  \begin{subfigure}[t]{0.48\linewidth}
    \centering
    \includegraphics[width=\linewidth]{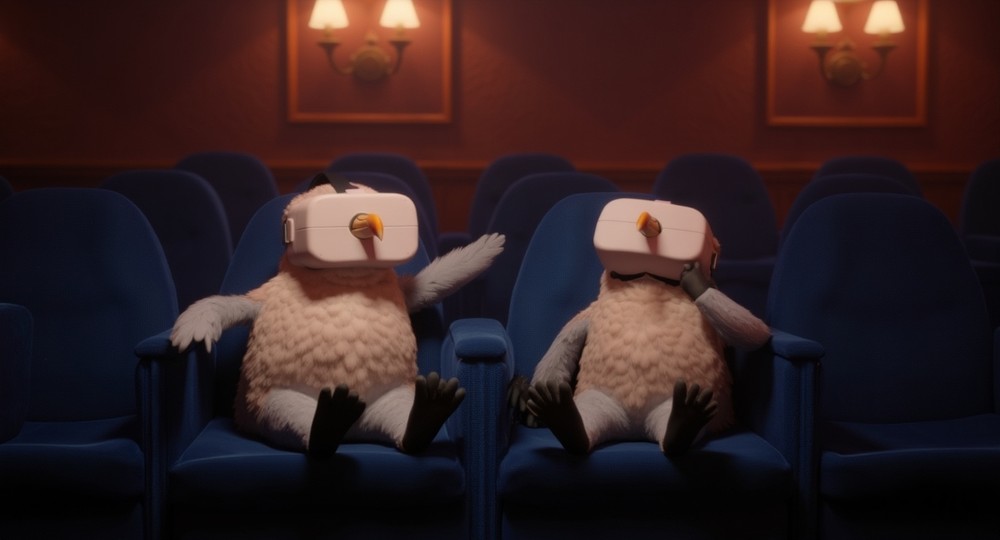}%
    \caption{From \Cref{fig:gull:c}:
    \emph{``The two bird characters are now sitting in a movie theater.''}
    }
    \label{fig:gull:e}
  \end{subfigure}
  \vspace{1pt}
  \begin{subfigure}[t]{0.48\linewidth}
    \centering
    \includegraphics[width=\linewidth]{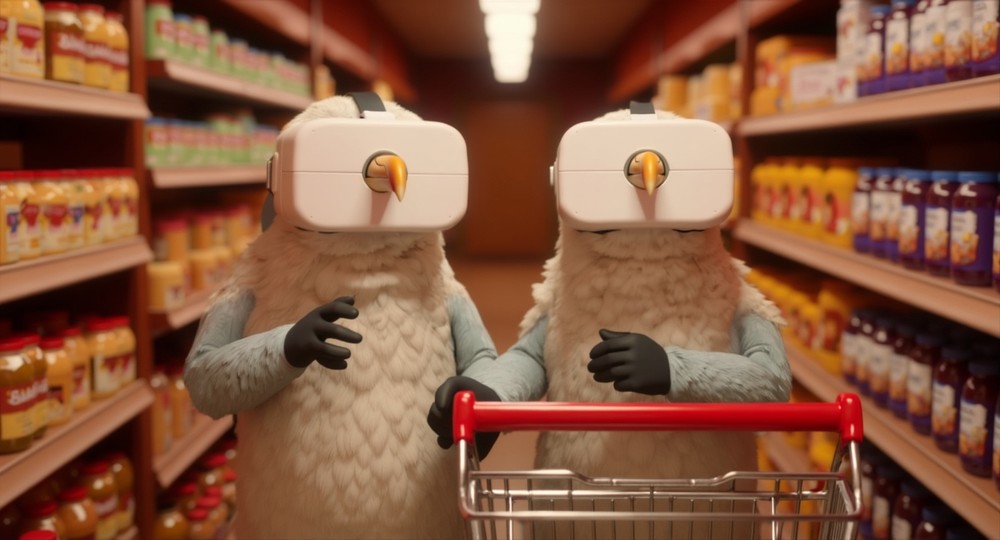}%
    \caption{From \Cref{fig:gull:c}:
    \emph{``The two bird characters are now grocery shopping.''}
    }
    \label{fig:gull:f}
  \end{subfigure}\hfill
  \begin{subfigure}[t]{0.48\linewidth}
    \centering
    \includegraphics[width=\linewidth]{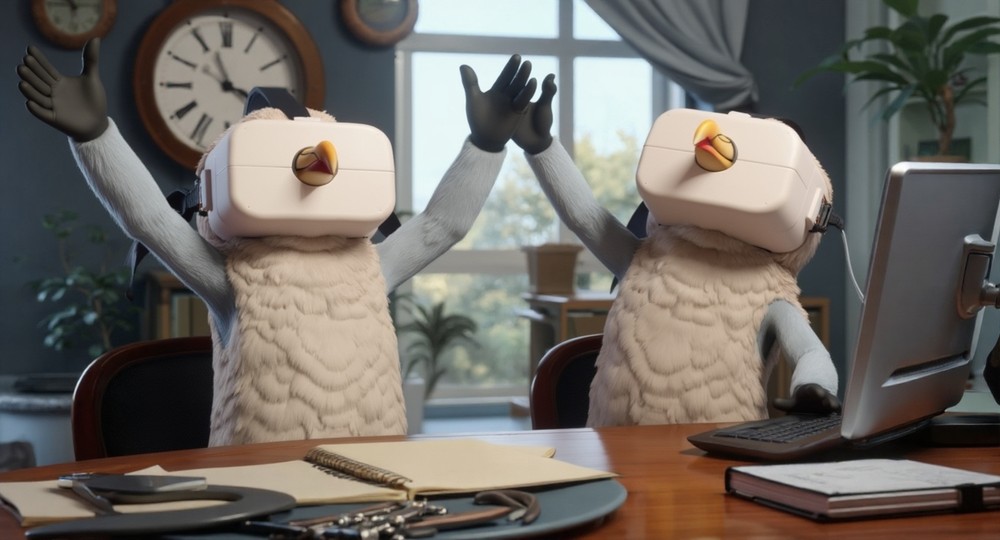}%
    \caption{From \Cref{fig:gull:f}:
    \emph{``The two bird characters are now celebrating a successful launch.''}
    }
    \label{fig:gull:g}
  \end{subfigure}
\vspace{-1pt}
\caption{Consistent character synthesis with \modelname{}. Generated images can be used iteratively as context for new generations, enabling applications such as storyboard generation and iterative narrative creation.
}
\label{fig:theultimategullteaser}
\vfill %
\end{figure}%
}

\newcommand{\snowtransform}{%
\begin{figure}[t]
  \centering
  \begin{subfigure}[t]{0.48\linewidth}
    \centering
    \includegraphics[width=\linewidth]{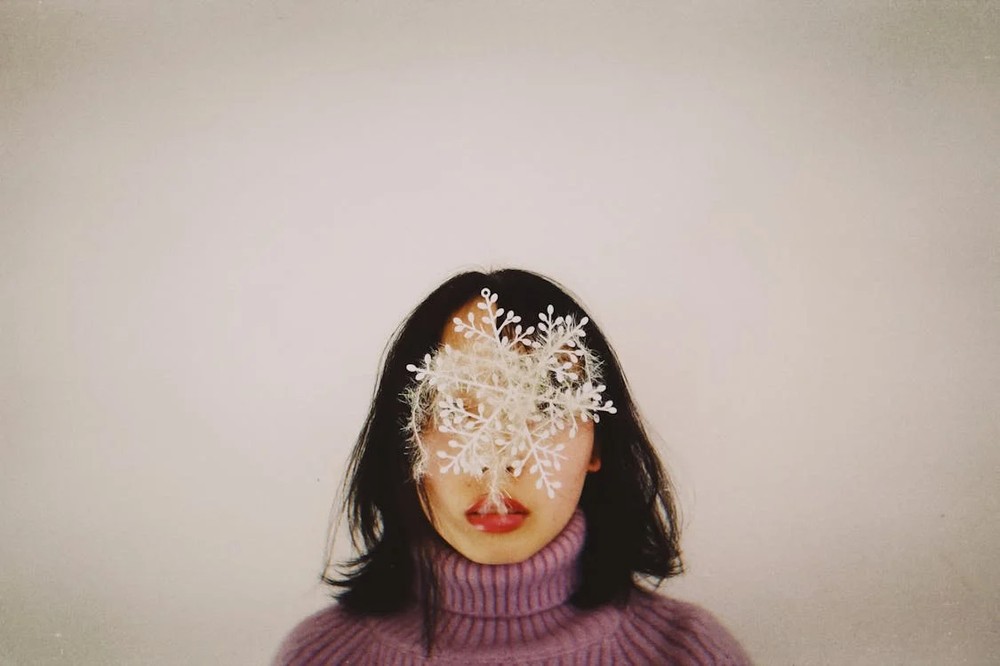}%
    \caption{Input image}
    \label{fig:selfie:a}
  \end{subfigure}\hfill
  \begin{subfigure}[t]{0.48\linewidth}
    \centering
    \includegraphics[width=\linewidth]{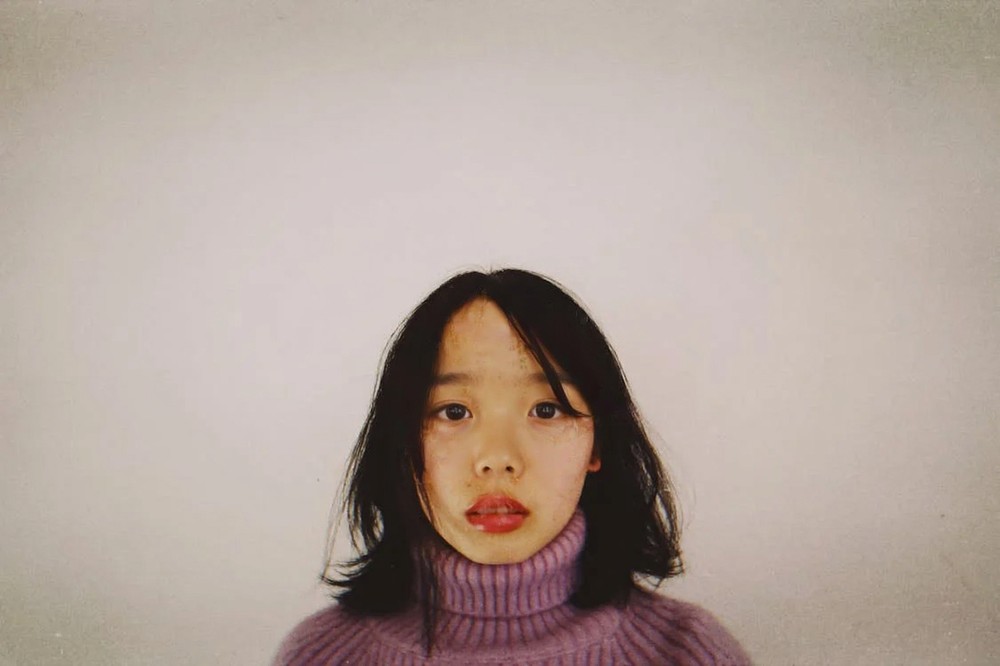}%
    \caption{\emph{``remove the thing from her face''}}
    \label{fig:selfie:b}
  \end{subfigure}

  \vspace{4pt}

  \begin{subfigure}[t]{0.48\linewidth}
    \centering
    \includegraphics[width=\linewidth]{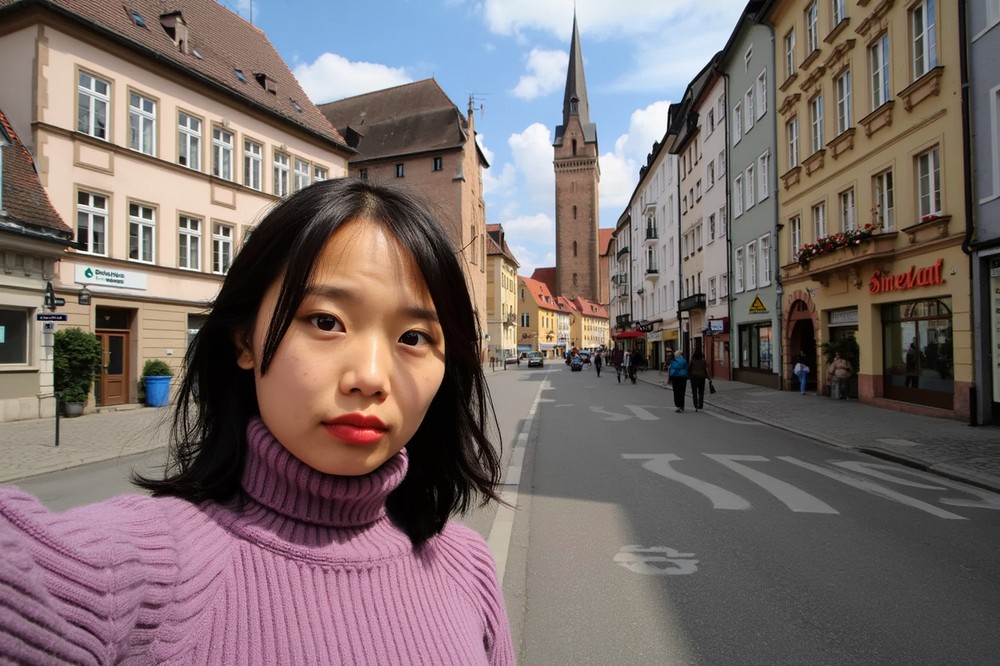}%
    \caption{\emph{``she is now taking a selfie in the streets of Freiburg,
                    it’s a lovely day out.''}}
    \label{fig:selfie:c}
  \end{subfigure}\hfill
  \begin{subfigure}[t]{0.48\linewidth}
    \centering
    \includegraphics[width=\linewidth]{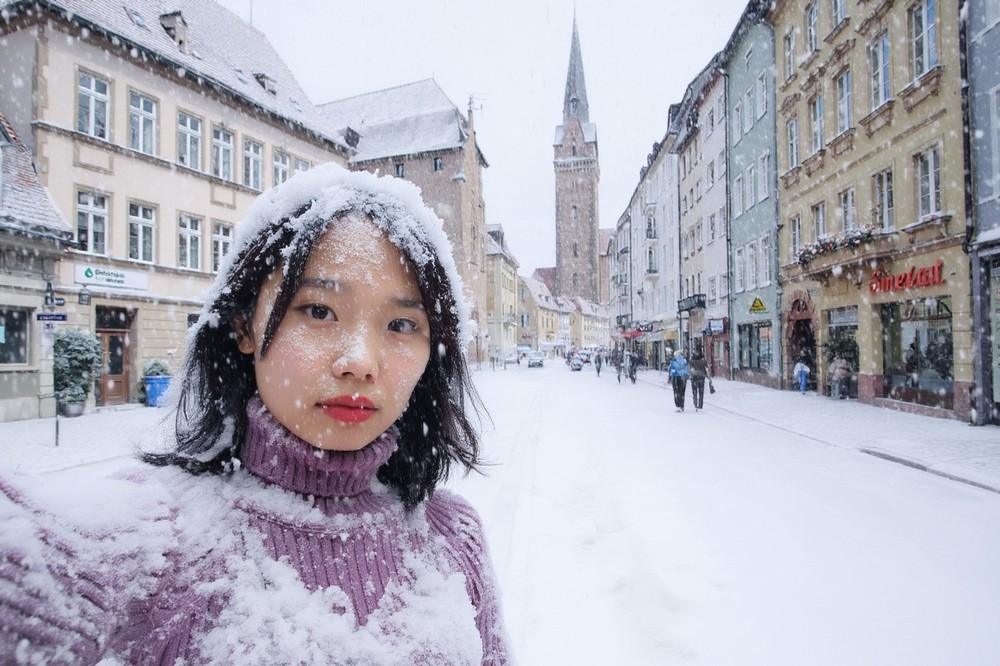}%
    \caption{\emph{``it’s now snowing, everything is covered in snow.''}}
    \label{fig:selfie:d}
  \end{subfigure}
  \caption{\textbf{Iterative, instruction-driven editing.}
  Starting from a reference photo~(\subref{fig:selfie:a}), %
  our model
  successively applies three natural-language edits—first removing an
  occlusion~(\subref{fig:selfie:b}), then relocating the subject to
  Freiburg~(\subref{fig:selfie:c}), and finally transforming the scene into
  snowy weather~(\subref{fig:selfie:d}). Character, pose, clothing, and overall
  photographic style remain consistent throughout the sequence.}
  \label{fig:iterative_edits}
\end{figure}}

\newcommand{\floatingfusedditblock}{
\begin{wrapfigure}[26]{r}{0.46\textwidth}
\vspace{4pt}              %
  \centering
  \includegraphics[width=\linewidth]{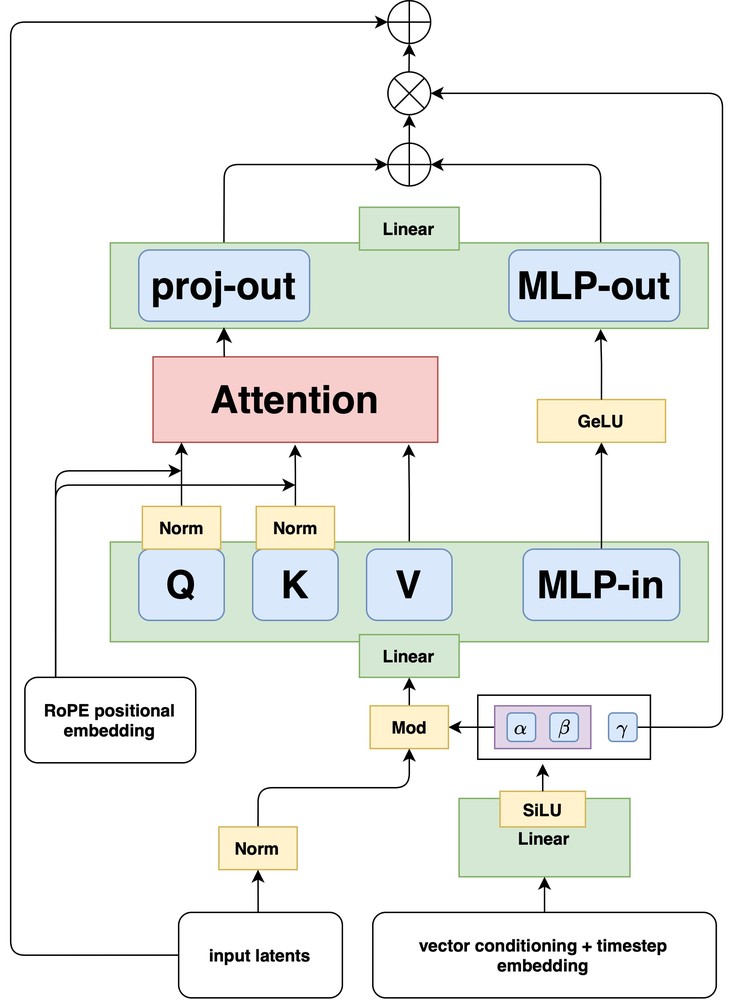}   %
  \caption{A fused DiT block equipped with rotary positional embeddings}
  \label{fig:floatingfusedditblock}
\end{wrapfigure}
}

\newcommand{\modelfigure}{
\begin{figure}
\begin{center}
\includegraphics[width=0.99\textwidth]{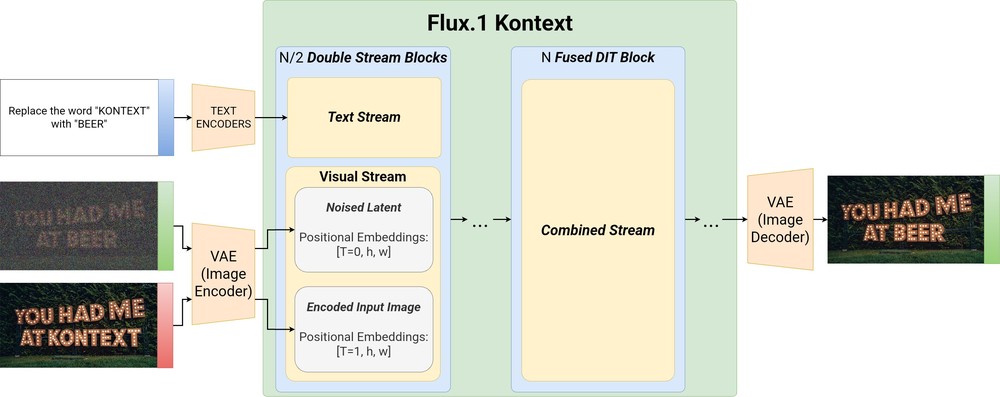}%
\caption{High-level overview of \modelname{}, with input and context image on the left. Details in \Cref{sec:method}.}%
\label{fig:modelfigure}%
\end{center}
\end{figure}
}

\newcommand{\ttoicherries}{
\begin{figure}[t]
\begin{center}
\includegraphics[width=\textwidth]{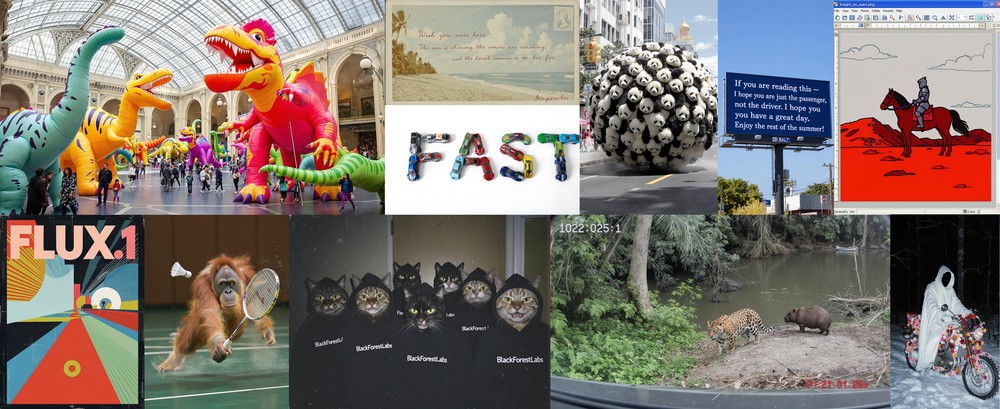}%
\caption{
Text-to-image samples by \modelname with low \textit{bakeyness}, diverse styles, and accurate typography. }%
\label{fig:ttoicherries}%
\end{center}
\end{figure}
}

\newcommand{\srefone}{%
\begin{figure}[t]
  \centering
  \begin{subfigure}[t]{0.24\linewidth}
    \centering
    \includegraphics[width=\linewidth]{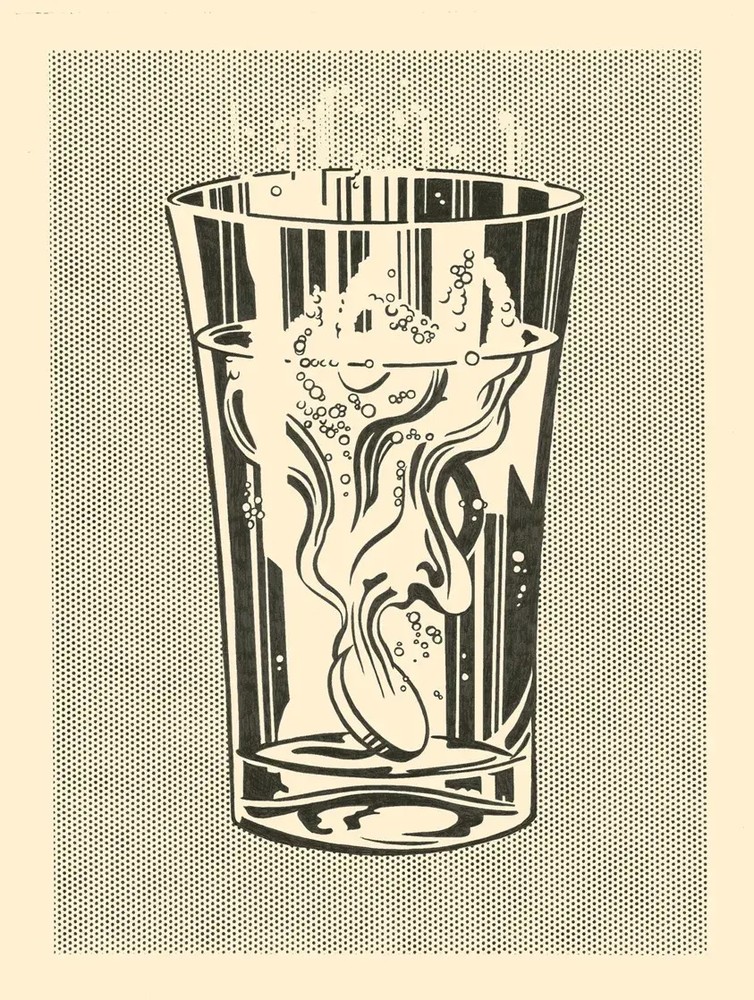}%
    \caption{Input image}
    \label{fig:sref1:a}
  \end{subfigure}\hfill
  \begin{subfigure}[t]{0.24\linewidth}
    \centering
    \includegraphics[width=\linewidth]{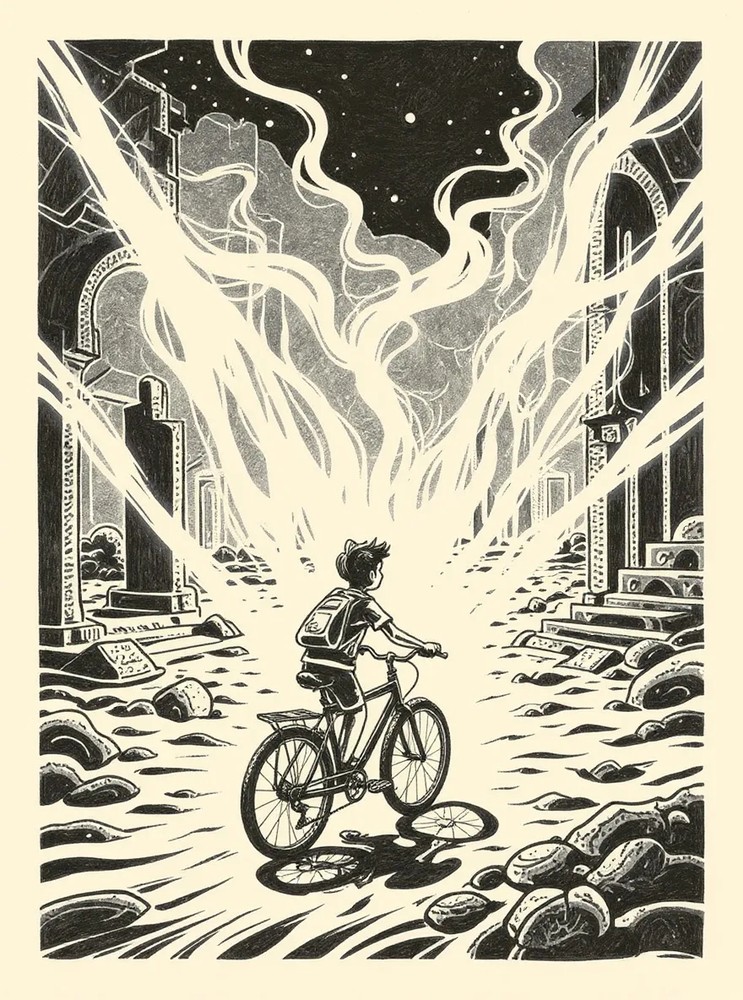}%
    \caption{\emph{``Using this style, a kid on a bicycles rolls through desert ruins, spotlights scanning ancient scrolls projected as holographic sandstorms.''}}
    \label{fig:sref1:b}
  \end{subfigure}\hfill
  \begin{subfigure}[t]{0.24\linewidth}
    \centering
    \includegraphics[width=\linewidth]{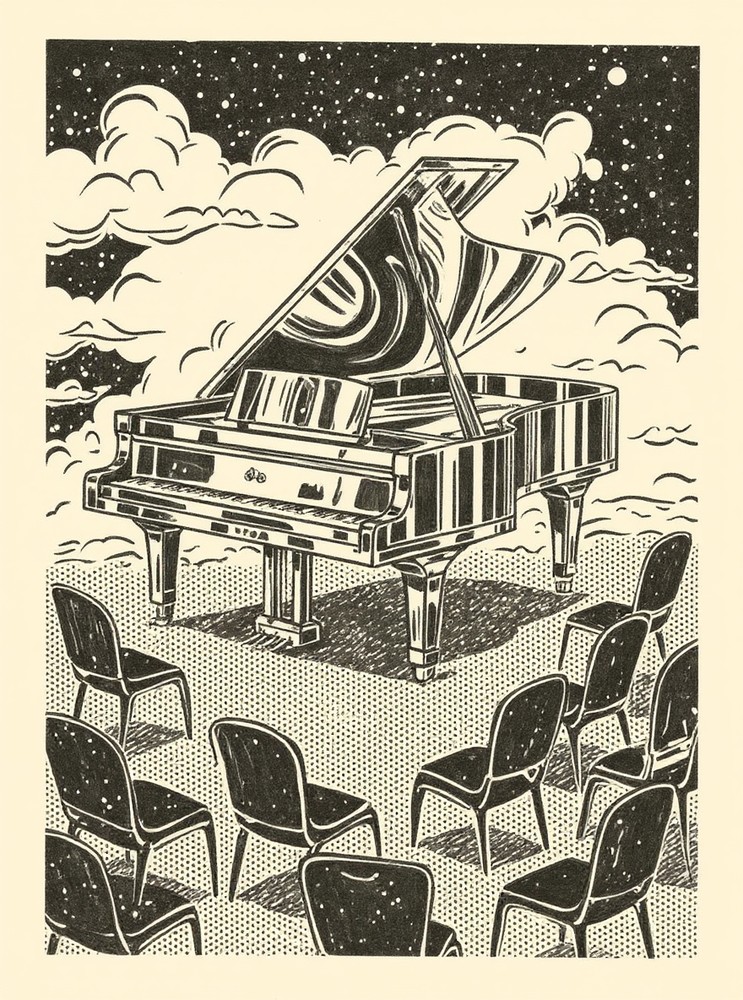}%
    \caption{\emph{``Using this style, a grand piano made of shifting mirrors performs itself for an audience of empty velvet chairs in zero-gravity.''}}
    \label{fig:sref1:c}
  \end{subfigure}\hfill
\begin{subfigure}[t]{0.24\linewidth}
    \centering
    \includegraphics[width=\linewidth]{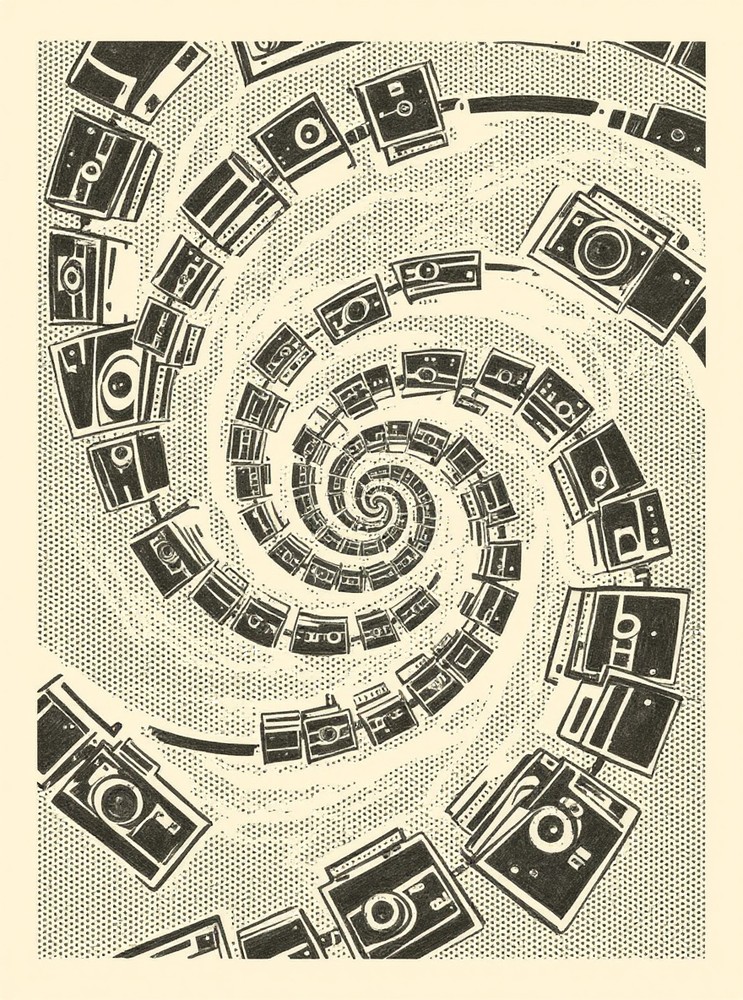}%
    \caption{\emph{``Using this style, a spiral of vintage cameras captures its own collapse, each flash freeze-framing a different timeline.''}}
    \label{fig:sref1:d}
  \end{subfigure}
  \begin{subfigure}[t]{0.24\linewidth}
    \centering
    \includegraphics[width=\linewidth]{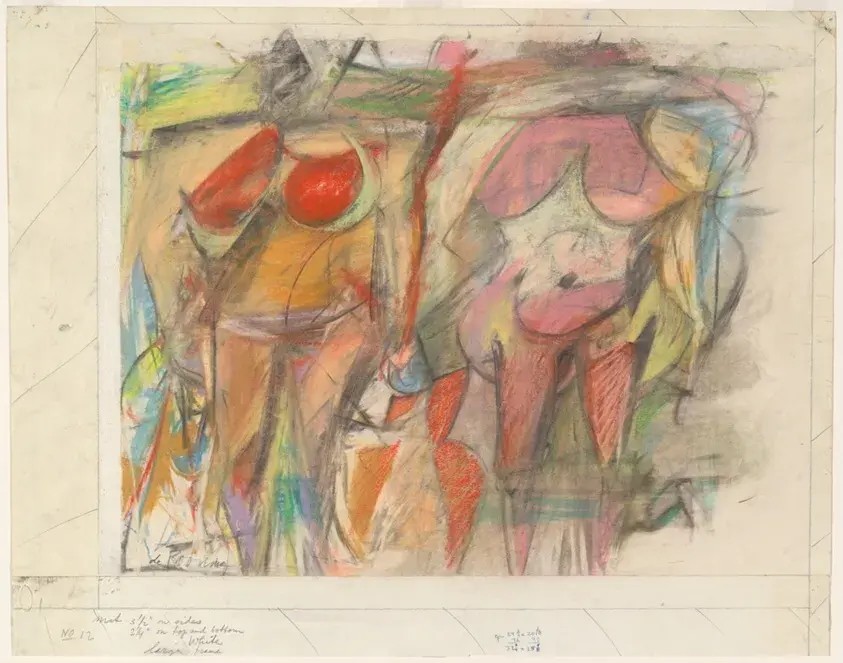}%
    \caption{Input image}
    \label{fig:sref1:e}
  \end{subfigure}\hfill
  \begin{subfigure}[t]{0.24\linewidth}
    \centering
    \includegraphics[width=\linewidth]{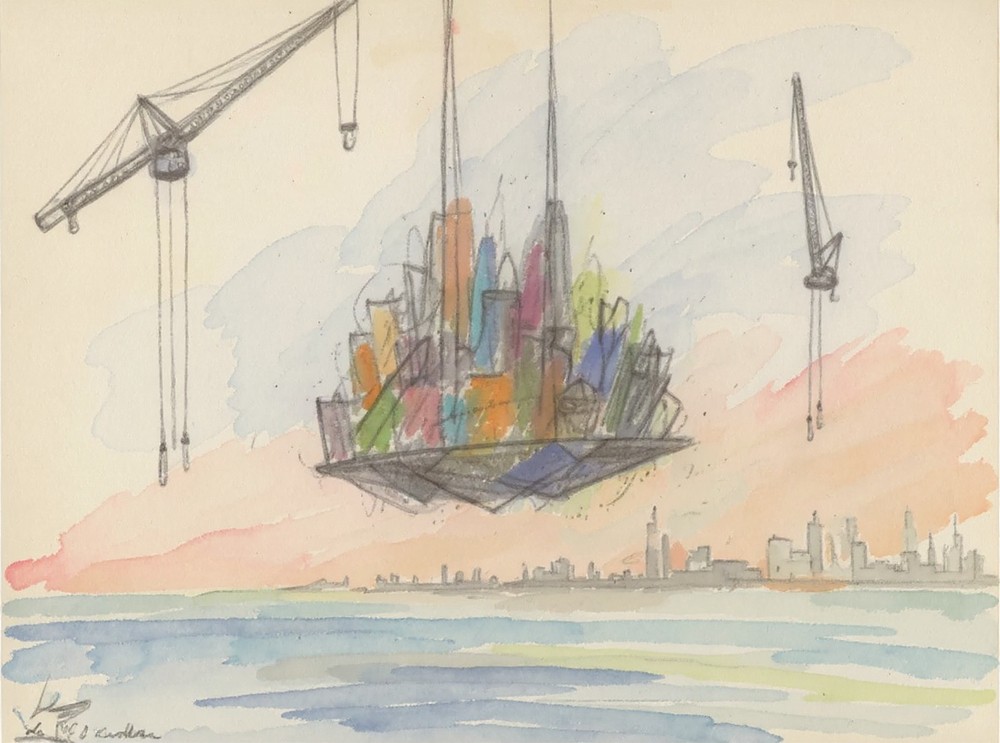}%
    \caption{\emph{``Using this style, a half-folded metropolis hangs from steel strings over an ink-wash ocean while cranes of light sketch new streets in mid-air.''}}
    \label{fig:sref1:f}
  \end{subfigure}\hfill
  \begin{subfigure}[t]{0.24\linewidth}
    \centering
    \includegraphics[width=\linewidth]{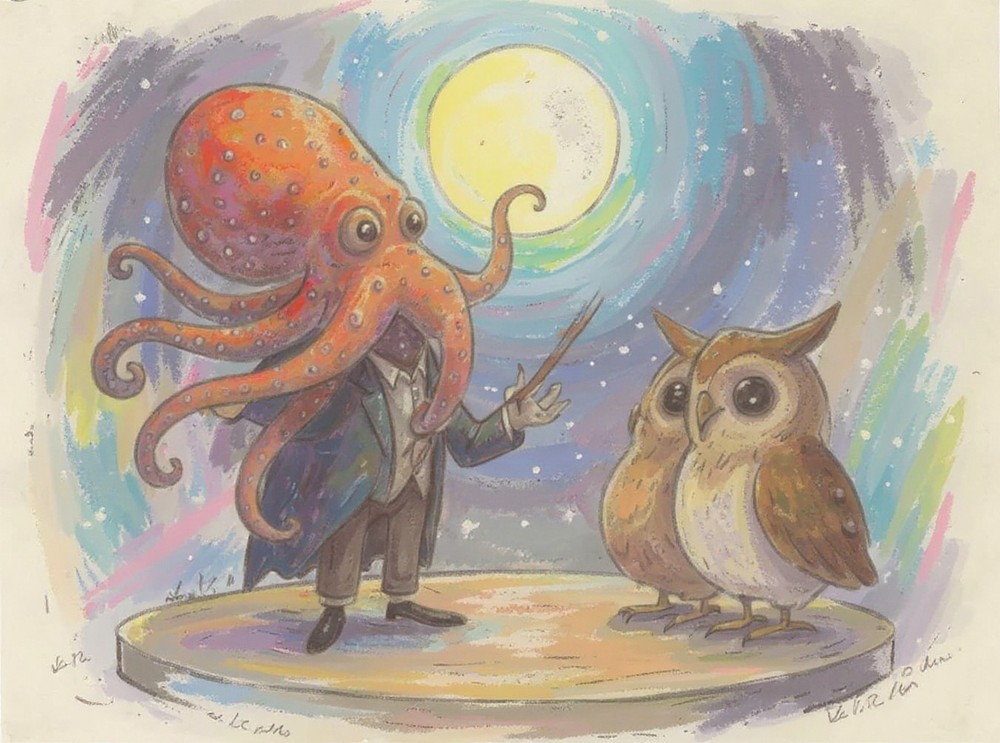}%
    \caption{\emph{``Using this style, a dapper octopus conducts a jazz duo of owls on a shimmering moonlit bandstand.''}}
    \label{fig:sref1:g}
  \end{subfigure}\hfill
\begin{subfigure}[t]{0.24\linewidth}
    \centering
    \includegraphics[width=\linewidth]{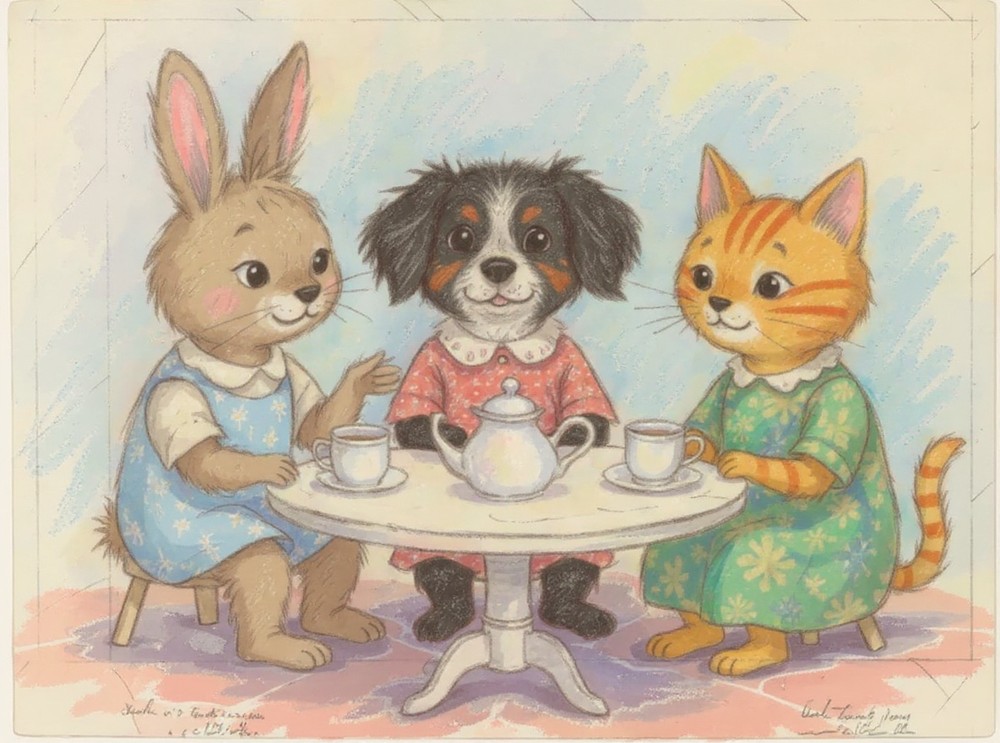}%
    \caption{\emph{``Using this style, a bunny, a dog and a cat are having a tea party seated around a small white table.''}}
    \label{fig:sref1:h}
  \end{subfigure}
\caption{\textbf{Style Reference.} Given an input image, the model extracts its artistic style and applies it to generate diverse new scenes while preserving the original stylistic characteristics.}
  \label{fig:srefone}
\end{figure}
}

\newcommand{\skirttransform}{%
\begin{figure}[t]
  \centering
  \begin{subfigure}[t]{0.32\linewidth}
    \centering
    \includegraphics[width=\linewidth]{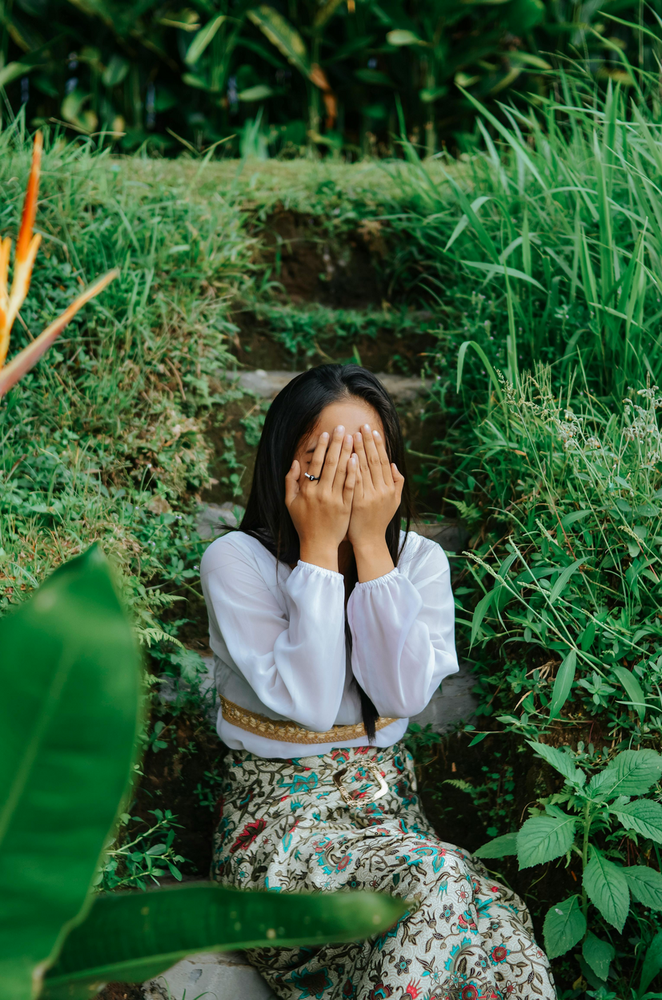}%
    \caption{Input image}
    \label{fig:skirt:a}
  \end{subfigure}\hfill
  \begin{subfigure}[t]{0.32\linewidth}
    \centering
    \includegraphics[width=\linewidth]{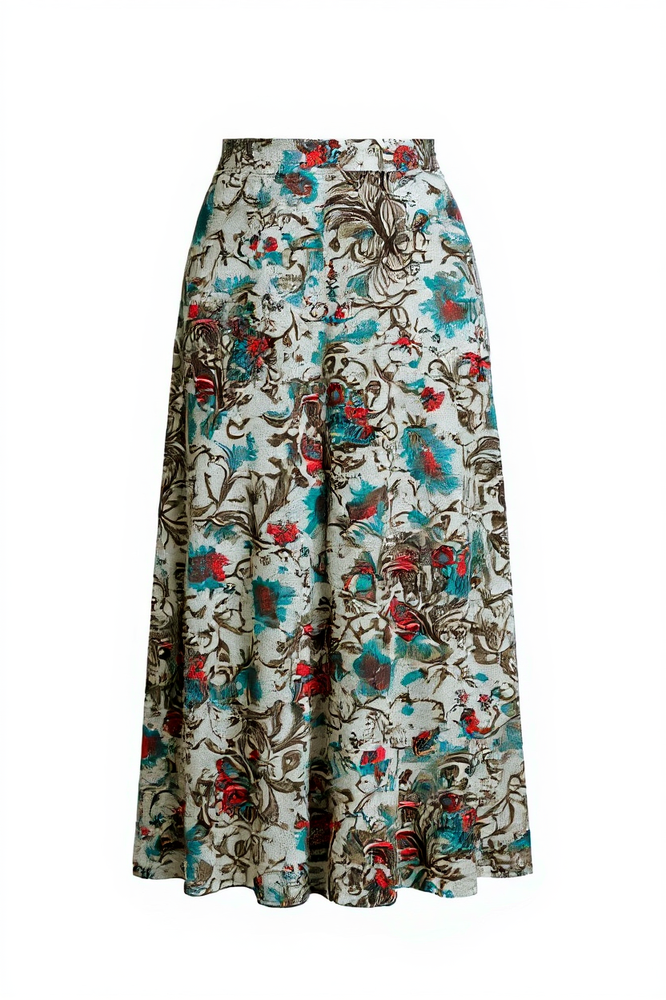}%
    \caption{\emph{``extract only the skirt over a white background, product photography style''}}
    \label{fig:skirt:b}
  \end{subfigure}\hfill
  \begin{subfigure}[t]{0.32\linewidth}
    \centering
    \includegraphics[width=\linewidth]{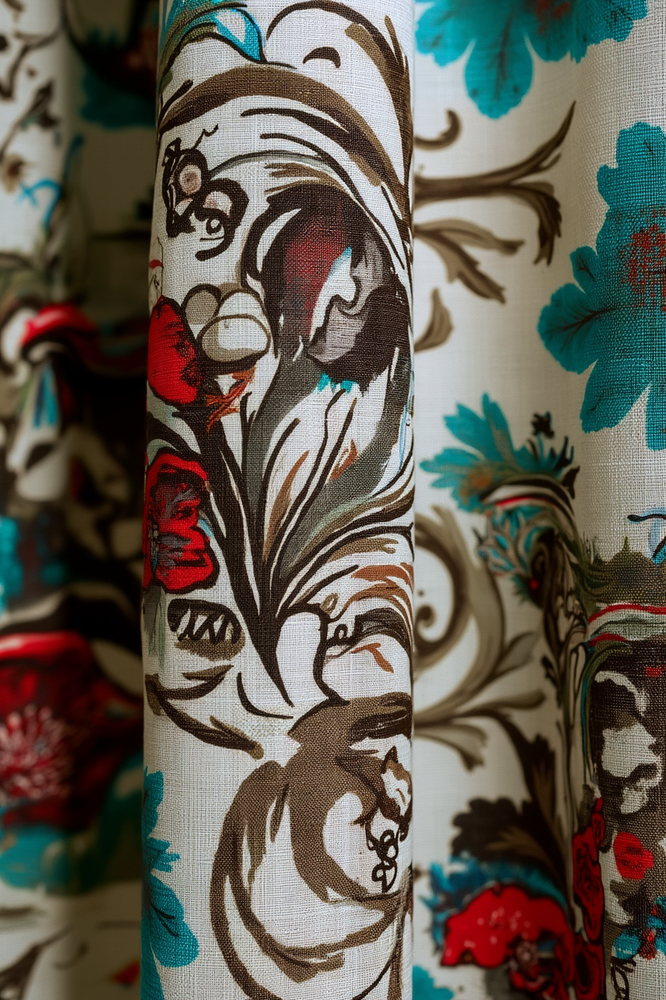}%
    \caption{\emph{``show me an extreme closeup of the fabric''}}
    \label{fig:skirt:c}
  \end{subfigure}

  \caption{\textbf{Product Photography.}
    (\subref{fig:skirt:a}) Input image showing the full outfit.
    (\subref{fig:skirt:b}) Extracted skirt on a white background in a product‐photography style.
    (\subref{fig:skirt:c}) Extreme close‐up of the skirt’s fabric, highlighting texture and pattern details.
  }
  \label{fig:skirttransform}
\end{figure}
}

\newcommand{\vasetransform}{%
\begin{figure}[t]
  \centering
  \begin{subfigure}[t]{0.58\linewidth}
    \centering
    \includegraphics[width=\linewidth]{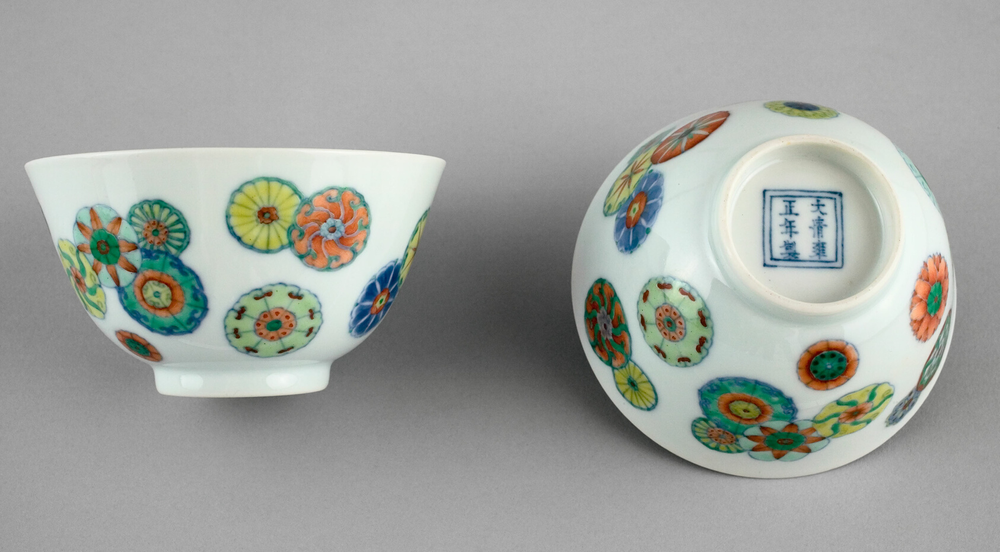}%
    \caption{Input image}
    \label{fig:vase:a}
  \end{subfigure}

  \vspace{6pt}

  \begin{subfigure}[t]{0.46\linewidth}
    \centering
    \includegraphics[width=\linewidth]{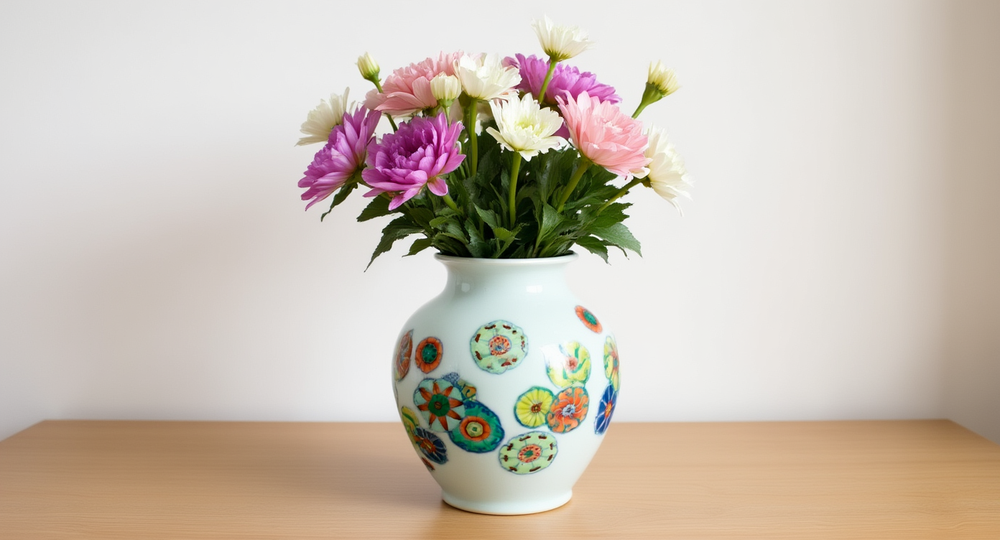}%
    \caption{\emph{``make me a matching flower vase, product photography set against a white wall, sitting on a wooden desk, put some nice flowers in it''}}
    \label{fig:vase:b}
  \end{subfigure}\hfill
  \begin{subfigure}[t]{0.46\linewidth}
    \centering
    \includegraphics[width=\linewidth]{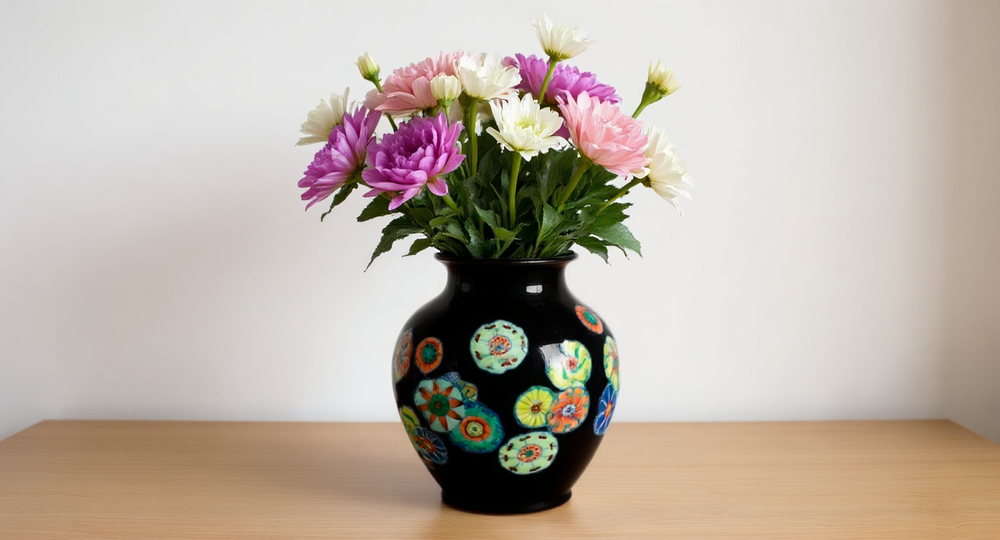}%
    \caption{\emph{``change the vase base color to black''}}
    \label{fig:vase:c}
  \end{subfigure}

  \caption{\textbf{Iterative, product-style editing.}
  Starting from the reference bowl~(\subref{fig:vase:a}), our model first
  generates a matching flower vase in a tabletop studio setting with fresh
  flowers~(\subref{fig:vase:b}), and subsequently changes the vase’s base
  color to black while preserving the floral pattern, lighting, and
  composition~(\subref{fig:vase:c}).}
  \label{fig:vase_edits}
\end{figure}
}

\newcommand{\laughtransform}{%
\begin{figure}[t]
  \centering
  \begin{subfigure}[t]{0.32\linewidth}
    \centering
    \includegraphics[width=\linewidth]{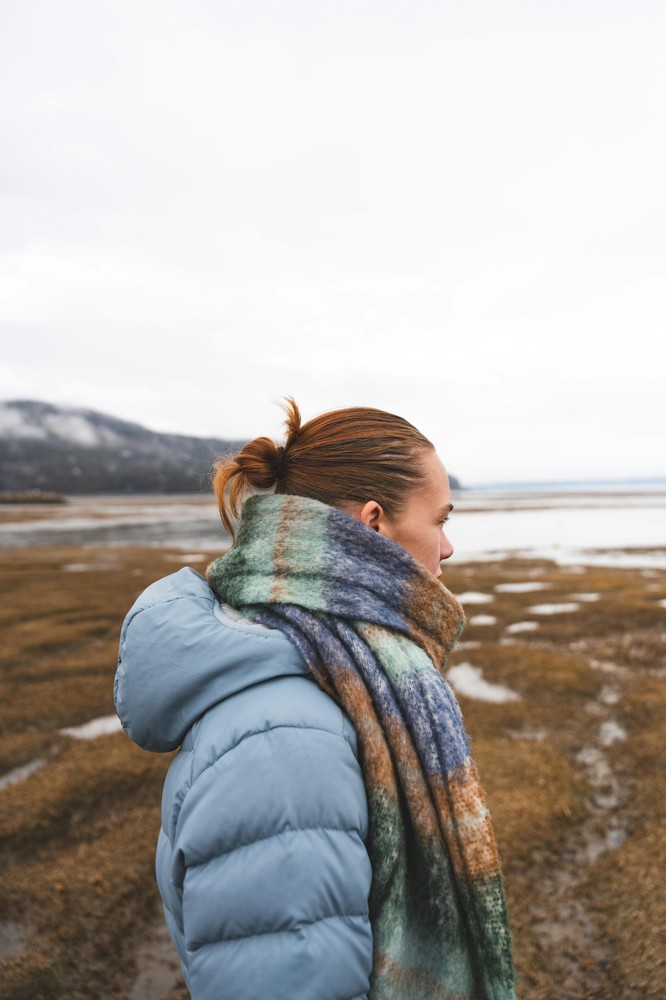}%
    \caption{Input image}
    \label{fig:laugh:a}
  \end{subfigure}\hfill
  \begin{subfigure}[t]{0.32\linewidth}
    \centering
    \includegraphics[width=\linewidth]{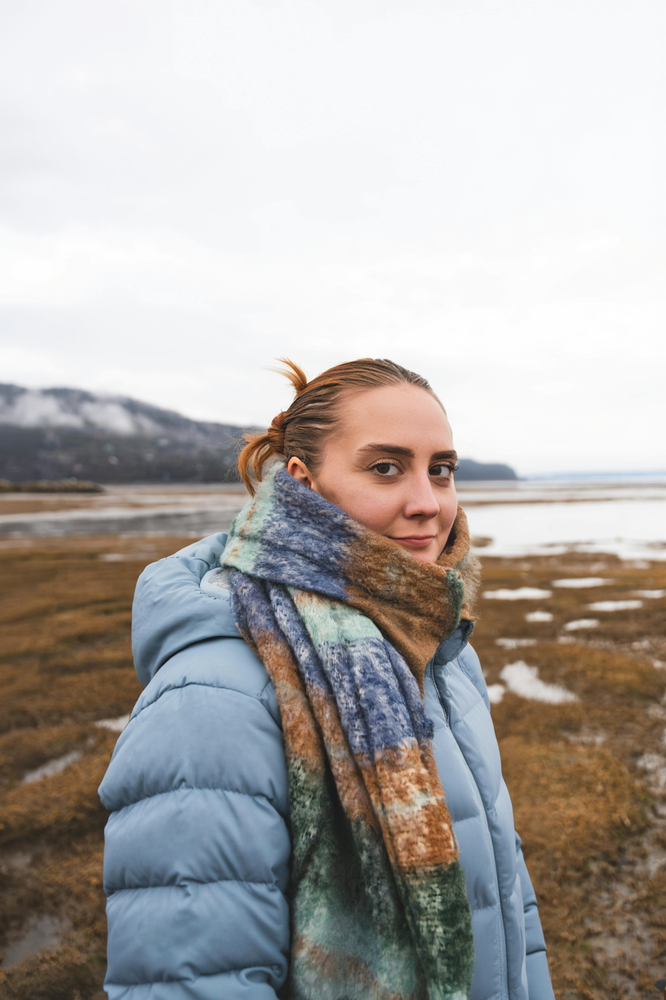}%
    \caption{\emph{``tilt her head towards the camera''}}
    \label{fig:laugh:b}
  \end{subfigure}\hfill
  \begin{subfigure}[t]{0.32\linewidth}
    \centering
    \includegraphics[width=\linewidth]{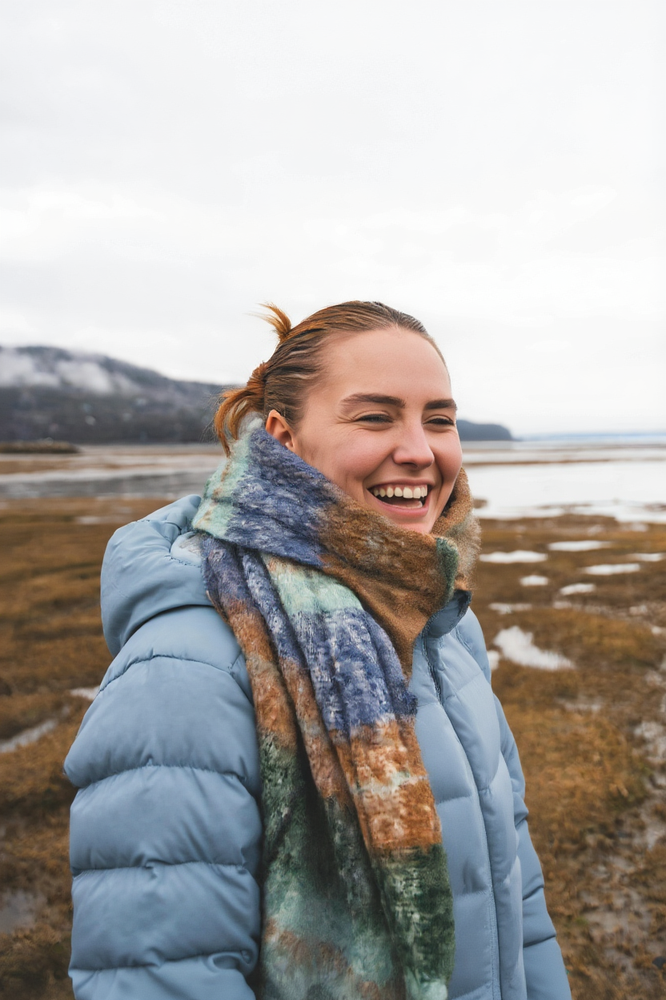}%
    \caption{\emph{``make her laugh''}}
    \label{fig:laugh:c}
  \end{subfigure}
  \caption{\textbf{Sequential, facial-expression editing.}
  Beginning with the profile reference~(\subref{fig:laugh:a}), our model first
  reorients the subject toward the camera~(\subref{fig:laugh:b}) and then
  changes her expression to a spontaneous laugh~(\subref{fig:laugh:c}),
  while preserving background, clothing and lighting.
  \label{fig:laugh_edits}}
\end{figure}}

\newcommand{\kontexttimings}{%
\begin{figure}[t]
  \centering
  \begin{subfigure}[t]{0.48\textwidth}
    \centering
    \includegraphics[width=\textwidth]{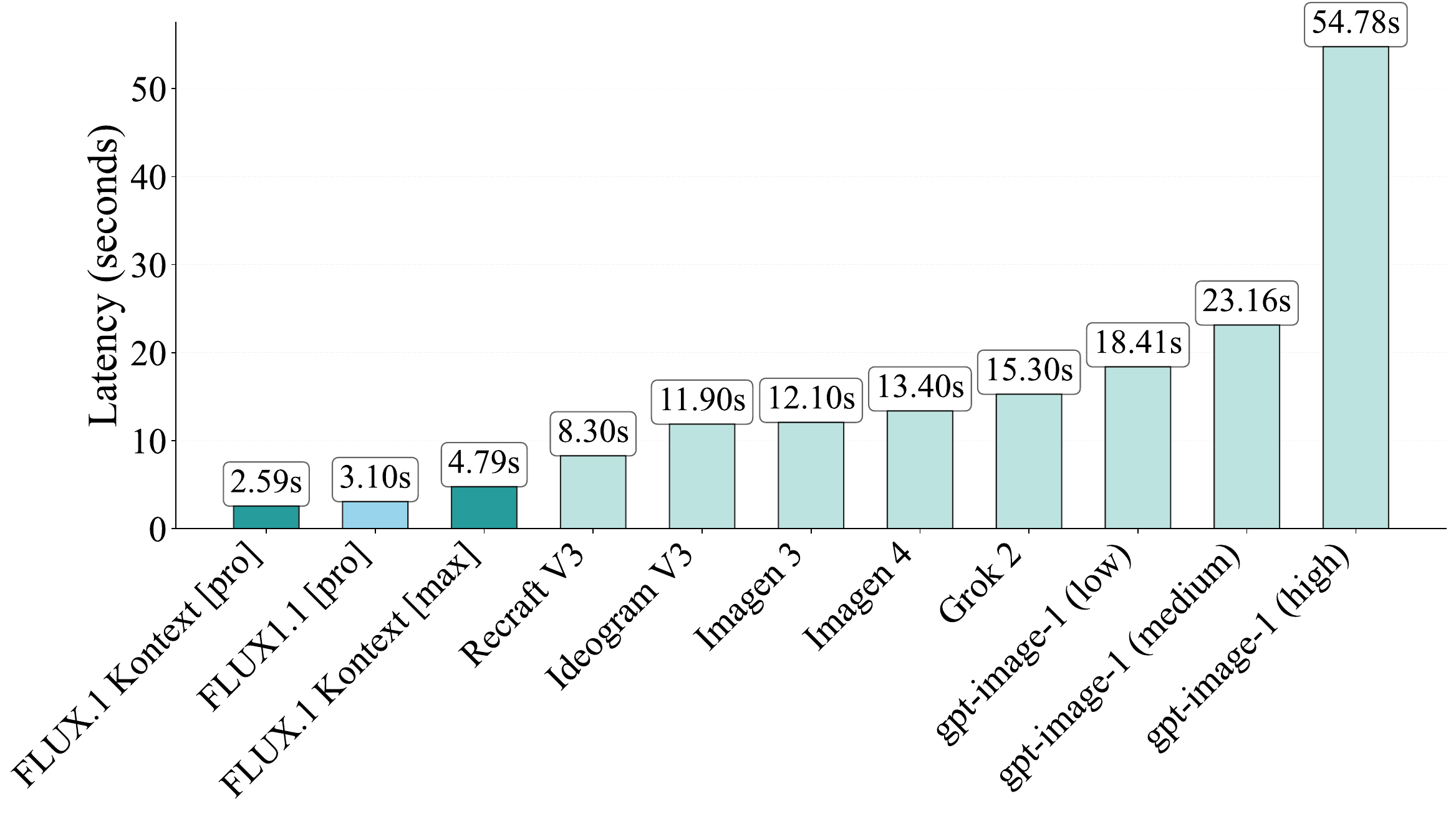}
    \caption{Text-to-image inference latency}
    \label{fig:kontexttimings:t2i}
  \end{subfigure}
  \hfill
  \begin{subfigure}[t]{0.48\textwidth}
    \centering
    \includegraphics[width=\textwidth]{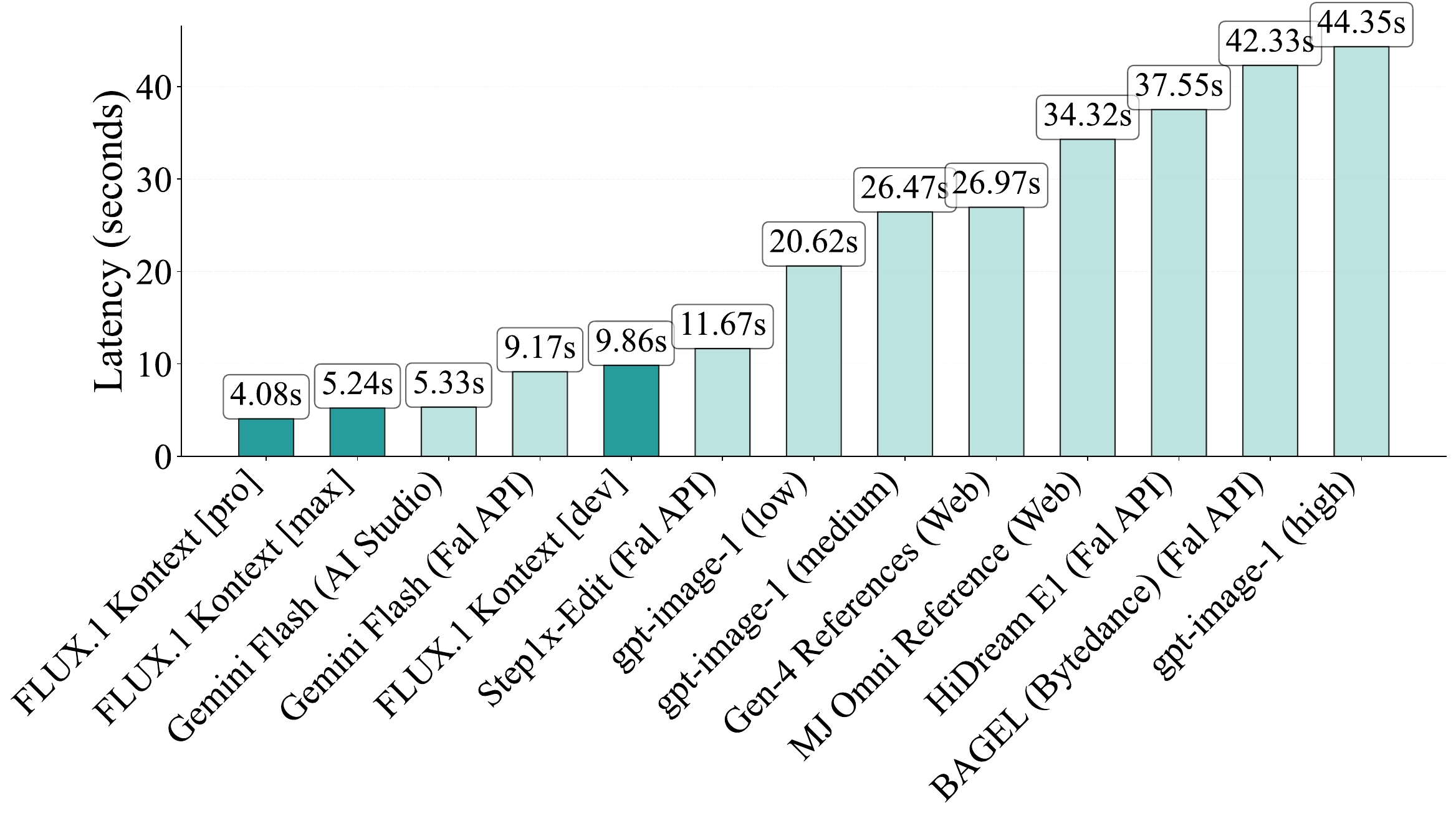}
    \caption{Image-to-image inference latency}
    \label{fig:kontexttimings:i2i}
  \end{subfigure}
  \caption{Median inference latency [seconds] for $1024 \times 1024$ generation across models (lower is better). FLUX.1 Kontext achieves competitive speeds for both text-to-image and image-to-image tasks.}
  \label{fig:kontexttimings}
\end{figure}}

\usepackage{subcaption}
\usepackage{pgfplots}
\pgfplotsset{compat=1.18}

\definecolor{kontextblue}{RGB}{70,130,180}
\definecolor{gptred}{RGB}{220,20,60}
\definecolor{genforange}{RGB}{255,140,0}
\definecolor{geminipurple}{RGB}{138,43,226}
\definecolor{kontextgreen}{RGB}{34,139,34}
\definecolor{hidreamteal}{RGB}{0,128,128}
\definecolor{stepbrown}{RGB}{160,82,45}
\definecolor{mjmagenta}{RGB}{199,21,133}

\pgfplotsset{
  KontextBar/.style={
    ybar,
    bar width=6pt,
    enlarge x limits=0.15,
    xlabel={},
    ylabel style={font=\small},
    x tick label style={font=\scriptsize, align=center},
    tick label style={font=\scriptsize},
    nodes near coords,
    nodes near coords style={font=\tiny, anchor=south},
    width=\linewidth,
    height=5cm,
  }
}

\pgfplotsset{compat=1.18}
\usetikzlibrary{pgfplots.groupplots}

\definecolor{kontextblue}{RGB}{70,130,180}
\definecolor{gptred}{RGB}{220,20,60}
\definecolor{genforange}{RGB}{255,140,0}
\definecolor{geminipurple}{RGB}{138,43,226}
\definecolor{kontextgreen}{RGB}{34,139,34}
\definecolor{hidreamteal}{RGB}{0,128,128}
\definecolor{stepbrown}{RGB}{160,82,45}
\definecolor{mjmagenta}{RGB}{199,21,133}

\pgfplotsset{
  kontextbar/.style={
    ybar,
    bar width=10pt,
    enlarge x limits=0.20,
    axis x line=bottom,
    axis y line=left,
    axis line style={gray!50},
    tick style       ={gray!50},
    ymajorgrids,
    grid style       ={dashed,gray!30},
    tick label style ={font=\footnotesize,/pgf/number format/1000 sep={}},
    x tick label style={font=\scriptsize,align=center},
    nodes near coords,
    nodes near coords style={font=\scriptsize,above},
  }
}

\newcommand{\kontextbenchresults}{%
\begin{figure}[t]
  \centering
  \begin{subfigure}[t]{0.32\textwidth}
    \centering
    \includegraphics[width=\textwidth]{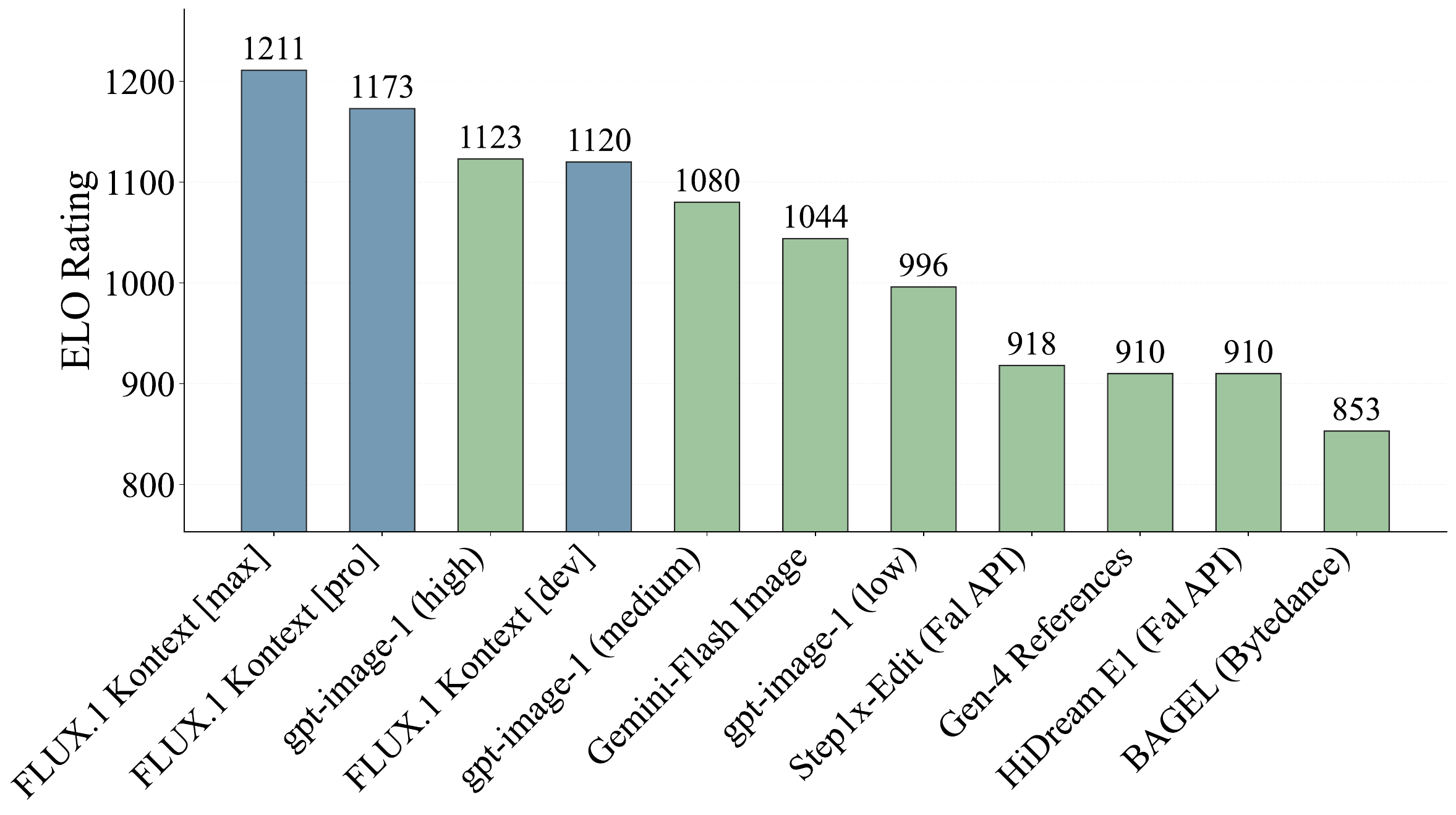}
    \caption{Text Editing}
    \label{fig:bench:textedit}
  \end{subfigure}
  \hfill
  \begin{subfigure}[t]{0.32\textwidth}
    \centering
    \includegraphics[width=\textwidth]{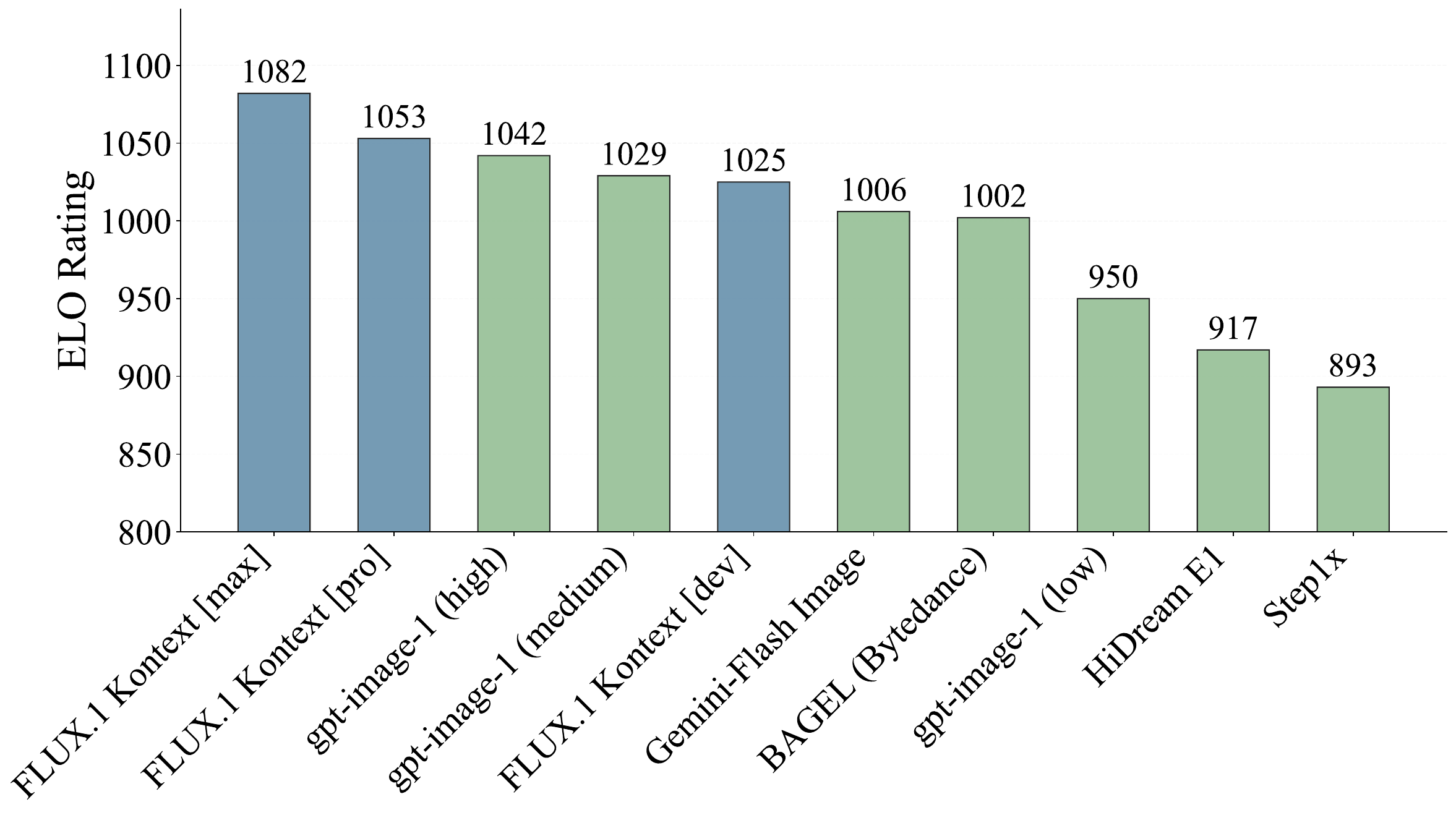}
    \caption{Local Editing}
    \label{fig:bench:localedit}
  \end{subfigure}
  \hfill
  \begin{subfigure}[t]{0.32\textwidth}
    \centering
    \includegraphics[width=\textwidth]{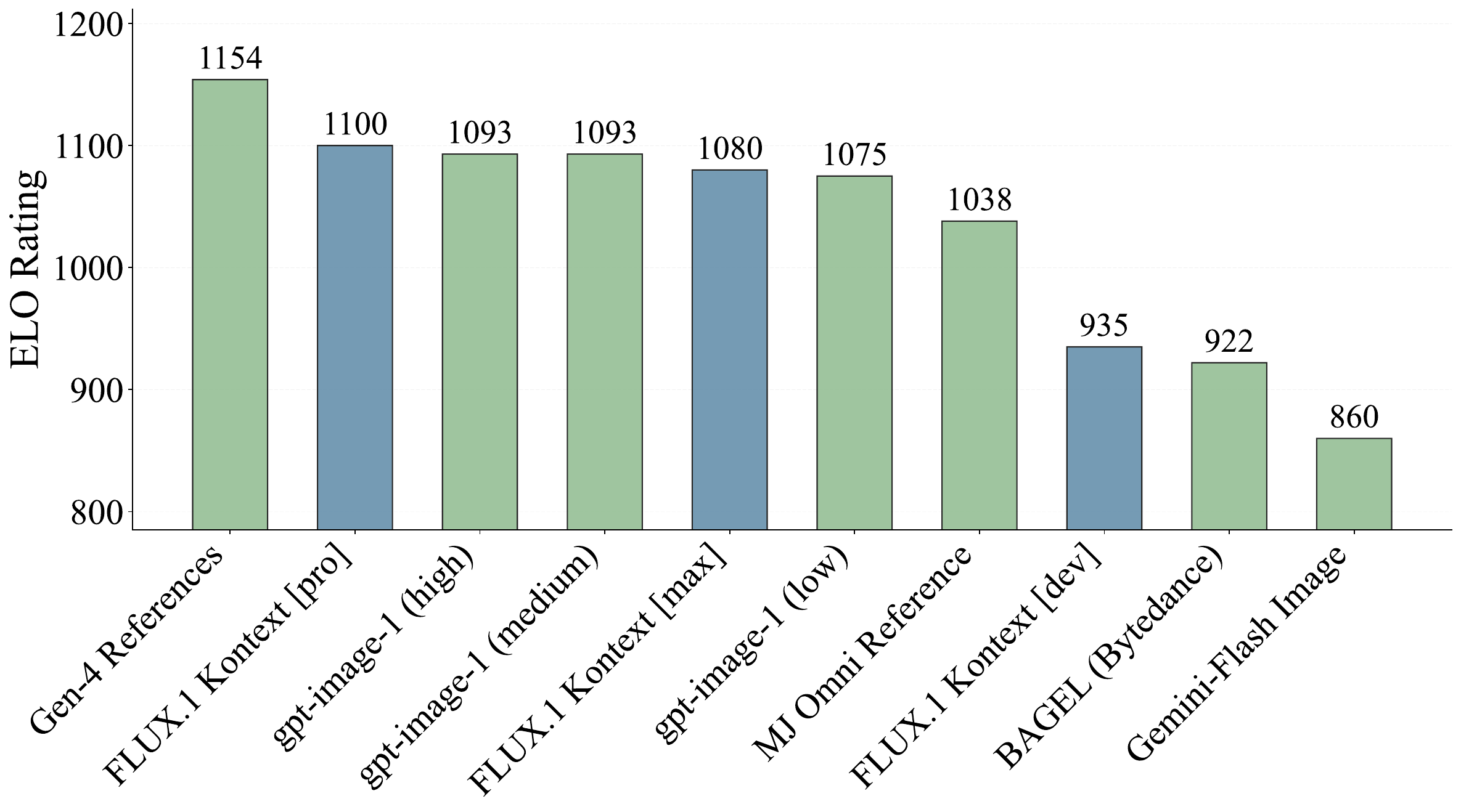}
    \caption{Style Reference}
    \label{fig:bench:styleref}
  \end{subfigure}

  \vspace{0.5cm}

  \begin{subfigure}[t]{0.32\textwidth}
    \centering
    \includegraphics[width=\textwidth]{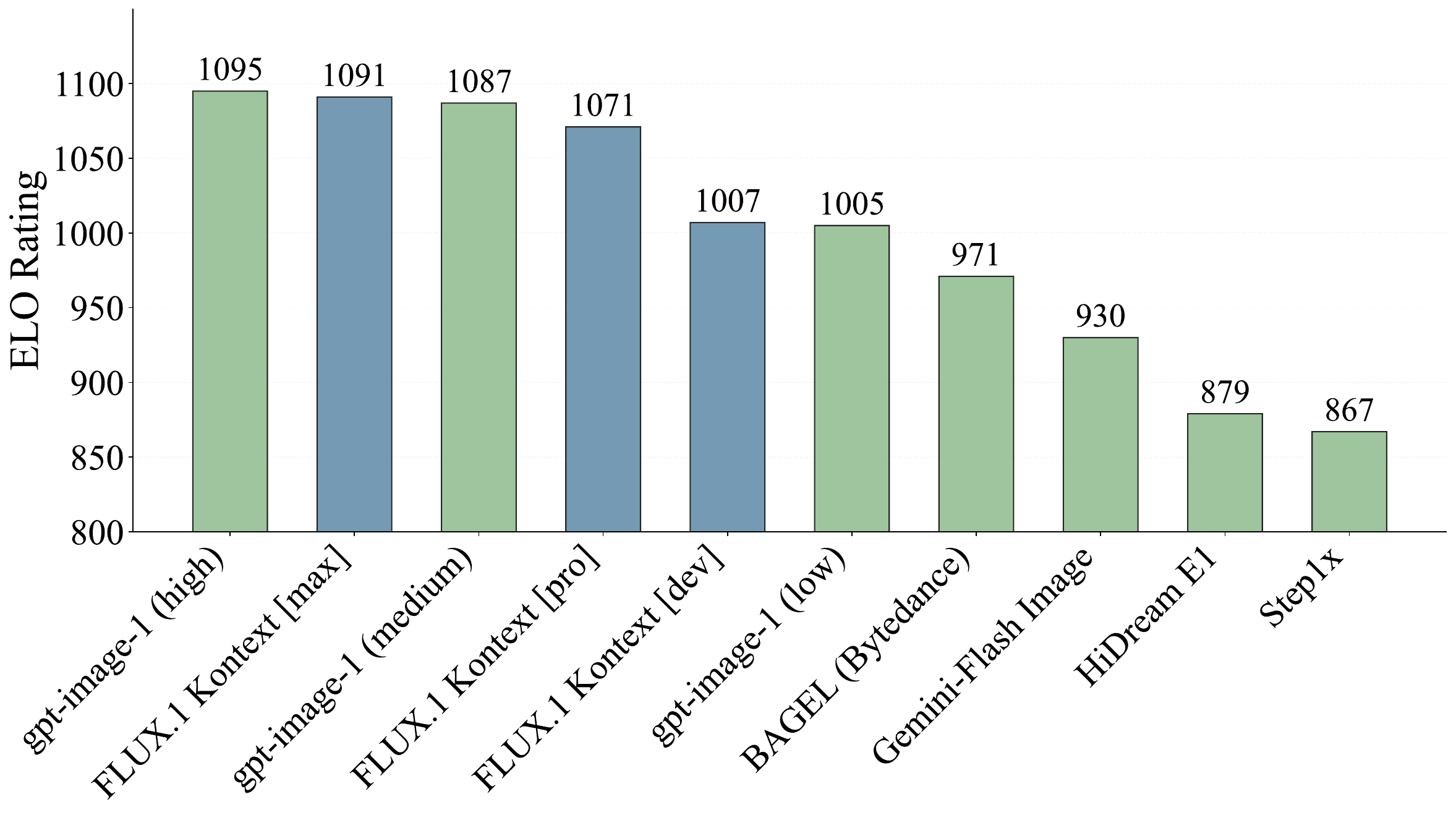}
    \caption{Global Editing}
    \label{fig:bench:globaledit}
  \end{subfigure}
  \hfill
  \begin{subfigure}[t]{0.32\textwidth}
    \centering
    \includegraphics[width=\textwidth]{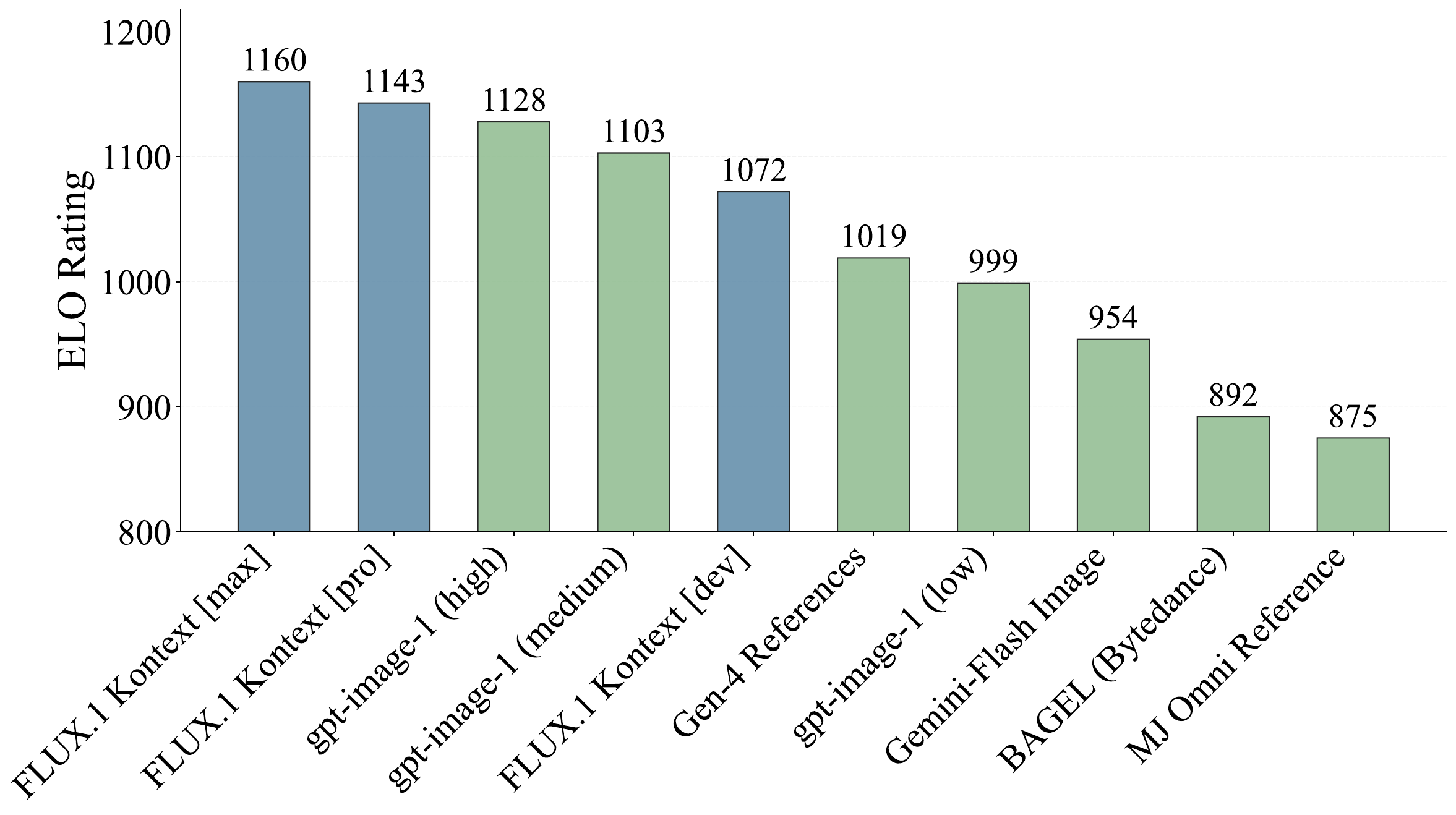}
    \caption{Character Reference}
    \label{fig:bench:charref}
  \end{subfigure}
  \hfill
  \begin{subfigure}[t]{0.32\textwidth}
    \centering
    \includegraphics[width=\textwidth]{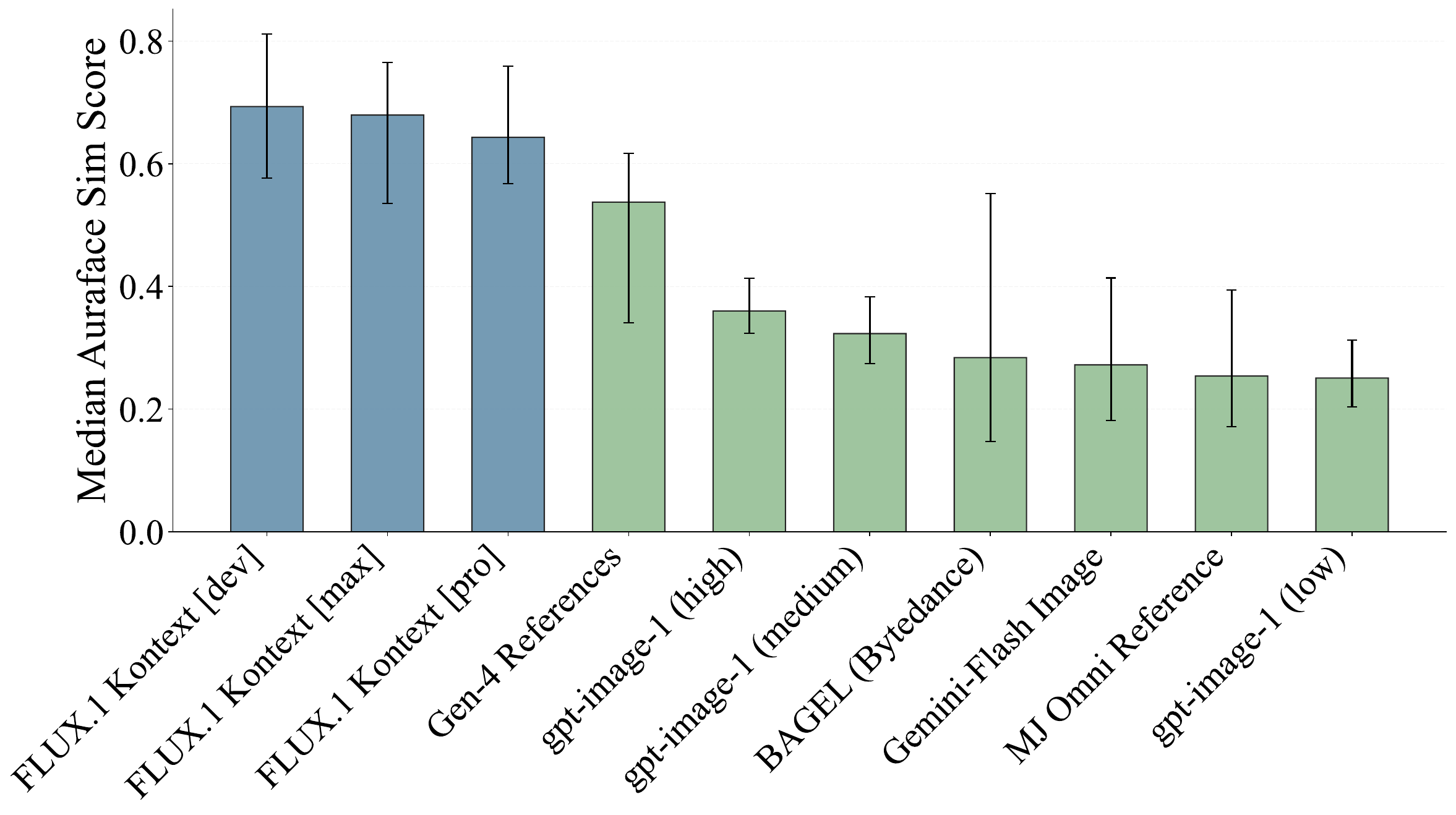}
    \caption{AuraFace Similarity}
    \label{fig:bench:auraface}
  \end{subfigure}

  \caption{\textbf{Image-to-image evaluation on KontextBench}. We show evaluation results across six in-context image generation tasks. \modelname [pro] consistently ranks among the top performers across all tasks, achieving the highest scores in Text Editing and Character Preservation.}
  \label{fig:kontextbenchresults}
\end{figure}}

\newcommand{\kontexttexttoimagebenchresults}{%
\begin{figure}[t]
  \centering
  \begin{subfigure}[t]{0.32\textwidth}
    \centering
    \includegraphics[width=\textwidth]{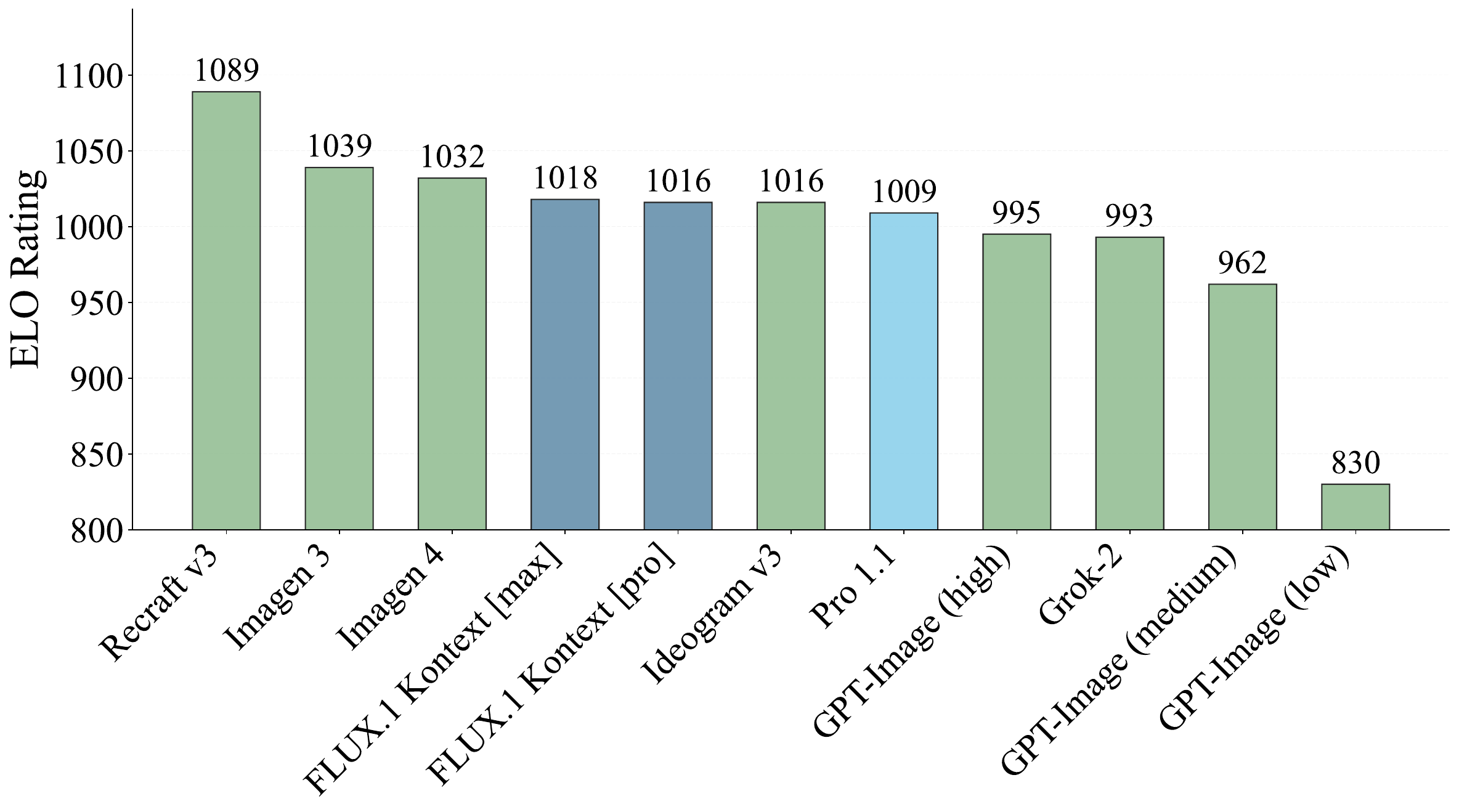}
    \caption{Aesthetics (Internal-T2I-Bench)}
    \label{fig:bench:t2iaesthetics}
  \end{subfigure}
  \hfill
  \begin{subfigure}[t]{0.32\textwidth}
    \centering
    \includegraphics[width=\textwidth]{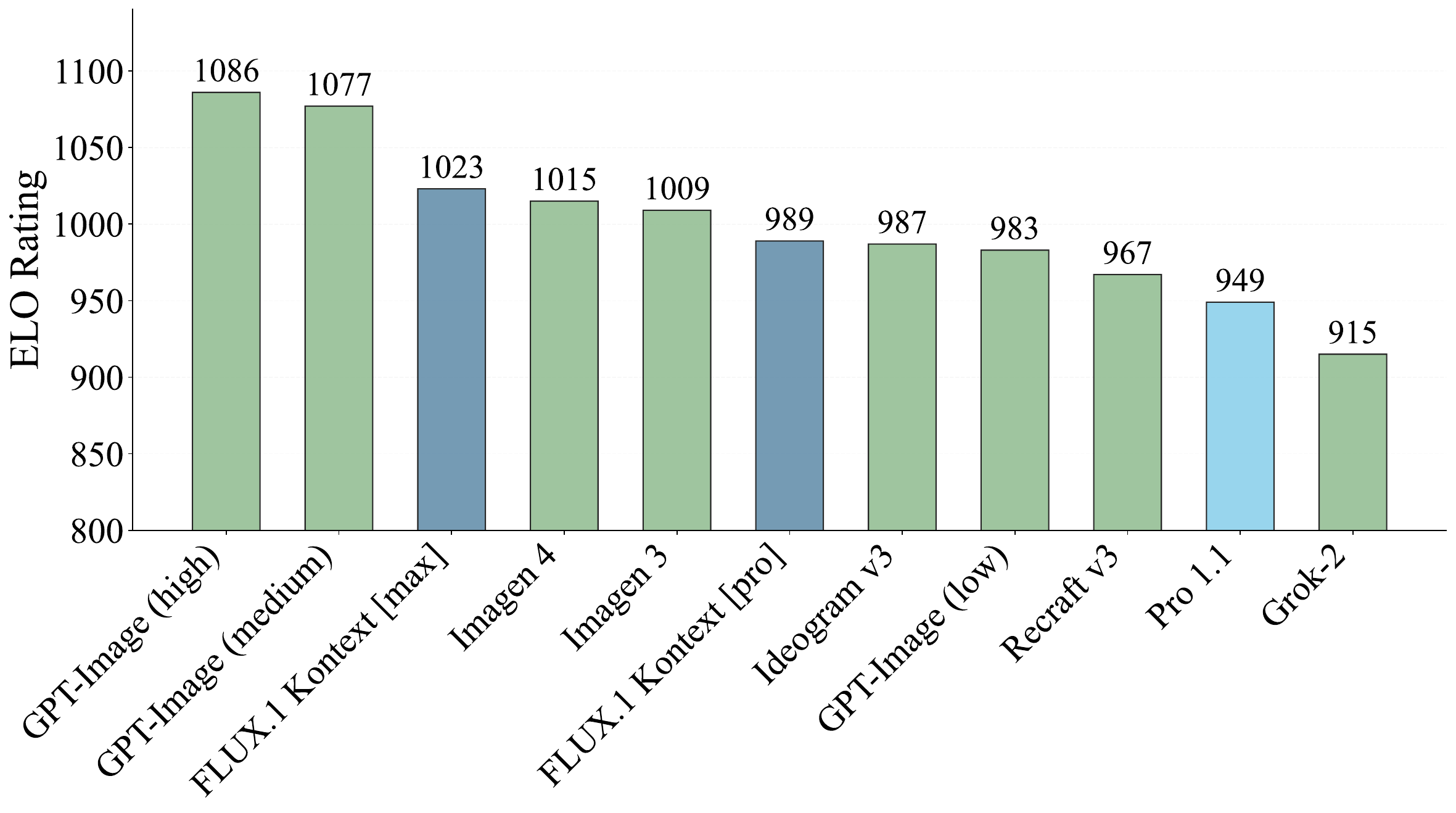}
    \caption{Prompt Following (Internal-T2I-Bench)}
    \label{fig:bench:t2iprompt}
  \end{subfigure}
  \hfill
  \begin{subfigure}[t]{0.32\textwidth}
    \centering
    \includegraphics[width=\textwidth]{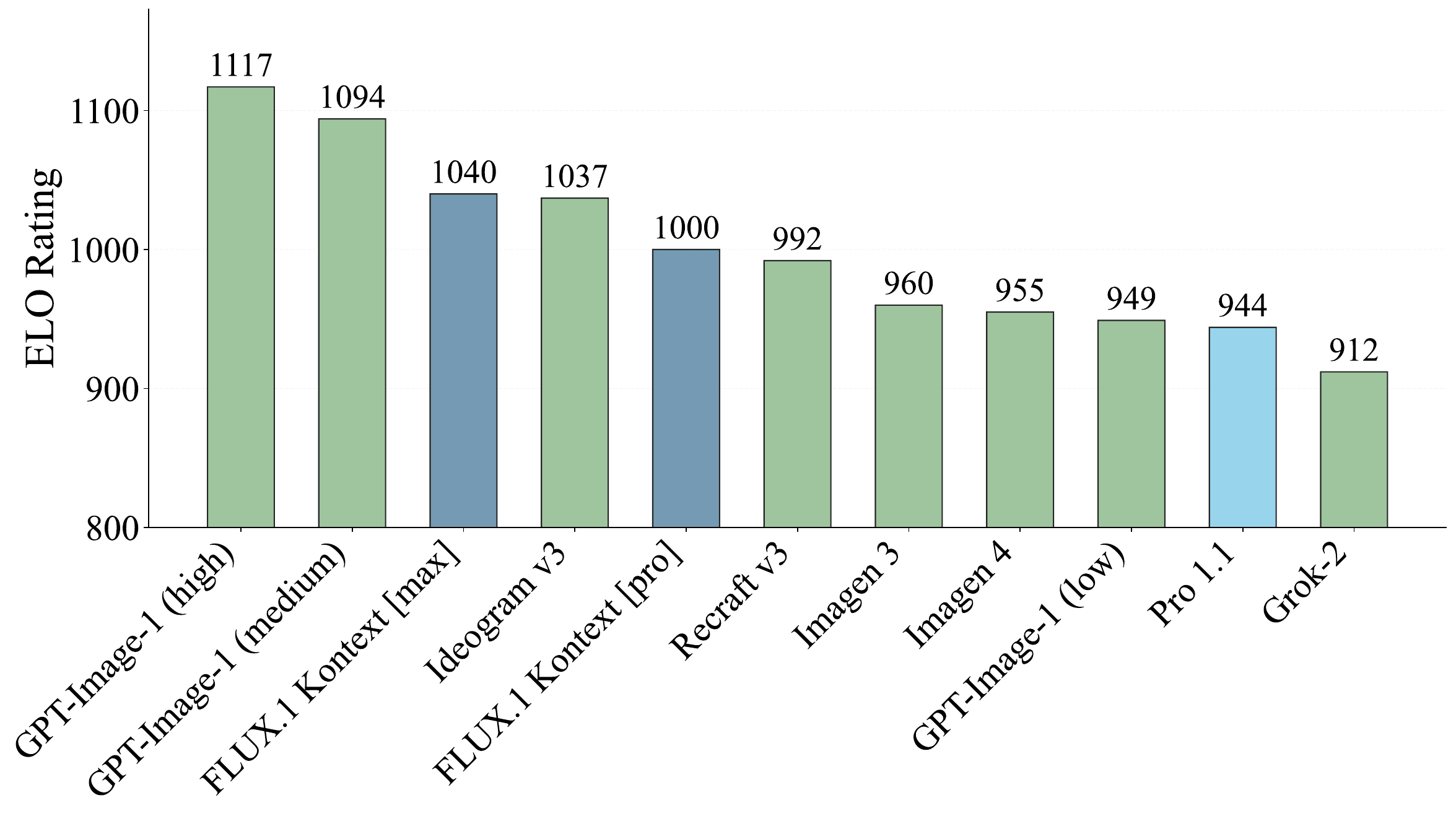}
    \caption{Typography (Internal-T2I-Bench)}
    \label{fig:bench:t2itypo}
  \end{subfigure}

  \vspace{0.5cm}

  \begin{subfigure}[t]{0.32\textwidth}
    \centering
    \includegraphics[width=\textwidth]{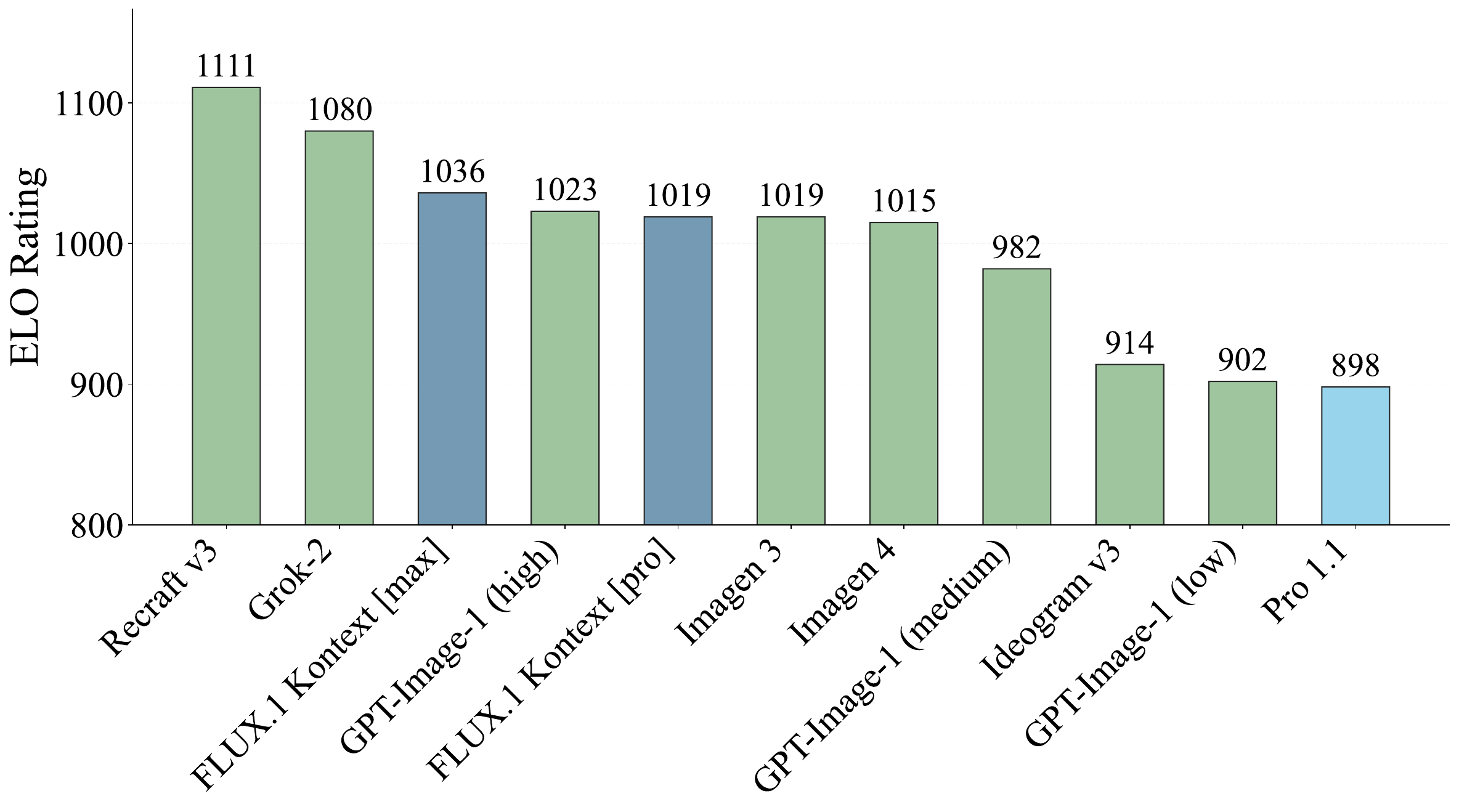}
    \caption{Realism (Internal-T2I-Bench)}
    \label{fig:bench:t2irealism}
  \end{subfigure}
  \hfill
  \begin{subfigure}[t]{0.32\textwidth}
    \centering
    \includegraphics[width=\textwidth]{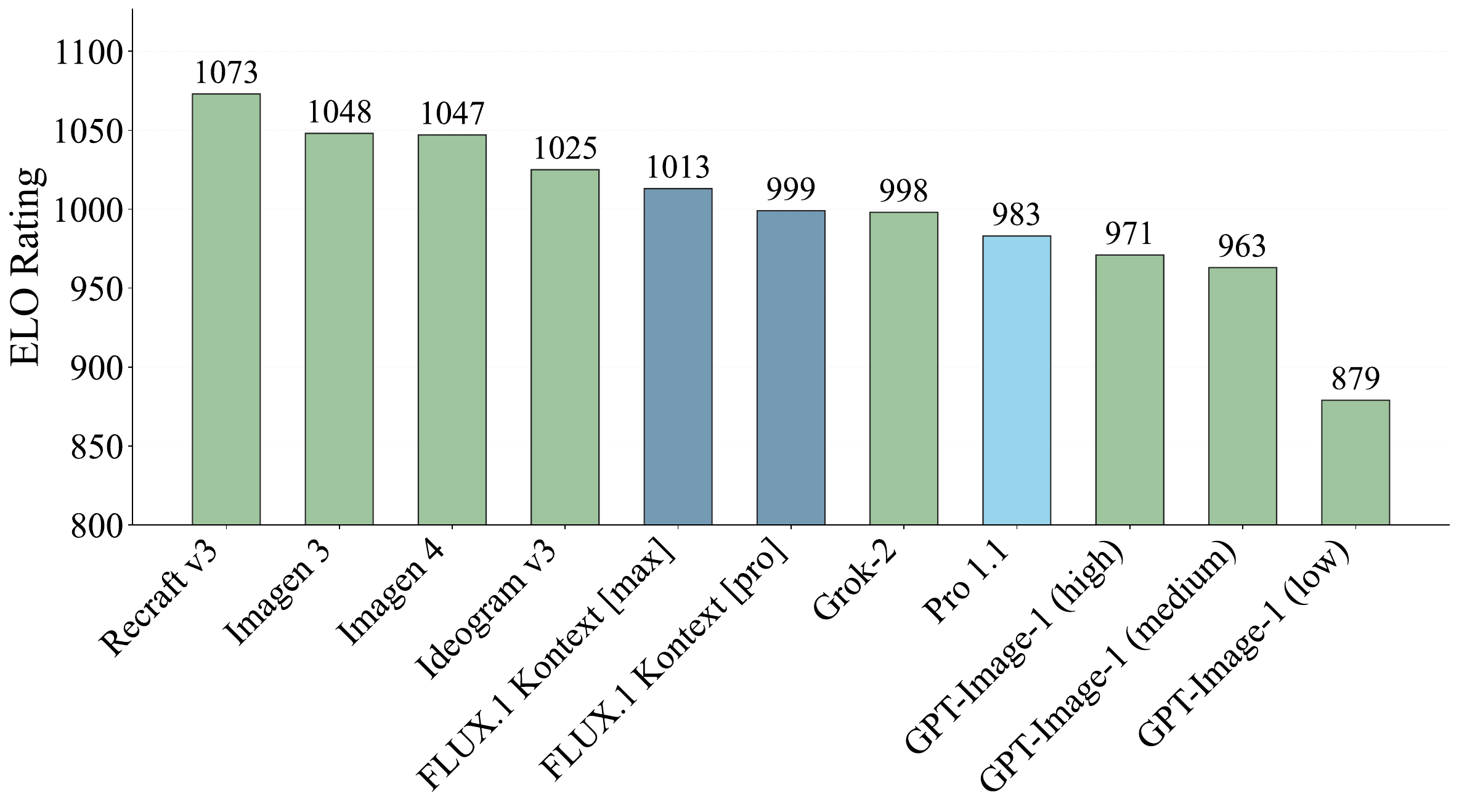}
    \caption{Aesthetics (GenAI)}
    \label{fig:bench:genaiaesthetics}
  \end{subfigure}
  \hfill
  \begin{subfigure}[t]{0.32\textwidth}
    \centering
    \includegraphics[width=\textwidth]{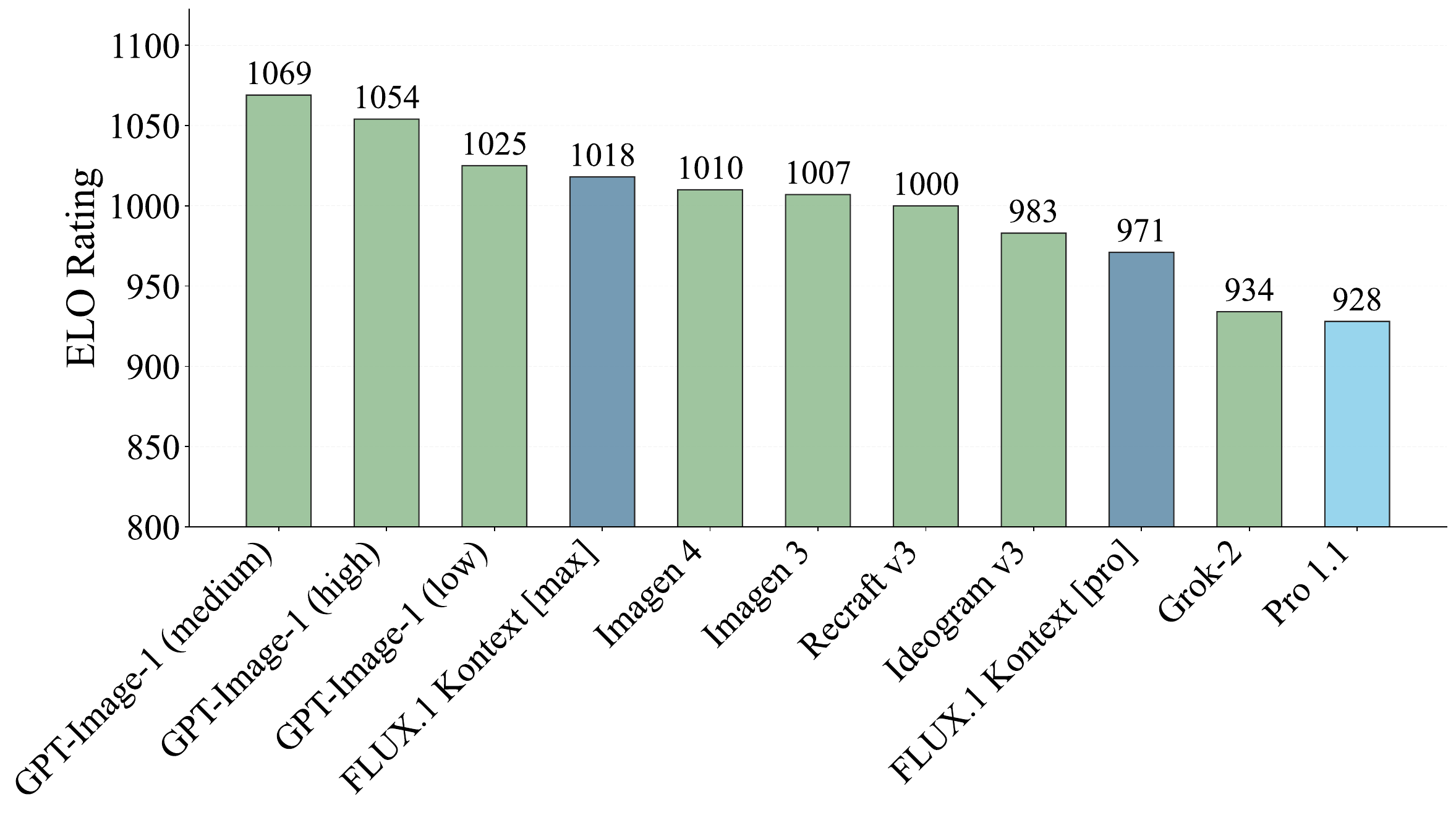}
    \caption{Prompt Following (GenAI)}
    \label{fig:bench:genaiprompt}
  \end{subfigure}

  \caption{
  \textbf{Text-to-image evaluation on Internal-t2i-bench.} We report evaluation results across multiple quality dimensions. \modelname models demonstrate competitive performance across aesthetics, prompt following, typography, and realism benchmarks.}
  \label{fig:kontextt2ibenchresults}
\end{figure}}

\newcommand{\failureone}{
\begin{figure}[t]
  \centering
  \begin{subfigure}[t]{0.32\linewidth}
    \centering
    \includegraphics[width=\linewidth,height=3.3cm,keepaspectratio]{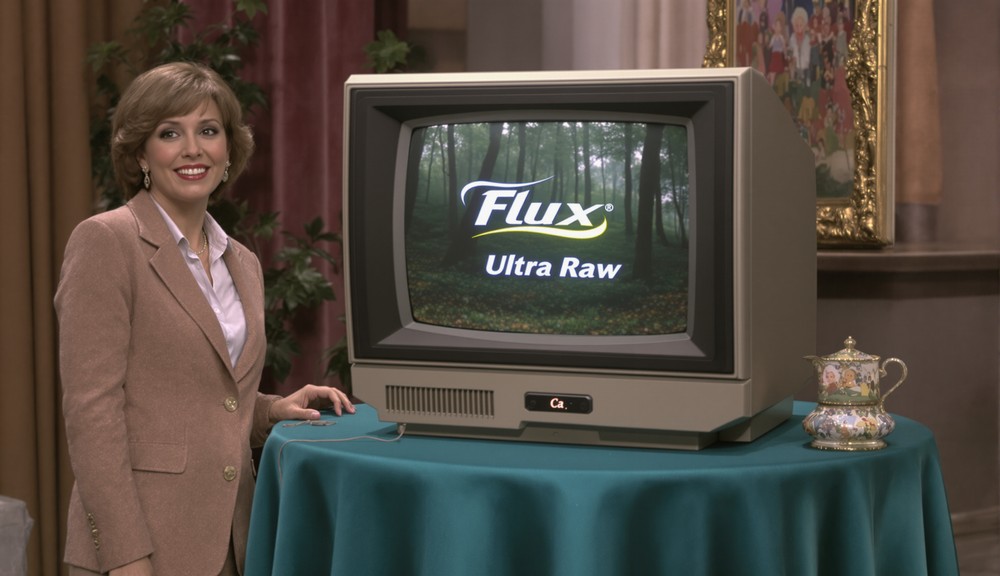}
    \caption{Input image}
    \label{fig:failure:a}
  \end{subfigure}\hfill
  \begin{subfigure}[t]{0.32\linewidth}
    \centering
    \includegraphics[width=\linewidth,height=3cm,keepaspectratio]{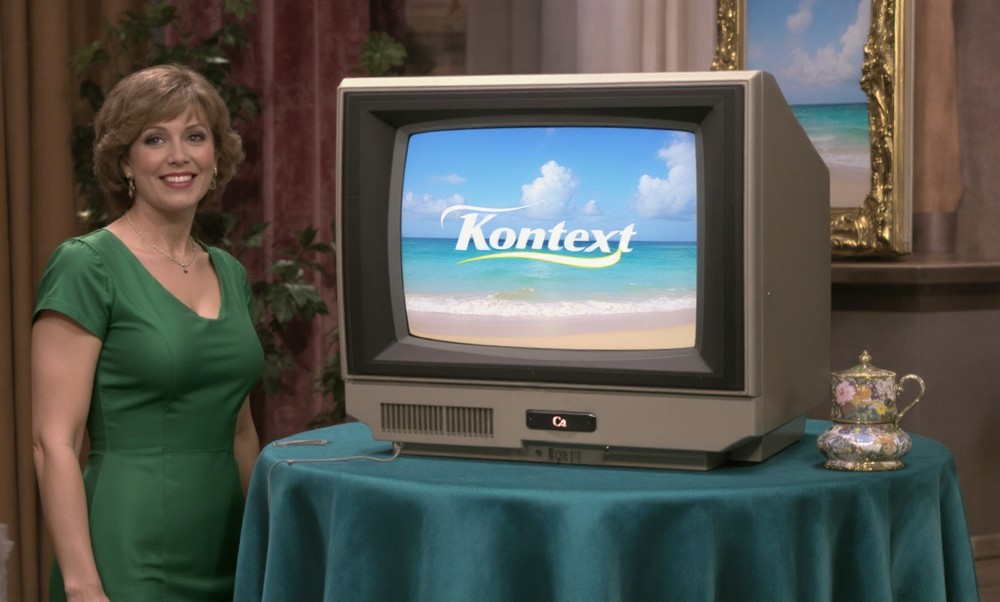}
    \caption{\emph{``The woman is now wearing a green dress, the painting in the back now shows a beach scene, the text on the TV says "Kontext" now''}}
    \label{fig:failure:c}
  \end{subfigure}\hfill
  \begin{subfigure}[t]{0.32\linewidth}
    \centering
    \includegraphics[width=\linewidth,height=3cm,keepaspectratio]{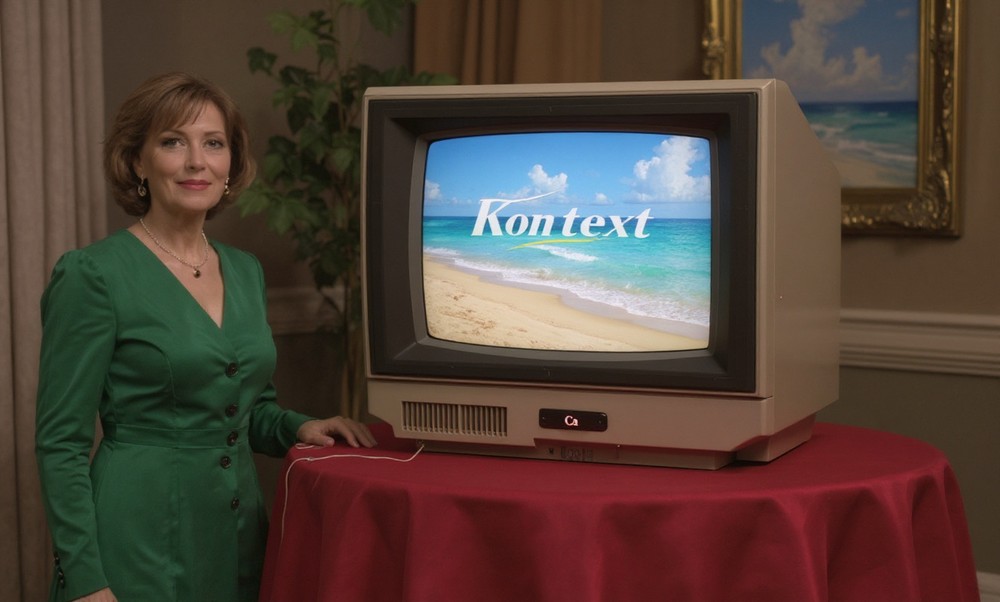}
    \caption{\emph{``...green dress...beach scene...tv says "Kontext"...table cloth is also now red and the lighting is a bit warmer''}}
    \label{fig:failure:d}
  \end{subfigure}

  \vspace{0.5cm}
  \begin{subfigure}[t]{0.48\linewidth}
    \centering
    \includegraphics[width=\linewidth,keepaspectratio]{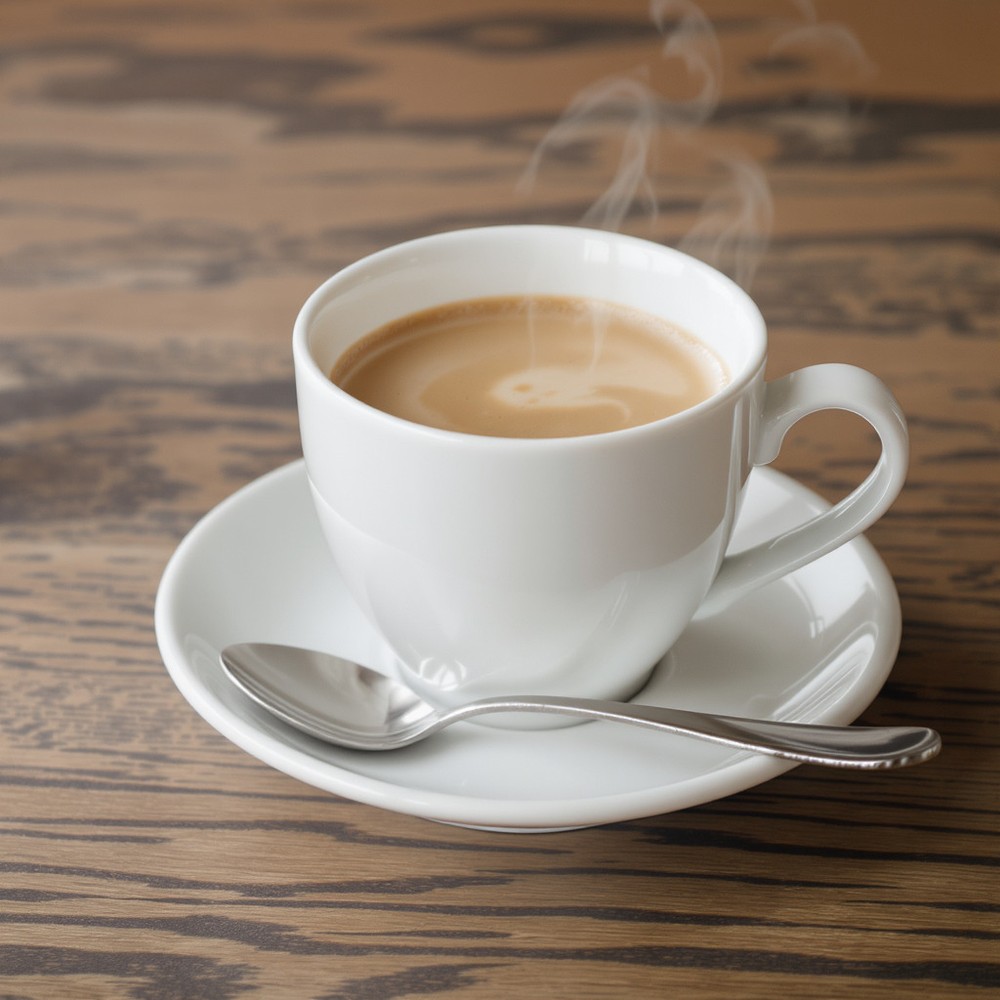}
    \caption{Input image}
    \label{fig:failure:e}
  \end{subfigure}\hfill
  \begin{subfigure}[t]{0.48\linewidth}
    \centering
    \includegraphics[width=\linewidth,keepaspectratio]{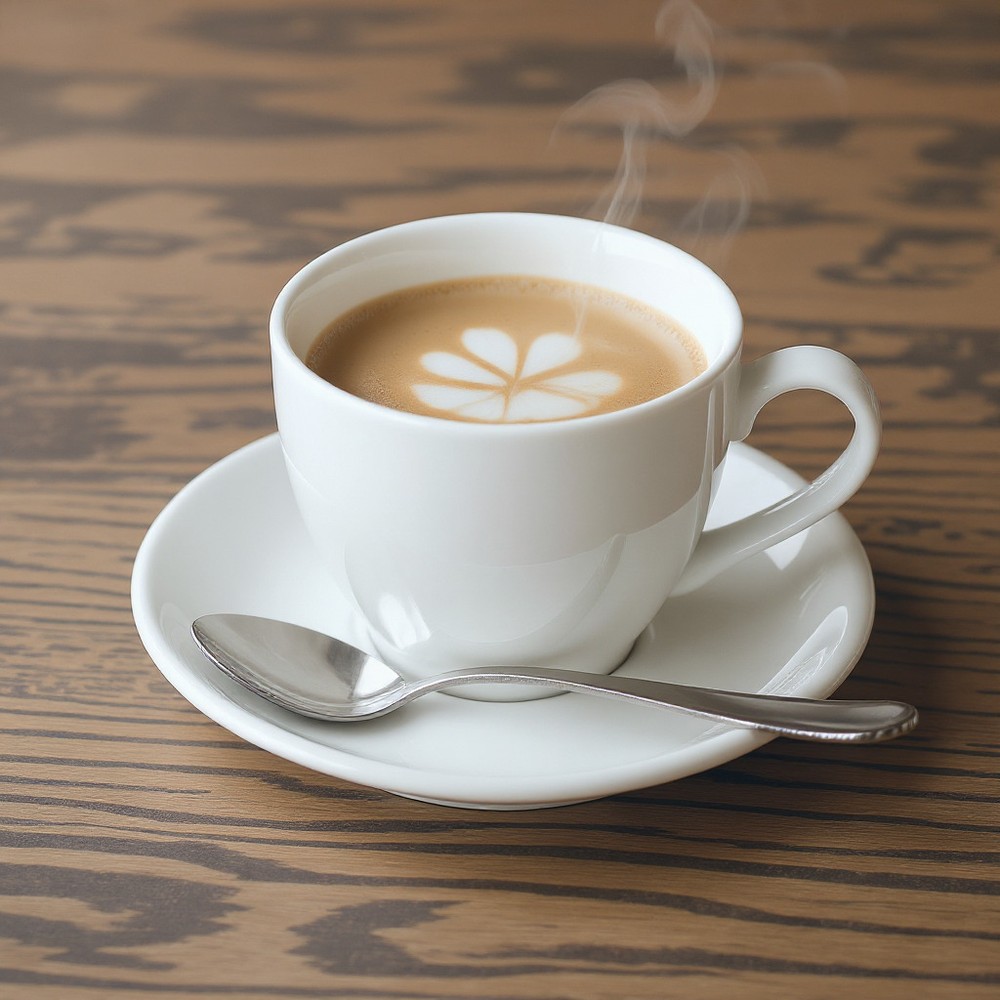}
    \caption{\emph{``move the coffee to the left''}}
    \label{fig:failure:f}
  \end{subfigure}

  \vspace{0.5cm}
  \begin{subfigure}[t]{0.48\linewidth}
    \centering
    \includegraphics[width=\linewidth,keepaspectratio]{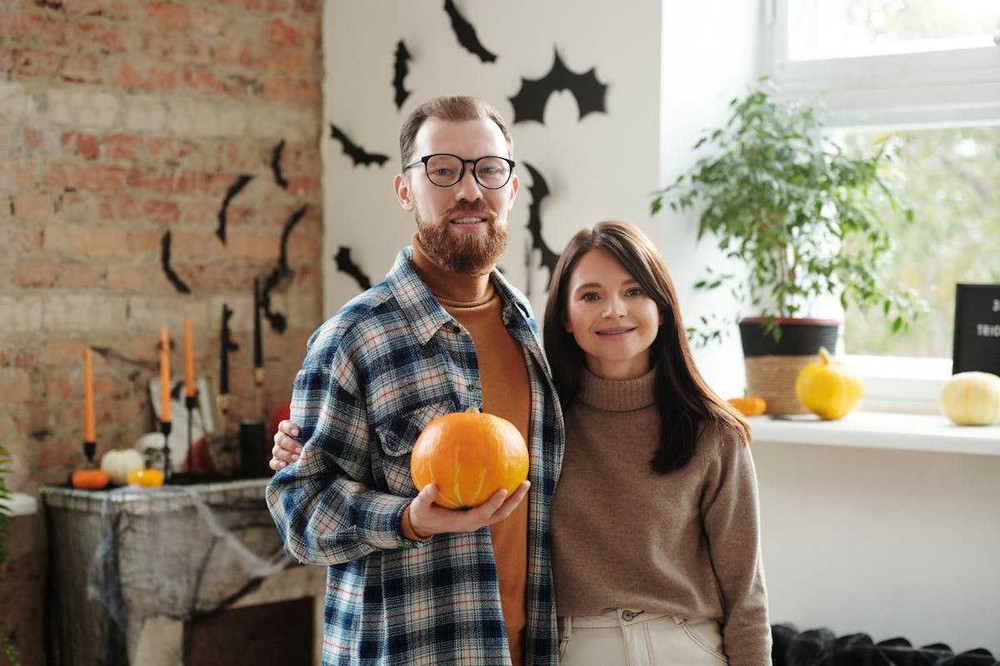}
    \caption{Input image}
    \label{fig:failure:e}
  \end{subfigure}\hfill
  \begin{subfigure}[t]{0.48\linewidth}
    \centering
    \includegraphics[width=\linewidth,keepaspectratio]{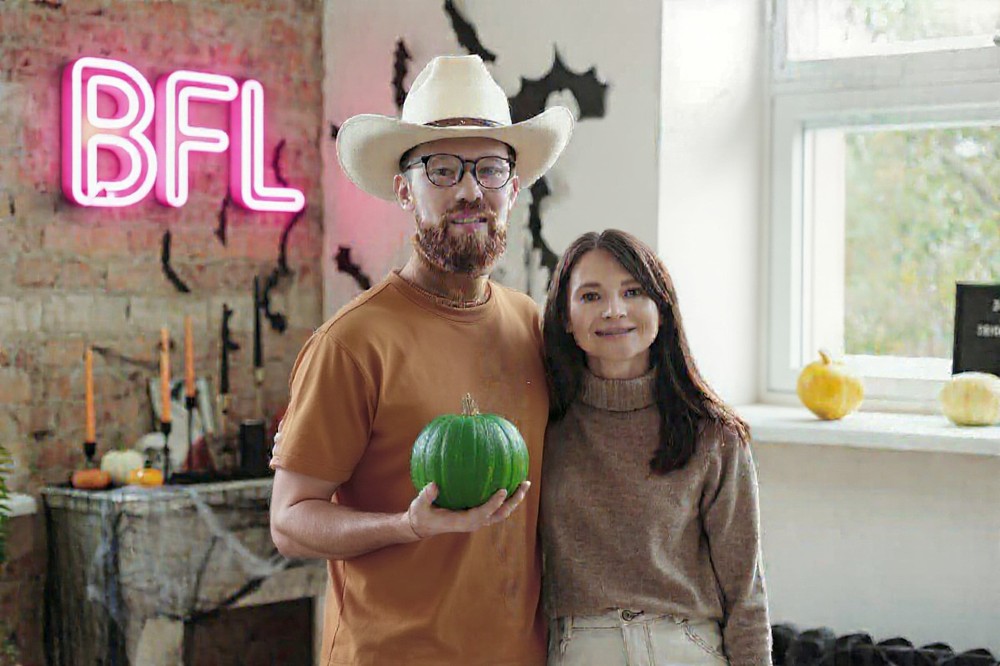}
    \caption{After six iterative edits.}
    \label{fig:failure:f}
  \end{subfigure}

  \caption{Editing failure cases. \emph{Top row:} An example for identity degradation: While the center image shows a good edit, preserving character identity the right one (using a slightly modified prompt) comes with significant identity loss. \emph{Middle row:} Scene modification instead of object movement: the model adds milk foam rather than repositioning the mug.
  \emph{Bottom row:} After six iterative edits, samples can exhibit visible artifacts.
  }
  \label{fig:failureone}
\end{figure}
}

\newcommand{\visualcues}{
\begin{figure}[t]
  \centering
  \begin{subfigure}[t]{0.16\linewidth}
    \centering
    \includegraphics[height=3cm]{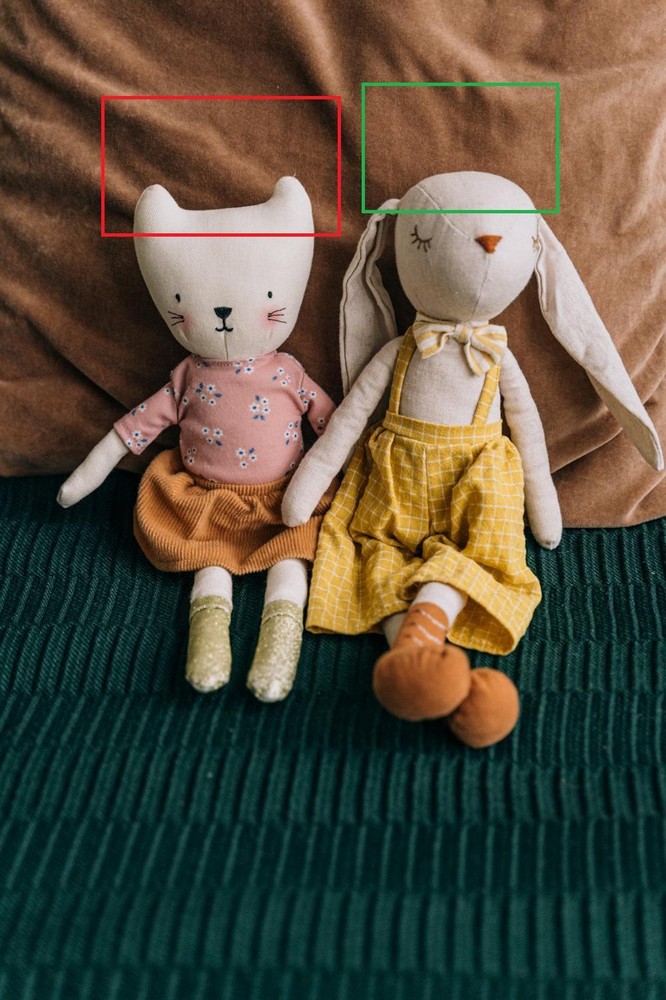}
    \caption{Input image}
    \label{fig:textedit:a}
  \end{subfigure}\hfill
  \begin{subfigure}[t]{0.16\linewidth}
    \centering
    \includegraphics[height=3cm]{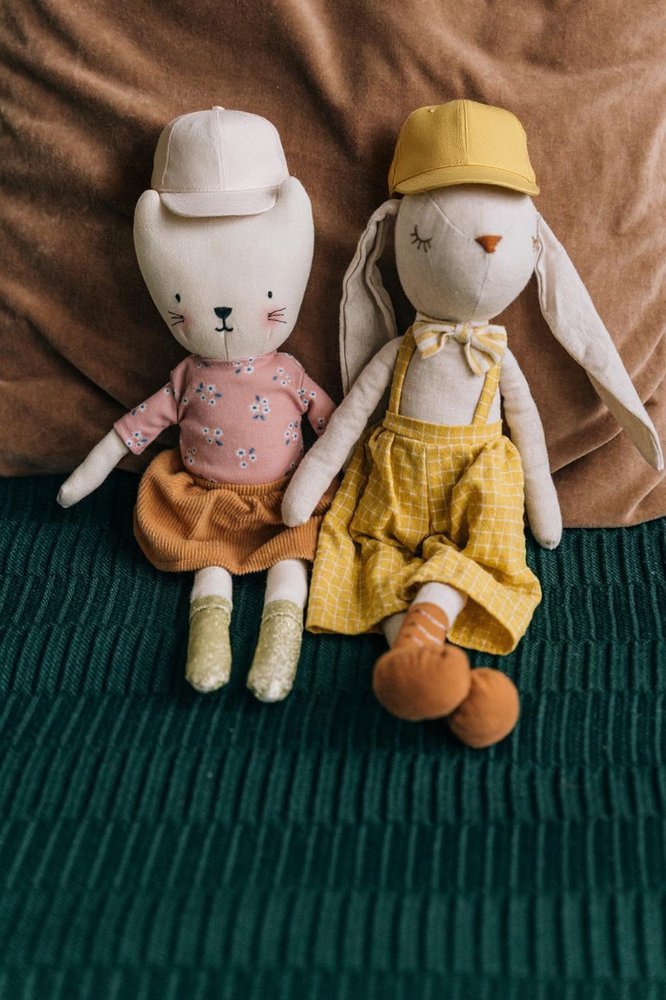}
    \caption{\emph{``Add hats in the boxes''}}
    \label{fig:textedit:b}
  \end{subfigure}\hfill
  \begin{subfigure}[t]{0.16\linewidth}
    \centering
    \includegraphics[height=3cm]{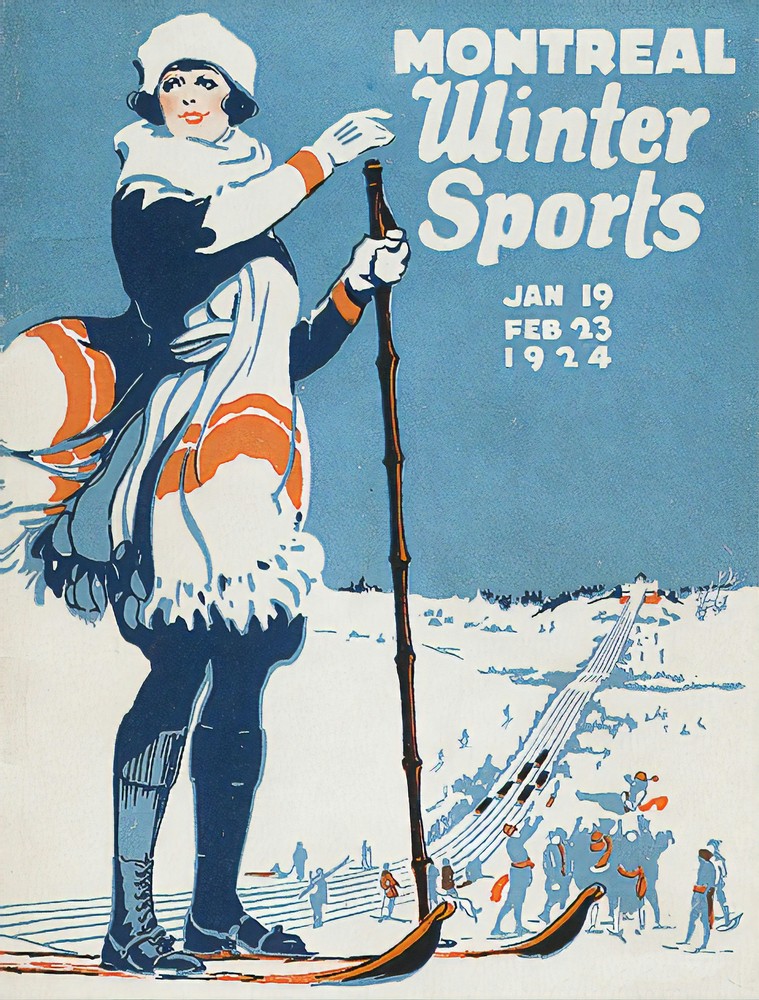}
    \caption{\emph{Input image}}
    \label{fig:textedit:c}
  \end{subfigure}\hfill
  \begin{subfigure}[t]{0.16\linewidth}
    \centering
    \includegraphics[height=3cm]{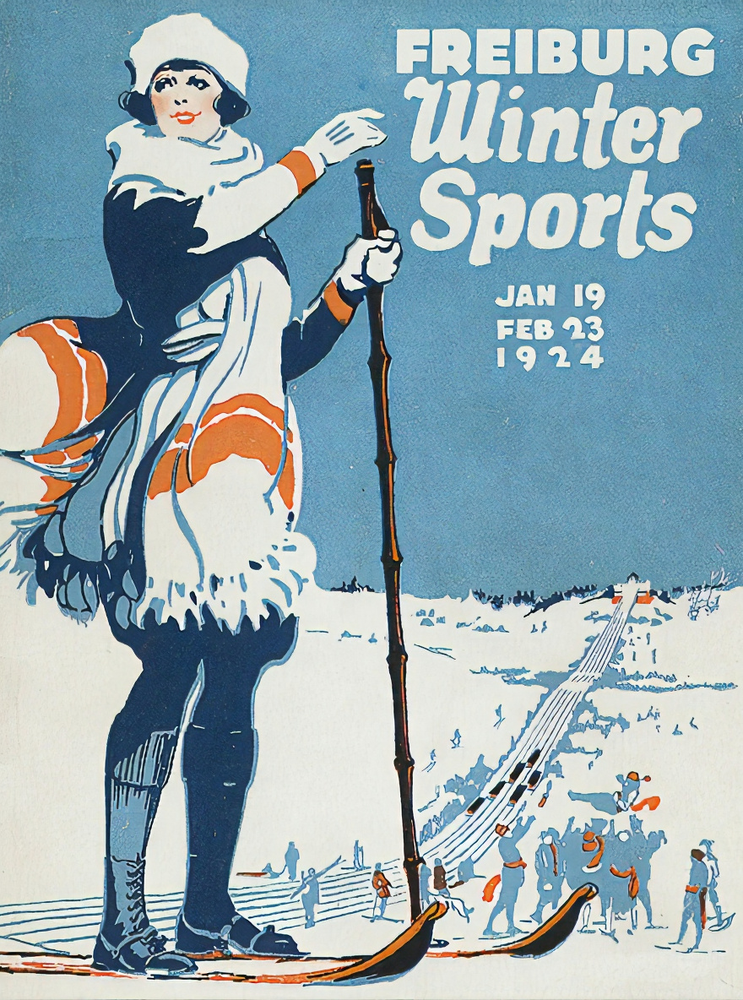}
    \caption{\emph{``Replace `MONTREAL' with `FREIBURG' ''}}
    \label{fig:textedit:d}
  \end{subfigure}\hfill
  \begin{subfigure}[t]{0.16\linewidth}
    \centering
    \includegraphics[height=3cm]{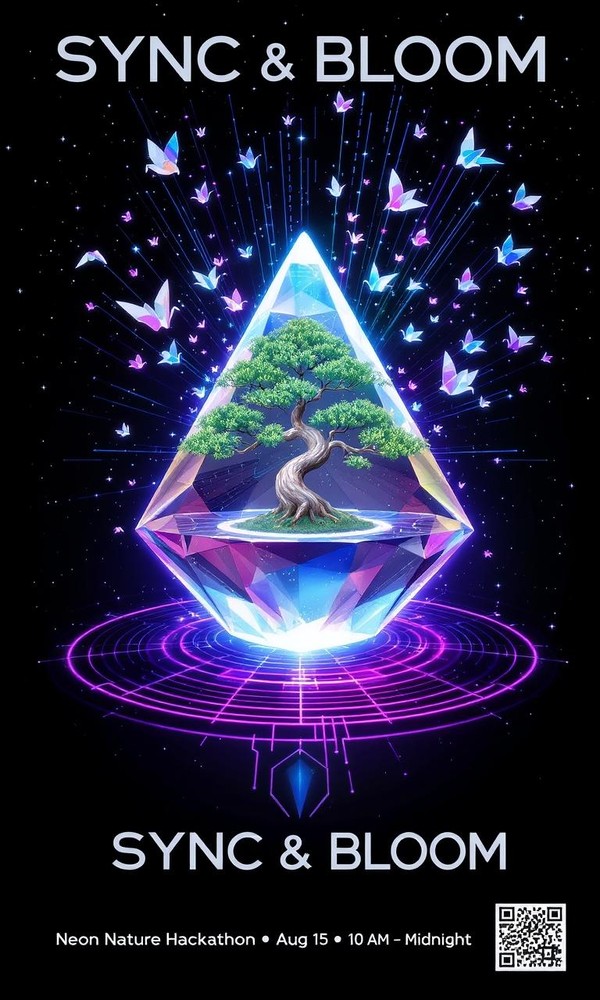}
    \caption{\emph{Input image}}
    \label{fig:textedit:e}
  \end{subfigure}\hfill
  \begin{subfigure}[t]{0.16\linewidth}
    \centering
    \includegraphics[height=3cm]{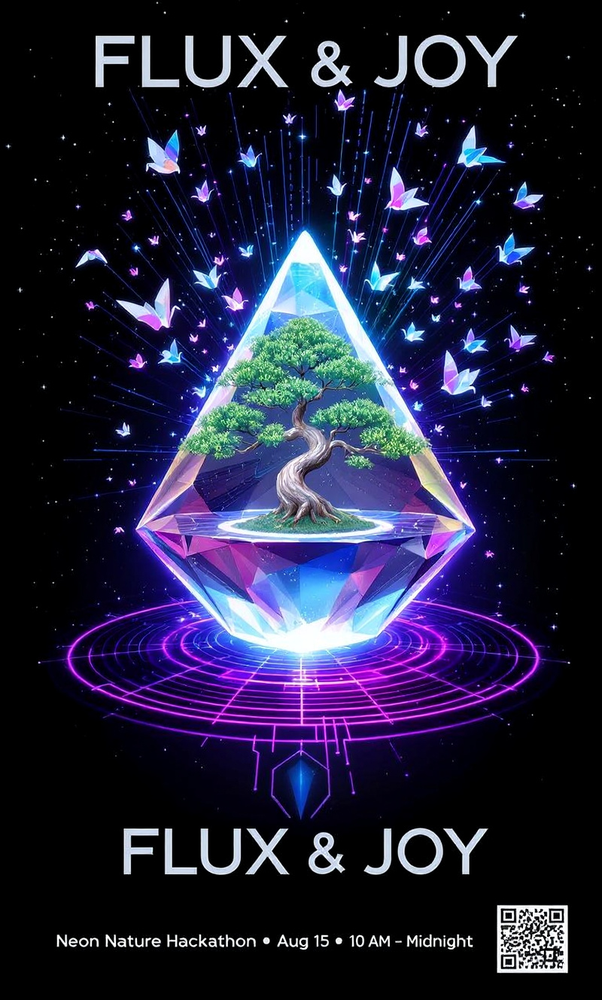}
    \caption{\emph{``Replace `SYNC \& BLOOM' with `FLUX \& JOY' ''}}
    \label{fig:textedit:f}
  \end{subfigure}
  \caption{\modelname is able to leverage visual cues like bounding boxes and edit text while keeping the style.}
  \label{fig:visualcues}
\end{figure}
}

\newcommand{\vaetable}{
\begin{table}[!t]
\centering
\label{tab:vaetable}
\begin{tabular}{@{}lccc@{}}
\toprule
\textbf{Model} & \textbf{PDist} $\downarrow$ & \textbf{SSIM} $\uparrow$ & \textbf{PSNR} $\uparrow$ \\
\midrule
\textbf{Flux-VAE} &
\textbf{0.332} $\pm$ 0.003 &
\textbf{0.896} $\pm$ 0.004 &
\textbf{31.1} $\pm$ 0.08 \\[2pt]
SD3-VAE~\citep{esser2024scalingrectifiedflowtransformers} &
0.452 $\pm$ 0.004 &
0.858 $\pm$ 0.005 &
29.6 $\pm$ 0.07 \\[2pt]
SD3-TAE\footnote{\url{https://huggingface.co/madebyollin/taesd3}} &
0.746 $\pm$ 0.004 &
0.774 $\pm$ 0.014 &
27.9 $\pm$ 0.06 \\[2pt]
SDXL-VAE~\citep{podell2023sdxl} &
0.890 $\pm$ 0.005 &
0.748 $\pm$ 0.006 &
25.9 $\pm$ 0.07 \\[2pt]
SD-VAE\footnote{\url{https://huggingface.co/stabilityai/sd-vae-ft-ema-original}} &
0.949 $\pm$ 0.005 &
0.720 $\pm$ 0.004 &
25.0 $\pm$ 0.07 \\
\bottomrule
\end{tabular}
\vspace{0.5em}
\caption{Reconstruction quality comparison across different VAE architectures.  
All metrics computed on 4096 ImageNet images. Values are mean ± standard error (rounded). See also~\Cref{sec:vaedetails}.}
\end{table}
}

\usepackage{cleveref}

\title{\modelname: Flow Matching for In-Context Image Generation and Editing in Latent Space}
\author{
\textbf{Black Forest Labs\thanks{Please cite this works as "Black Forest Labs (2025)"}} \\[3pt]
\textbf{Stephen Batifol} \quad
\textbf{Andreas Blattmann} \quad
\textbf{Frederic Boesel} \quad
\textbf{Saksham Consul} \quad
\textbf{Cyril Diagne} \\[3pt]
\textbf{Tim Dockhorn} \quad
\textbf{Jack English} \quad
\textbf{Zion English} \quad
\textbf{Patrick Esser} \quad
\textbf{Sumith Kulal} \\[3pt]
\textbf{Kyle Lacey} \quad
\textbf{Yam Levi} \quad
\textbf{Cheng Li} \quad
\textbf{Dominik Lorenz} \quad
\textbf{Jonas M{\"u}ller} \\[3pt]
\textbf{Dustin Podell} \quad
\textbf{Robin Rombach} \quad
\textbf{Harry Saini} \quad
\textbf{Axel Sauer} \quad
\textbf{Luke Smith} \\[3pt]
}

\makeatletter
\renewcommand{\@toptitlebar}{
  \hrule height 4\p@
  \vskip 0.1in  %
  \vskip -\parskip%
}

\renewcommand{\@bottomtitlebar}{
  \vskip 0.15in  %
  \vskip -\parskip
  \hrule height 1\p@
  \vskip 0.05in%
}

\renewcommand{\@maketitle}{%
  \vbox{%
    \hsize\textwidth
    \linewidth\hsize
    \vskip -0.2in  %
    \@toptitlebar
    \centering
    {\LARGE\bf \@title\par}
    \vskip 0.1in  %
    \@bottomtitlebar
    \if@submission
      \begin{tabular}[t]{c}\bf\rule{\z@}{24\p@}
        \vspace{-2em} Anonymous Author(s) \\
        Affiliation \\
        Address \\
        \texttt{email} \\
      \end{tabular}%
    \else
      \def\And{%
        \end{tabular}\hfil\linebreak[0]\hfil%
        \begin{tabular}[t]{c}\bf\rule{\z@}{24\p@}\ignorespaces%
      }
      \def\AND{%
        \end{tabular}\hfil\linebreak[4]\hfil%
        \begin{tabular}[t]{c}\bf\rule{\z@}{24\p@}\ignorespaces%
      }
      \begin{tabular}[t]{c}\bf\rule{\z@}{24\p@}\@author\end{tabular}%
    \fi
    \vskip 0.1in \@minus 0.05in  %
  }
}
\makeatother

\begin{document}
\maketitle
\enlargethispage{\baselineskip}

\begin{abstract}

We present evaluation results for \modelname{}, a generative flow matching model that unifies image generation and editing. The model generates novel output views by incorporating semantic context from text and image inputs.
Using a simple sequence concatenation approach, \modelname{} handles both local editing and generative in-context tasks within a single unified architecture.
Compared to current editing models that exhibit degradation in character consistency and stability across multiple turns, we observe that \modelname{} improved preservation of objects and characters, leading to greater robustness in iterative workflows.
The model achieves competitive performance with current state-of-the-art systems while delivering significantly faster generation times, enabling interactive applications and rapid prototyping workflows.
To validate these improvements, we introduce
\emph{KontextBench}, a comprehensive benchmark with 1026 image-prompt pairs covering five task categories: local editing, global editing, character reference, style reference and text editing. Detailed evaluations show the superior performance of \modelname{} in terms of both single-turn quality and multi-turn consistency, setting new standards for unified image processing models.

\end{abstract}

\theultimategullteaser

\section{Introduction}
Images are a foundation of modern communication and form the basis for areas as diverse as social media, e-commerce, scientific visualization, entertainment, and memes. As the volume and speed of visual content increases, so does the demand for intuitive but faithful and accurate image editing. Professional and casual users expect tools that preserve fine detail, maintain semantic coherence, and respond to increasingly natural language commands. The advent of large-scale generative models has changed this landscape, enabling purely text-driven image synthesis and modifications that were previously impractical or impossible~\citep{esser2021imagebart,ho2020denoising, ramesh2022hierarchical, Rombach_2022, podell2023sdxl, betker2023improving, esser2024scalingrectifiedflowtransformers, sauer2024fasthighresolutionimagesynthesis, imagenteamgoogle2024imagen3}. 

Traditional image processing pipelines work by directly manipulating pixel values or by applying geometric and photometric transformations under explicit user control~\citep{gonzalez2009digital, szeliski2022computer}. In contrast, generative processing uses deep learning models and their learned representations to synthesize content that seamlessly fits into the new scene. Two complementary capabilities are central to this paradigm 

\begin{itemize}
    \item \textbf{Local editing.} Local, limited modifications that keep the surrounding context intact (e.g. changing the color of a car while preserving the background or replacing the background while keeping the subject in the foreground). Generative inpainting systems such as LaMa~\citep{suvorov2022resolution}, Latent Diffusion inpainting~\citep{Rombach_2022}, RePaint~\citep{lugmayr2022repaint}, Stable Diffusion Inpainting variants\footnote{\url{https://huggingface.co/runwayml/stable-diffusion-inpainting}}, and FLUX.1 Fill\footnote{\url{https://huggingface.co/black-forest-labs/FLUX.1-Fill-dev}} make such context-aware edits instantaneous; see also Palette~\citep{saharia2022palette} and Paint-by-Example~\citep{yang2023paint}. Beyond inpainting, ControlNet~\citep{zhang2023adding} enables mask-guided background replacement, while DragGAN~\citep{pan2023drag} offers interactive point-based geometric manipulation.
    \item \textbf{Generative editing.} Extraction of a visual concept (\eg a particular figure or logo), followed by its faithful reproduction in new environments, potentially synthesized under a new viewpoint or rendering in a new visual context. Similarly to \emph{in-context learning} in large language models, where the network learns a task from the examples provided in the prompt without any parameter updates \citep{brown2020language}, the generator adapts its output to the conditioning context on the fly. This property enables personalization of generative image and video models without the need for finetuning~\citep{ruiz2023dreamboothfinetuningtexttoimage} or LoRA training~\cite{hu2022lora, kumari2023multi, huang2024context}. Early works on such training-free subject-driven image synthesis include \emph{IP-Adapter}~\citep{ye2023ip} or retrieval-augmented diffusion variants~\citep{chen2022re, blattmann2022retrieval}.
\end{itemize}

\snowtransform

\textbf{Recent Advances.} InstructPix2Pix~\citep{brooks2023instructpix2pix} and subsequent work~\citep{Boesel2024ImprovingIE} demonstrated the promise of synthetic instruction-response pairs for fine-tuning a diffusion model for image editing, while learning-free methods for personalized text-to-image synthesis\citep{gal2022image,ruiz2023dreamboothfinetuningtexttoimage,kawar2023imagictextbasedrealimage} enable image modification with off-the-shelf, high-performance image generation models~\citep{Rombach_2022, podell2023sdxl}. Subsequent instruction-driven editors such as Emu Edit~\cite{sheynin2023emu}, OmniGen~\citep{xiao2024omnigen}, HiDream-E1\citep{hidream2025_e1} and ICEdit~\citep{zhang2025context} -- extend these ideas to refined datasets and model architectures. 
\citet{huang2024context} introduce in-context LoRAs for diffusion transformers on specific tasks, where each task needs to train dedicated LoRA weights. Novel proprietary systems embedded in multimodal LLMs (e.g., GPT-Image\citep{openai2025_4o_image} and Gemini Native Image Gen~\citep{google2025_gemini_flash_image}) further blur the line between dialog and editing. Generative platforms such as Midjourney\citep{midjourney2025_home} and RunwayML\citep{runway2025_home} integrate these advances into end-to-end creative workflows.

\textbf{Shortcomings of recent approaches.} In terms of results, current approaches struggle with three major shortcomings: (i) instruction-based methods trained on synthetic pairs inherit the shortcomings of their generation pipelines, limiting the variety and realism of achievable edits; (ii) maintaining the accurate appearance of characters and objects across multiple edits remains an open problem, hindering story-telling and brand-sensitive applications; (iii) in addition to lower quality compared to denoising-based approaches, autoregressive editing models integrated into large multimodal systems often come with long runtimes that are incompatible with interactive use.

\textbf{Our Solution.} We introduce \modelname, 
a flow-based generative image processing model that matches or exceeds the quality of state-of-the-art black-box systems while overcoming the above limitations.  
\modelname is a simple flow matching model trained using only a velocity prediction target on a concatenated sequence of context and instruction tokens. %

In particular, \modelname offers:

\begin{itemize}
\item \textbf{Character consistency:} \modelname excels at character preservation, including multiple, iterative edit turns.
\item \textbf{Interactive speed:} \modelname is \emph{fast}. Both text-to-image and image-to-image application reach speeds for synthesising an image at $1024 \times 1024$ of 3--5 seconds. 
\item \textbf{Iterative application:} Fast inference and robust consistency allow users to refine an image through multiple successive edits with minimal visual drift.
\end{itemize}

\section{FLUX.1}
\label{sec:fluxone}

\floatingfusedditblock FLUX.1 is a rectified flow transformer~\citep{esser2024scalingrectifiedflowtransformers,lipman2023flow,liu2022flow} trained in the latent space of an image autoencoder~\citep{Rombach_2022}. 
We follow~\citet{Rombach_2022} and train a convolutional autoencoder with an adversarial objective from scratch. By scaling up the training compute and using 16 latent channels, we improve the reconstruction capabilities compared to related models; see \Cref{tab:vaetable}. Furthermore, 
FLUX.1 is 
built from 
a mix of double stream and single stream~\citep{Peebles_2023} blocks.
Double stream blocks employ separate weights for image and text tokens, and \textit{mixing} is done by applying the attention operation over the concatenation of tokens. 
After passing the sequences through the double stream blocks, we concatenate them and apply 38 single stream blocks to the image and text tokens. Finally, we discard the text tokens and decode the image tokens.

To improve GPU utilization of single stream blocks, we leverage \textit{fused} feed-forward blocks inspired by~\citet{dehghani2023scaling}, which i) reduce the number of modulation parameters in a feedforward block by a factor of 2 and ii) fuse the attention input- and output linear layers with that of the MLP, leading to larger matrix-vector multiplications and thus more efficient training and inference.  We utilize factorized three--dimensional Rotary Positional Embeddings (3D RoPE)~\citep{su2024roformer}. Every latent token is indexed by its space-time coordinates $(t, h,w)$ (with $t\equiv 0$ for single image inputs).
See~\Cref{fig:floatingfusedditblock} for a visualization. 
\vaetable

\modelfigure

\section{FLUX.1 Kontext}
\label{sec:method}

Our goal is to learn a model that can generate images conditioned jointly on a text prompt and a reference images. 
More formally, we aim to approximate the conditional distribution
\begin{equation}
p(x\mid y,c)     
\end{equation}
where \(x\) is the target image, \(y\) is a context image (or \(\varnothing\)), and \(c\) is a natural-language instruction.  
Unlike classic text-to-image generation, this objective entails learning \emph{relations between images themselves}, mediated by \(c\), so that the same network can >i) perform image-driven edits when \(y \neq \varnothing\), and (ii) create new content from scratch when \(y = \varnothing\). 

To that end, let $x\in\mathcal X$ be an output (target) image, $y\in\mathcal X\cup\{\varnothing\}$ an optional \emph{context} image, and $c\in\mathcal C$ a text prompt.  
We model the conditional distribution $ p_\theta(x\mid y,c) $ such that the same network handles \emph{in-context and local edits} when $y\neq\varnothing$ and free \emph{text-to-image generation} when $y=\varnothing$.
Training starts from a \textsc{FLUX.1} text-to-image checkpoint, and we collect and curate millions of relational pairs $(x\,|\,y,c)$ for optimization. In practice, we do not model images in pixel space but instead encode them into a token sequence as discussed in the following paragraph.

\paragraph{Token sequence construction.}
Images are encoded into latent tokens by the frozen FLUX auto-encoder. 
These context image tokens $y$ are then appended to the image tokens $x$ and fed into the visual stream of the model.
This simple \emph{sequence concatenation} (i) supports different input/output resolutions and aspect ratios, and (ii) readily extends to multiple images $y_1,y_2,\dots, y_N$.
Channel-wise concatenation of $x$ and $y$ was also tested but in initial experiments we found this design choice to perform worse.

We encode positional information via 3D RoPE embeddings, where the embeddings for the context $y$ receive a constant offset for all context tokens. We treat the offset as a \emph{virtual time step} that cleanly separates the context and target blocks while leaving their internal spatial structure intact. 
Concretely, if a token position is denoted by the triplet $\mathbf u=(t,h,w)$, then we set $\mathbf u_x=(0,h,w)$ for the target tokens and for context tokens we set
\begin{equation}
\mathbf u_{y_i} \;=\; (\,i,\,h,\,w\,), 
\qquad i=1,\dots,N,
\end{equation}

\paragraph{Rectified-flow objective.}
We train with a rectified flow–matching loss
\begin{equation}
\mathcal L_\theta 
   =\mathbb E_{\,t\sim p(t),\,x,y,c}\bigl[\,
      \lVert v_\theta(z_t,t,y,c)-(\varepsilon - x)\rVert_2^2
   \bigr],
\label{eq:flowm}
\end{equation}

where $z_t$ is the linearly interpolated latent between $x$ and noise $\varepsilon \sim \mathcal{N}(0, 1)$; $z_t = (1-t) x + t \varepsilon$. 
We use a logit normal shift schedule (see \Cref{sec:explicit_shift}) for $p(t; \mu, \sigma = 1.0)$, where we change the mode $\mu$ depending on the resolution of the data during training. When sampling pure text–image pairs ($y=\varnothing$) we omit all tokens $y$, preserving the text-to-image generation capability of the model.

\paragraph{Adversarial Diffusion Distillation}
Sampling of a flow matching model obtained by optimizing \Cref{eq:flowm} typically involves solving an ordinary or stochastic differential equation~\citep{lipman2023flow, albergo2022building}, using 50--250 guided~\citep{ho2022classifierfree} network evaluations. While samples obtained through such a procedure are of good quality for a well-trained model $v_\Theta$, this comes with a few potential drawbacks: First, such multi-step sampling is slow, rendering model-serving at scale expensive and hindering low-latency, interactive applications. Moreover, guidance may occasionally introduce visual artifacts such as over-saturated samples. We tackle both challenges using latent \emph{adversarial diffusion distillation} (LADD)~\citep{sauer2021projected,sauer2023adversarial, sauer2024fasthighresolutionimagesynthesis}, reducing the number of sampling steps while increasing the quality of the samples through adversarial training.

\srefone

\paragraph{Implementation details.}
Starting from a pure text-to-image checkpoint, we jointly fine-tune the model on image-to-image and text-to-image tasks following \Cref{eq:flowm}. 
While our formulation naturally covers multiple input images, we focus on single context images for conditioning at this time.
\modelname [pro] is trained with the flow objective followed by LADD~\citep{sauer2023adversarial}. We obtain \modelname [dev] through guidance-distillation into a 12B diffusion transformer following the techniques outlined in~\citet{meng2023distillation}. To optimize \modelname [dev] performance on edit tasks, we focus exclusively on image-to-image training, \ie do not train on the pure text-to-image task for \modelname [dev].
We incorporate safety training measures including classifier-based filtering and adversarial training to prevent the generation of non-consensual intimate imagery (NCII) and child sexual abuse material (CSAM).

We use FSDP2~\citep{liang2024torchtitan} with mixed precision: all-gather operations are performed in \texttt{bfloat16} while gradient reduce-scatter uses \texttt{float32} for improved numerical stability. We use selective activation checkpointing~\citep{korthikanti2023reducing} to reduce maximum VRAM usage. To improve throughput, we use \emph{Flash Attention~3}~\citep{shah2024flashattention} and regional compilation of individual Transformer blocks.

\skirttransform

\section{Evaluations \& Applications}
\label{sec:applications}
In this section, we evaluate \modelname's performance and demonstrate its capabilities. We first introduce KontextBench, a novel benchmark featuring real-world image editing challenges crowd-sourced from users. We then present our primary evaluation: a systematic comparison of \modelname against state-of-the-art text-to-image and image-to-image synthesis methods, where we demonstrate competitive performance across diverse editing tasks. Finally, we explore \modelname's practical applications, including iterative editing workflows, style transfer, visual cue editing, and text editing.

\definecolor{steelblue}{RGB}{70,120,180}

\subsection{KontextBench -- Crowd-sourced Real-World Benchmark for In-Context Tasks}
Existing benchmarks for editing models are often limited when it comes to capturing real-world usage. InstructPix2Pix~\cite{brooks2023instructpix2pix} relies on synthetic Stable Diffusion samples and GPT-generated instructions, creating inherent bias. MagicBrush~\citep{zhang2024magicbrushmanuallyannotateddataset}, while using authentic MS-COCO images, is constrained by DALLE-2's~\citep{ramesh2022hierarchical} capabilities during data collection. Other benchmarks like Emu-Edit~\citep{sheynin2023emu} use lower-resolution images with unrealistic distributions and focus solely on editing tasks, while DreamBench~\citep{peng2025dreambenchhumanalignedbenchmarkpersonalized} lacks broad coverage and GEdit-bench~\citep{liu2025step1x} does not represent the full scope of modern multimodal models. IntelligentBench~\citep{deng2025emerging} remains unavailable with only 300 examples of uncertain task coverage.

To address these gaps, we compile \emph{KontextBench} from crowd-sourced real-world use cases.
The benchmark comprises $1026$ unique image-prompt pairs derived from 108 base images including personal photos, CC-licensed art, public domain images, and AI-generated content. It spans five core tasks: local instruction editing (416 examples), global instruction editing (262), text editing (92), style reference (63), and character reference (193). We found that the scale of the benchmark provides a good balance between reliable human evaluation and comprehensive coverage of real-world applications. We will publish this benchmark including the samples of \modelname and all reported baselines.

\subsection{State-of-the-Art Comparison} \label{subsec:eval}

\kontexttimings
\modelname is designed to perform both text-to-image (T2I) and image-to-image (I2I) synthesis. We evaluate our approach against the strongest proprietary and open-weight models in both domains. 
We evaluate \textit{FLUX.1 Kontext [pro]} and \textit{[dev]}. As stated above, for \textit{[dev]} we exclusively focus on image-to-image tasks.
Additionally, we introduce \textit{FLUX.1 Kontext [max]}, which uses more compute to improve generative performance.

\textbf{Image-to-Image Results.}
For image editing evaluation, we assess performance across multiple editing tasks: image quality, local editing, \emph{character reference} (CREF), \emph{style reference} (SREF), text editing, and computational efficiency.
CREF enables consistent generation of specific characters or objects across novel settings, whereas SREF allows style transfer from reference images while maintaining semantic control. 
We compare different APIs and find that our models offer the fastest latency (cf. \Cref{fig:kontexttimings:i2i}), outperforming related models by up to an order of magnitude in speed difference.
In our human evaluation (\Cref{fig:kontextbenchresults}), we find that \textit{FLUX.1 Kontext [max]} and \textit{[pro]} are the best solution in the categories local and text editing, and for general CREF. We also calculate quantitative scores for CREF, to asses changes in facial characteristics between input and output images we use AuraFace\footnote{\url{https://huggingface.co/fal/AuraFace-v1}} to extract facial embeddings before and after and edit and compare both, see \Cref{fig:bench:auraface}. In alignment with our human evaluations, \textit{FLUX.1 Kontext} outperforms all other models. 
For global editing and SREF, \textit{FLUX.1 Kontext} is second only to gpt-image-1, and Gen-4 References, respectively.

Overall, \modelname offers state-of-the-art character consistency, and editing capabilities, while outperforming competing models such as GPT-Image-1 by up to an order of magnitude in speed.

\textbf{Text-to-Image Results.} Current T2I benchmarks predominantly focus on general preference, typically asking questions like \emph{``which image do you prefer?''}. We observe that this broad evaluation criterion often favors a characteristic ``AI aesthetic'' meaning over-saturated colors, excessive focus on central subjects, pronounced bokeh effects, and convergence toward homogeneous styles. We term this phenomenon \textit{bakeyness}. 
To address this limitation, we decompose T2I evaluation into five distinct dimensions: prompt following, aesthetic (\emph{``which image do you find more aesthetically pleasing''}), realism (\emph{``which image looks more real''}), typography accuracy, and inference speed. We evaluate on 1\,000 diverse test prompts compiled from academic benchmarks (DrawBench~\citep{saharia2022photorealistic}, PartiPrompts~\citep{yu2022scalingautoregressivemodelscontentrich}) and real user queries. We refer to this benchmark as Internal-T2I-Bench in the following. In addition, we complement this benchmark with additional evaluations on GenAI bench~\citep{li2024genai}.

In T2I, FLUX.1 Kontext demonstrates balanced performance across evaluation categories (see \Cref{fig:kontextt2ibenchresults}). Although competing models excel in certain domains, this often comes at the expense of other categories. For instance, Recraft delivers strong aesthetic quality but limited prompt adherence, whereas GPT-Image-1 shows the inverse overall performance pattern. \modelname consistently improves performance across categories over its predecessor FLUX1.1 [pro]. We also observe progressive gains from FLUX.1 Kontext [pro] to FLUX.1 Kontext [max]. We highlight samples in \figref{fig:ttoicherries}.

\kontextbenchresults
\kontexttexttoimagebenchresults

\subsection{Iterative Workflows}

\begin{figure}[t]
    \centering
    \includegraphics[width=\linewidth]{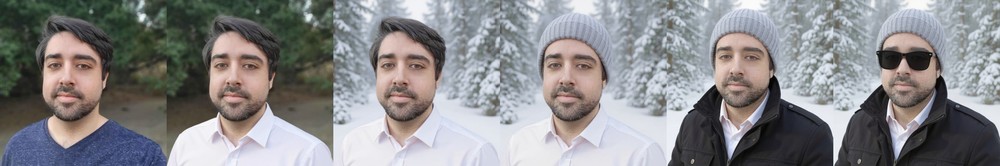} \\
    \includegraphics[width=\linewidth]{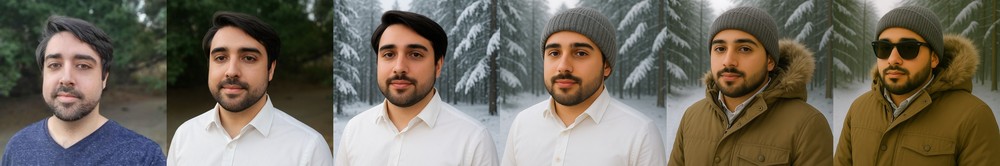} \\
    \includegraphics[width=\linewidth]{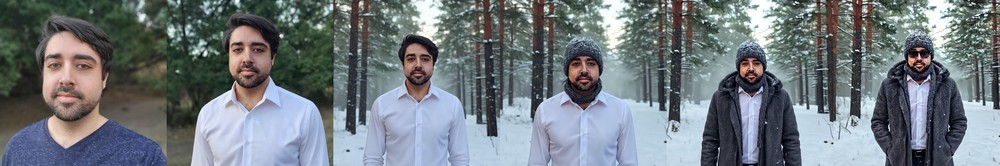} \\
    \includegraphics[width=0.7\linewidth]{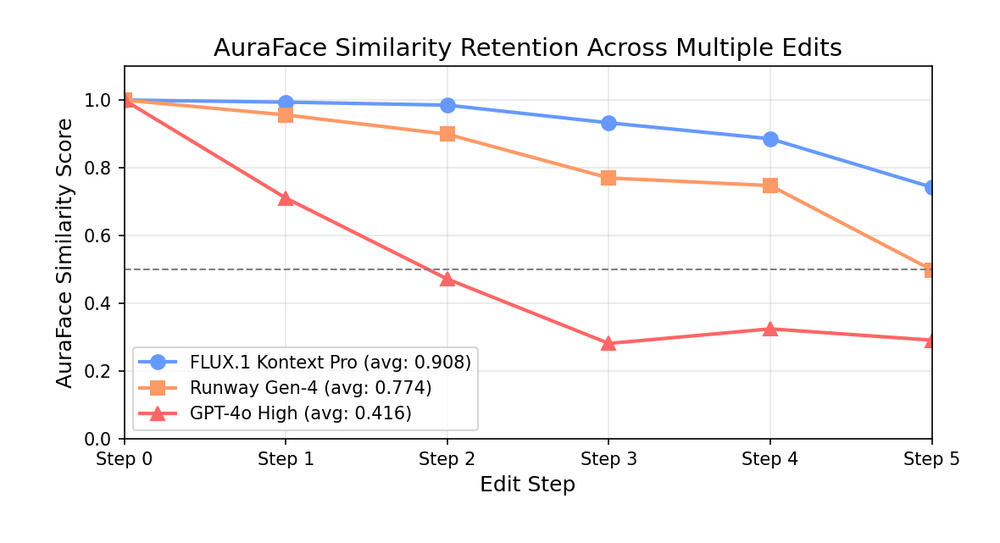}
    \caption{Iterative editing based on the same starting image with the same prompts and different models (top: \textit{FLUX.1 Kontext}, middle: gpt-image-1, bottom: Runway Gen4). Below are face similarity scores between the edited images and the edited images at different steps. For the last edit ("Add sunglasses"), a relative drop is expected due to partial exclusion of the face.}
    \label{fig:qualitative-iterative-editing}
\end{figure}

Maintaining character and object consistency across multiple edits is crucial for brand-sensitive and storytelling applications. Current state-of-the-art approaches suffer from noticeable visual drift: characters lose identity and objects lose defining features with each edit. In \figref{fig:qualitative-iterative-editing}, we demonstrate character identity drift across edit sequences produced by \modelname, Gen-4, and GPT-Image-high. We additionally compute the cosine similarity of AuraFace~\citep{deng2019arcface,isidentical2024auraface} embeddings between the input and images generated via successive edits, highlighting the slower drift of \modelname relative to competing methods. Consistency is essential: marketing needs stable brand characters, media production demands asset continuity, and e-commerce must preserve product details. %
Applications enabled by \modelname's reliable consistency are shown in \figref{fig:vase_edits} and \figref{fig:laugh_edits}.

\vasetransform
\laughtransform

\visualcues
\clearpage
\ttoicherries
\subsection{Specialized Applications}
\modelname supports several applications beyond standard generation. Style reference (SREF), first popularized by Midjourney~\citep{midjourney2025_home} and commonly implemented via IP-Adapters~\citep{ye2023ip}, enables style transfer from reference images while maintaining semantic control (see \secref{subsec:eval}). Additionally, the model supports intuitive editing through visual cues, responding to geometric markers like red ellipses to guide targeted modifications. It also provides sophisticated text editing capabilities, including logo refinement, spelling corrections, and style adaptations while preserving surrounding context. We demonstrate style reference in~\figref{fig:srefone} and visual cue-based editing in~\figref{fig:visualcues}.
\section{Discussion}\label{sec:conclusion}

We introduced \modelname, a flow matching model that combines in-context image generation and editing in a single framework. Through simple sequence concatenation and training recipes, \modelname achieves state-of-the-art performance while addressing key limitations such as character drift during multi-turn edits, slow inference, and low output quality.
Our contributions include a unified architecture that handles multiple processing tasks, superior character consistency across iterations, interactive speed, and KontextBench: A real-world benchmark with 1\,026 image-prompt pairs. Our extensive evaluations reveal that \modelname is comparable to proprietary systems while enabling fast, multi-turn creative workflows.

\textbf{Limitations.} \modelname exhibits a few limitations in its current implementation. Excessive multi-turn editing can introduce visual artifacts that degrade image quality, see \Cref{fig:failureone}. The model occasionally fails to follow instructions accurately, ignoring specific prompt requirements. 
In addition, the distillation process can introduce visual artifacts that impact the fidelity of the output.

\textbf{Future work} should focus on extending to multiple image inputs, further scaling, and reducing inference latency to unlock real-time applications. A natural extension of our approach is to include edits in the video domain. Most importantly, reducing degradation during multi-turn editing would enable infinitely fluid content creation. 
The release of \modelname and KontextBench provides a solid foundation and a comprehensive evaluation framework to drive unified image generation and editing.

\failureone

\clearpage
\appendix
\newpage

\section{Image Generation using Flow Matching}

\subsection{Primer on Rectified Flow Matching}
\label{sec:rflow}

For training our models, we construct forward noising processes in the latent space of an image autoencoder as

\begin{equation}
z_t = a_t x_0 + b_t \varepsilon, 
\label{eq:forward_proc}
\end{equation}

with $x_0 \sim p_{data}$, $\varepsilon \sim \mathcal{N}(0, 1)$, 
and the coefficients $a_t$ and $b_t$ define the log signal-to-noise ratio (log-SNR)~\citep{Kingma2023UnderstandingDO}

\begin{equation}
\lambda_t = \log \frac{a_t^2}{b_t^2}
\label{eq:logsnr}
\end{equation}

Further, we use the conditional flow matching loss~\citep{lipman2023flow}

\begin{equation}
\mathcal{L}_{\text{CFM}} = \mathbb{E}_{t\sim p(t),\varepsilon\sim\mathcal{N}(0,1)} || v_{\Theta}(z_t, t) - \frac{a_t'}{a_t} z_t + \frac{b_t}{2} \lambda_t' \varepsilon ||_2^2
\end{equation}

For rectified flow models~\citep{liu2022flow}, $a_t = 1-t$ and $b_t = t$, and thus

\begin{equation}
\mathcal{L}_{\text{CFM}} = \mathbb{E}_{t\sim p(t),\varepsilon\sim\mathcal{N}(0,1),x_0\sim p_{data}} || v_{\Theta}(z_t, t) + x_0 - \varepsilon ||_2^2
\end{equation}
 
and we sample $t$ from a \emph{Logit-Normal Distribution} \citep{esser2024scalingrectifiedflowtransformers}: $p(t) = \frac{\exp{(-0.5\cdot(\mathrm{logit}(t) - \mu)^2 / \sigma^2)}}{\sigma \sqrt{2\pi} \cdot (1-t) \cdot t }$, where $\mathrm{logit}(t) = \log\frac{t}{1-t}$. From the definition of the Logit-Normal Distribution, it follows that a random variable $Y = \mathrm{logit}(t) \sim \mathcal{N}(\mu,\sigma)$. 

\subsection{Expressing shifting of the timestep schedule via the Logit-Normal Distribution}
\label{sec:explicit_shift}
Previous work on high-resolution image synthesis introduced an additional shift of the timestep sampling (and, equivalently, the log-SNR schedule) via a parameter $\alpha$~\citep{esser2024scalingrectifiedflowtransformers, hoogeboom2023simple}. \citet{esser2024scalingrectifiedflowtransformers} empirically demonstrated that $\alpha=3.0$ worked best when increasing the image resolution from $256^2$ to $1024^2$. In the following, we show that this shifting can be expressed via the Logit-Normal Distribution.

Consider the log-SNR of a rectified flow forward process with $\mu=0$ and $\sigma=1$: 
\begin{equation}
\lambda_{t}^{0,1} = 2 \log \frac{1-t}{t} = -2 \mathrm{logit}(t), 
\end{equation}

where 
$\mathrm{logit}(t) \sim \mathcal{N}(0, 1)$. Expressing the log-SNR for arbitrary $\mu$ and $\sigma$ gives 

\begin{equation}
\lambda_{t}^{\mu,\sigma} = -2 (\sigma \cdot \text{logit}(t) + \mu) = \sigma \cdot \lambda_{t}^{0,1} - 2\mu \, .
\label{eq:shift_mu_sigma}
\end{equation}

The $\alpha$-shifted log-SNR~\citep{esser2024scalingrectifiedflowtransformers, hoogeboom2023simple} is obtained as 
\begin{equation}
\lambda_{t}^{\alpha}= \lambda_{t}^{0,1} - 2 \log \alpha \, .
\label{eq:shift_alpha}
\end{equation}

Comparing \Cref{eq:shift_mu_sigma} and \Cref{eq:shift_alpha}, we identify $\mu = \log \alpha$ for $\sigma=1.0$, i.e. a shift of $\alpha = 3.0$ would correspond to a logit-normal distribution with $\mu = \log 3.0 = 1.0986$ and $\sigma=1.0$. 

We can further express the shifted log-SNR as a function of shifted timesteps $t'$
\begin{equation}
\lambda_{t'} = 2 \log \frac{1-t'}{t'} = \sigma \lambda_{t}^{0,1} - 2 \mu = 2 \sigma \log \frac{1-t}{t} - 2 \mu
\end{equation}

and solve for $t'$:
\begin{equation}
t' = \frac{e^\mu}{e^\mu + (1/t - 1)^\sigma}
\label{eq:timestep_redist}
\end{equation}

For $\sigma=1.0$ and $\mu=\log \alpha$ this recovers the redistribution function for the timesteps proposed in~\citep{esser2024scalingrectifiedflowtransformers} $t' = \frac{\alpha t}{1 + (\alpha -1)t}$, as expected. This generalized shifting formula \ref{eq:shift_mu_sigma} can be useful both for training and via \ref{eq:timestep_redist} for inference. %

\section{VAE Evaluation Details}
\label{sec:vaedetails}

We compare our VAE with related models using three reconstruction metrics, namely, SSIM, PSNR, and the \emph{Perceptual Distance} (PDist) in VGG~\citep{simonyan2014very} feature space. All metrics are computed over 4\,096 random ImageNet evaluation images at resolution $256 \times 256$. \Cref{tab:vaetable} shows the mean and the standard deviation over the 4\,096 inputs.

\bibliography{refs}

\begin{thebibliography}{62}
\providecommand{\natexlab}[1]{#1}
\providecommand{\url}[1]{\texttt{#1}}
\expandafter\ifx\csname urlstyle\endcsname\relax
  \providecommand{\doi}[1]{doi: #1}\else
  \providecommand{\doi}{doi: \begingroup \urlstyle{rm}\Url}\fi

\bibitem[Albergo and Vanden-Eijnden(2022)]{albergo2022building}
Michael~S. Albergo and Eric Vanden-Eijnden.
\newblock Building normalizing flows with stochastic interpolants, 2022.

\bibitem[Betker et~al.(2023)Betker, Goh, Jing, Brooks, Wang, Li, Ouyang, Zhuang, Lee, Guo, et~al.]{betker2023improving}
James Betker, Gabriel Goh, Li~Jing, Tim Brooks, Jianfeng Wang, Linjie Li, Long Ouyang, Juntang Zhuang, Joyce Lee, Yufei Guo, et~al.
\newblock Improving image generation with better captions.
\newblock \emph{Computer Science. https://cdn. openai. com/papers/dall-e-3. pdf}, 2\penalty0 (3), 2023.

\bibitem[Blattmann et~al.(2022)Blattmann, Rombach, Oktay, M{\"u}ller, and Ommer]{blattmann2022retrieval}
Andreas Blattmann, Robin Rombach, Kaan Oktay, Jonas M{\"u}ller, and Bj{\"o}rn Ommer.
\newblock Retrieval-augmented diffusion models.
\newblock \emph{Advances in Neural Information Processing Systems}, 35:\penalty0 15309--15324, 2022.

\bibitem[Boesel and Rombach(2024)]{Boesel2024ImprovingIE}
Frederic Boesel and Robin Rombach.
\newblock Improving image editing models with generative data refinement.
\newblock In \emph{Tiny Papers @ ICLR}, 2024.
\newblock URL \url{https://api.semanticscholar.org/CorpusID:271461432}.

\bibitem[Brooks et~al.(2023)Brooks, Holynski, and Efros]{brooks2023instructpix2pix}
Tim Brooks, Aleksander Holynski, and Alexei~A Efros.
\newblock Instructpix2pix: Learning to follow image editing instructions.
\newblock In \emph{Proceedings of the IEEE/CVF Conference on Computer Vision and Pattern Recognition}, pages 18392--18402, 2023.

\bibitem[Brown et~al.(2020)Brown, Mann, Ryder, Subbiah, Kaplan, Dhariwal, Neelakantan, Shyam, Sastry, Askell, et~al.]{brown2020language}
Tom Brown, Benjamin Mann, Nick Ryder, Melanie Subbiah, Jared~D Kaplan, Prafulla Dhariwal, Arvind Neelakantan, Pranav Shyam, Girish Sastry, Amanda Askell, et~al.
\newblock Language models are few-shot learners.
\newblock \emph{Advances in neural information processing systems}, 33:\penalty0 1877--1901, 2020.

\bibitem[Chen et~al.(2022)Chen, Hu, Saharia, and Cohen]{chen2022re}
Wenhu Chen, Hexiang Hu, Chitwan Saharia, and William~W Cohen.
\newblock Re-imagen: Retrieval-augmented text-to-image generator.
\newblock \emph{arXiv preprint arXiv:2209.14491}, 2022.

\bibitem[Dehghani et~al.(2023)Dehghani, Djolonga, Mustafa, Padlewski, Heek, Gilmer, Steiner, Caron, Geirhos, Alabdulmohsin, et~al.]{dehghani2023scaling}
Mostafa Dehghani, Josip Djolonga, Basil Mustafa, Piotr Padlewski, Jonathan Heek, Justin Gilmer, Andreas~Peter Steiner, Mathilde Caron, Robert Geirhos, Ibrahim Alabdulmohsin, et~al.
\newblock Scaling vision transformers to 22 billion parameters.
\newblock In \emph{International Conference on Machine Learning}, pages 7480--7512. PMLR, 2023.

\bibitem[Deng et~al.(2025)Deng, Zhu, Li, Gou, Li, Wang, Zhong, Yu, Nie, Song, et~al.]{deng2025emerging}
Chaorui Deng, Deyao Zhu, Kunchang Li, Chenhui Gou, Feng Li, Zeyu Wang, Shu Zhong, Weihao Yu, Xiaonan Nie, Ziang Song, et~al.
\newblock Emerging properties in unified multimodal pretraining.
\newblock \emph{arXiv preprint arXiv:2505.14683}, 2025.

\bibitem[Deng et~al.(2019)Deng, Guo, Xue, and Zafeiriou]{deng2019arcface}
Jiankang Deng, Jia Guo, Niannan Xue, and Stefanos Zafeiriou.
\newblock Arcface: Additive angular margin loss for deep face recognition.
\newblock In \emph{Proceedings of the IEEE/CVF conference on computer vision and pattern recognition}, pages 4690--4699, 2019.

\bibitem[Esser et~al.(2021)Esser, Rombach, Blattmann, and Ommer]{esser2021imagebart}
Patrick Esser, Robin Rombach, Andreas Blattmann, and Bjorn Ommer.
\newblock Imagebart: Bidirectional context with multinomial diffusion for autoregressive image synthesis.
\newblock \emph{Advances in neural information processing systems}, 34:\penalty0 3518--3532, 2021.

\bibitem[Esser et~al.(2024)Esser, Kulal, Blattmann, Entezari, Müller, Saini, Levi, Lorenz, Sauer, Boesel, Podell, Dockhorn, English, Lacey, Goodwin, Marek, and Rombach]{esser2024scalingrectifiedflowtransformers}
Patrick Esser, Sumith Kulal, Andreas Blattmann, Rahim Entezari, Jonas Müller, Harry Saini, Yam Levi, Dominik Lorenz, Axel Sauer, Frederic Boesel, Dustin Podell, Tim Dockhorn, Zion English, Kyle Lacey, Alex Goodwin, Yannik Marek, and Robin Rombach.
\newblock Scaling rectified flow transformers for high-resolution image synthesis, 2024.
\newblock URL \url{https://arxiv.org/abs/2403.03206}.

\bibitem[Gal et~al.(2022)Gal, Alaluf, Atzmon, Patashnik, Bermano, Chechik, and Cohen-Or]{gal2022image}
Rinon Gal, Yuval Alaluf, Yuval Atzmon, Or~Patashnik, Amit~H Bermano, Gal Chechik, and Daniel Cohen-Or.
\newblock An image is worth one word: Personalizing text-to-image generation using textual inversion.
\newblock \emph{arXiv preprint arXiv:2208.01618}, 2022.

\bibitem[Gonzalez(2009)]{gonzalez2009digital}
Rafael~C Gonzalez.
\newblock \emph{Digital image processing}.
\newblock Pearson education india, 2009.

\bibitem[HiDream-ai(2025)]{hidream2025_e1}
HiDream-ai.
\newblock Hidream-e1: Instruction-based image editing model, 2025.
\newblock URL \url{https://github.com/HiDream-ai/HiDream-E1}.

\bibitem[Ho and Salimans(2022)]{ho2022classifierfree}
Jonathan Ho and Tim Salimans.
\newblock Classifier-free diffusion guidance, 2022.

\bibitem[Ho et~al.(2020)Ho, Jain, and Abbeel]{ho2020denoising}
Jonathan Ho, Ajay Jain, and Pieter Abbeel.
\newblock Denoising diffusion probabilistic models, 2020.

\bibitem[Hoogeboom et~al.(2023)Hoogeboom, Heek, and Salimans]{hoogeboom2023simple}
Emiel Hoogeboom, Jonathan Heek, and Tim Salimans.
\newblock Simple diffusion: End-to-end diffusion for high resolution images, 2023.

\bibitem[Hu et~al.(2022)Hu, Shen, Wallis, Allen-Zhu, Li, Wang, Wang, Chen, et~al.]{hu2022lora}
Edward~J Hu, Yelong Shen, Phillip Wallis, Zeyuan Allen-Zhu, Yuanzhi Li, Shean Wang, Lu~Wang, Weizhu Chen, et~al.
\newblock Lora: Low-rank adaptation of large language models.
\newblock \emph{ICLR}, 1\penalty0 (2):\penalty0 3, 2022.

\bibitem[Huang et~al.(2024)Huang, Wang, Wu, Shi, Dou, Liang, Feng, Liu, and Zhou]{huang2024context}
Lianghua Huang, Wei Wang, Zhi-Fan Wu, Yupeng Shi, Huanzhang Dou, Chen Liang, Yutong Feng, Yu~Liu, and Jingren Zhou.
\newblock In-context lora for diffusion transformers.
\newblock \emph{arXiv preprint arXiv:2410.23775}, 2024.

\bibitem[Imagen-Team-Google et~al.(2024)Imagen-Team-Google, :, Baldridge, Bauer, Bhutani, Brichtova, Bunner, Castrejon, Chan, Chen, Dieleman, Du, Eaton-Rosen, Fei, de~Freitas, Gao, Gladchenko, Colmenarejo, Guo, Haig, Hawkins, Hu, Huang, Igwe, Kaplanis, Khodadadeh, Kim, Konyushkova, Langner, Lau, Lawton, Luo, Mokrá, Nandwani, Onoe, van~den Oord, Parekh, Pont-Tuset, Qi, Qian, Ramachandran, Rane, Rashwan, Razavi, Riachi, Srinivasan, Srinivasan, Strudel, Uria, Wang, Wang, Waters, Wolff, Wright, Xiao, Xiong, Xu, van Zee, Zhang, Zhang, Zhou, Zolna, Aboubakar, Akbulut, Akerlund, Albuquerque, Anderson, Andreetto, Aroyo, Bariach, Barker, Ben, Berman, Biles, Blok, Botadra, Brennan, Brown, Buckley, Bunel, Bursztein, Butterfield, Caine, Carpenter, Casagrande, Chang, Chang, Chaudhuri, Chen, Choi, Churbanau, Clement, Cohen, Cole, Dektiarev, Du, Dutta, Eccles, Elue, Feden, Fruchter, Garcia, Garg, Ge, Ghazy, Gipson, Goodman, Górny, Gowal, Gupta, Halpern, Han, Hao, Hayes, Heek, Hertz, Hirst, Hoogeboom, Hou, Howard, Ibrahim,
  Ike-Njoku, Iljazi, Ionescu, Isaac, Jana, Jennings, Jenson, Jia, Jones, Ju, Kajic, Kaplanis, Ayan, Kelly, Kothawade, Kouridi, Ktena, Kumakaw, Kurniawan, Lagun, Lavitas, Lee, Li, Liang, Li-Calis, Liu, Alberca, Lorrain, Lu, Lum, Ma, Malik, Mellor, Mensink, Mosseri, Murray, Nematzadeh, Nicholas, Nørly, Oliveira, Ortiz-Jimenez, Paganini, Paine, Paiss, Parrish, Peckham, Peswani, Petrovski, Pfaff, Pirozhenko, Poplin, Prabhu, Qi, Rahtz, Rashtchian, Rastogi, Raul, Razavi, Rebuffi, Ricco, Riedel, Robinson, Rohatgi, Rosgen, Rumbley, Ryu, Salgado, Salimans, Singla, Schroff, Schumann, Shah, Shaw, Shaw, Shillingford, Shivakumar, Shtatnov, Singer, Sluzhaev, Sokolov, Sottiaux, Stimberg, Stone, Stutz, Su, Tabellion, Tang, Tao, Thomas, Thornton, Toor, Udrescu, Upadhyay, Vasconcelos, Vasiloff, Voynov, Walker, Wang, Wang, Wang, Wang, Wang, Wang, Ágoston Weisz, Wiles, Wu, Xu, Xue, Yang, Yu, Yurtoglu, Zand, Zhang, Zhang, Zhao, Zhaxybay, Zhou, Zhu, Zhu, Bloxwich, Bordbar, Cobo, Collins, Dai, Doshi, Dragan, Eck, Hassabis, Hsiao,
  Hume, Kavukcuoglu, King, Krawczyk, Li, Meier-Hellstern, Orban, Pinsky, Subramanya, Vinyals, Yu, and Zwols]{imagenteamgoogle2024imagen3}
Imagen-Team-Google, :, Jason Baldridge, Jakob Bauer, Mukul Bhutani, Nicole Brichtova, Andrew Bunner, Lluis Castrejon, Kelvin Chan, Yichang Chen, Sander Dieleman, Yuqing Du, Zach Eaton-Rosen, Hongliang Fei, Nando de~Freitas, Yilin Gao, Evgeny Gladchenko, Sergio~Gómez Colmenarejo, Mandy Guo, Alex Haig, Will Hawkins, Hexiang Hu, Huilian Huang, Tobenna~Peter Igwe, Christos Kaplanis, Siavash Khodadadeh, Yelin Kim, Ksenia Konyushkova, Karol Langner, Eric Lau, Rory Lawton, Shixin Luo, Soňa Mokrá, Henna Nandwani, Yasumasa Onoe, Aäron van~den Oord, Zarana Parekh, Jordi Pont-Tuset, Hang Qi, Rui Qian, Deepak Ramachandran, Poorva Rane, Abdullah Rashwan, Ali Razavi, Robert Riachi, Hansa Srinivasan, Srivatsan Srinivasan, Robin Strudel, Benigno Uria, Oliver Wang, Su~Wang, Austin Waters, Chris Wolff, Auriel Wright, Zhisheng Xiao, Hao Xiong, Keyang Xu, Marc van Zee, Junlin Zhang, Katie Zhang, Wenlei Zhou, Konrad Zolna, Ola Aboubakar, Canfer Akbulut, Oscar Akerlund, Isabela Albuquerque, Nina Anderson, Marco Andreetto, Lora
  Aroyo, Ben Bariach, David Barker, Sherry Ben, Dana Berman, Courtney Biles, Irina Blok, Pankil Botadra, Jenny Brennan, Karla Brown, John Buckley, Rudy Bunel, Elie Bursztein, Christina Butterfield, Ben Caine, Viral Carpenter, Norman Casagrande, Ming-Wei Chang, Solomon Chang, Shamik Chaudhuri, Tony Chen, John Choi, Dmitry Churbanau, Nathan Clement, Matan Cohen, Forrester Cole, Mikhail Dektiarev, Vincent Du, Praneet Dutta, Tom Eccles, Ndidi Elue, Ashley Feden, Shlomi Fruchter, Frankie Garcia, Roopal Garg, Weina Ge, Ahmed Ghazy, Bryant Gipson, Andrew Goodman, Dawid Górny, Sven Gowal, Khyatti Gupta, Yoni Halpern, Yena Han, Susan Hao, Jamie Hayes, Jonathan Heek, Amir Hertz, Ed~Hirst, Emiel Hoogeboom, Tingbo Hou, Heidi Howard, Mohamed Ibrahim, Dirichi Ike-Njoku, Joana Iljazi, Vlad Ionescu, William Isaac, Reena Jana, Gemma Jennings, Donovon Jenson, Xuhui Jia, Kerry Jones, Xiaoen Ju, Ivana Kajic, Christos Kaplanis, Burcu~Karagol Ayan, Jacob Kelly, Suraj Kothawade, Christina Kouridi, Ira Ktena, Jolanda Kumakaw, Dana
  Kurniawan, Dmitry Lagun, Lily Lavitas, Jason Lee, Tao Li, Marco Liang, Maggie Li-Calis, Yuchi Liu, Javier~Lopez Alberca, Matthieu~Kim Lorrain, Peggy Lu, Kristian Lum, Yukun Ma, Chase Malik, John Mellor, Thomas Mensink, Inbar Mosseri, Tom Murray, Aida Nematzadeh, Paul Nicholas, Signe Nørly, João~Gabriel Oliveira, Guillermo Ortiz-Jimenez, Michela Paganini, Tom~Le Paine, Roni Paiss, Alicia Parrish, Anne Peckham, Vikas Peswani, Igor Petrovski, Tobias Pfaff, Alex Pirozhenko, Ryan Poplin, Utsav Prabhu, Yuan Qi, Matthew Rahtz, Cyrus Rashtchian, Charvi Rastogi, Amit Raul, Ali Razavi, Sylvestre-Alvise Rebuffi, Susanna Ricco, Felix Riedel, Dirk Robinson, Pankaj Rohatgi, Bill Rosgen, Sarah Rumbley, Moonkyung Ryu, Anthony Salgado, Tim Salimans, Sahil Singla, Florian Schroff, Candice Schumann, Tanmay Shah, Eleni Shaw, Gregory Shaw, Brendan Shillingford, Kaushik Shivakumar, Dennis Shtatnov, Zach Singer, Evgeny Sluzhaev, Valerii Sokolov, Thibault Sottiaux, Florian Stimberg, Brad Stone, David Stutz, Yu-Chuan Su, Eric
  Tabellion, Shuai Tang, David Tao, Kurt Thomas, Gregory Thornton, Andeep Toor, Cristian Udrescu, Aayush Upadhyay, Cristina Vasconcelos, Alex Vasiloff, Andrey Voynov, Amanda Walker, Luyu Wang, Miaosen Wang, Simon Wang, Stanley Wang, Qifei Wang, Yuxiao Wang, Ágoston Weisz, Olivia Wiles, Chenxia Wu, Xingyu~Federico Xu, Andrew Xue, Jianbo Yang, Luo Yu, Mete Yurtoglu, Ali Zand, Han Zhang, Jiageng Zhang, Catherine Zhao, Adilet Zhaxybay, Miao Zhou, Shengqi Zhu, Zhenkai Zhu, Dawn Bloxwich, Mahyar Bordbar, Luis~C. Cobo, Eli Collins, Shengyang Dai, Tulsee Doshi, Anca Dragan, Douglas Eck, Demis Hassabis, Sissie Hsiao, Tom Hume, Koray Kavukcuoglu, Helen King, Jack Krawczyk, Yeqing Li, Kathy Meier-Hellstern, Andras Orban, Yury Pinsky, Amar Subramanya, Oriol Vinyals, Ting Yu, and Yori Zwols.
\newblock Imagen 3, 2024.
\newblock URL \url{https://arxiv.org/abs/2408.07009}.

\bibitem[isidentical(2024)]{isidentical2024auraface}
isidentical.
\newblock Introducing auraface: Open-source face recognition and identity preservation models.
\newblock \url{https://huggingface.co/blog/isidentical/auraface}, 2024.
\newblock Accessed: 2025-05-26.

\bibitem[Kampf and Brichtova(2025)]{google2025_gemini_flash_image}
Kat Kampf and Nicole Brichtova.
\newblock Experiment with gemini 2.0 flash native image generation, 2025.
\newblock URL \url{https://developers.googleblog.com/en/experiment-with-gemini-20-flash-native-image-generation/}.

\bibitem[Kawar et~al.(2023)Kawar, Zada, Lang, Tov, Chang, Dekel, Mosseri, and Irani]{kawar2023imagictextbasedrealimage}
Bahjat Kawar, Shiran Zada, Oran Lang, Omer Tov, Huiwen Chang, Tali Dekel, Inbar Mosseri, and Michal Irani.
\newblock Imagic: Text-based real image editing with diffusion models, 2023.
\newblock URL \url{https://arxiv.org/abs/2210.09276}.

\bibitem[Kingma and Gao(2023)]{Kingma2023UnderstandingDO}
Diederik~P Kingma and Ruiqi Gao.
\newblock Understanding diffusion objectives as the elbo with simple data augmentation.
\newblock In \emph{Thirty-seventh Conference on Neural Information Processing Systems}, 2023.

\bibitem[Korthikanti et~al.(2023)Korthikanti, Casper, Lym, McAfee, Andersch, Shoeybi, and Catanzaro]{korthikanti2023reducing}
Vijay~Anand Korthikanti, Jared Casper, Sangkug Lym, Lawrence McAfee, Michael Andersch, Mohammad Shoeybi, and Bryan Catanzaro.
\newblock Reducing activation recomputation in large transformer models.
\newblock \emph{Proceedings of Machine Learning and Systems}, 5:\penalty0 341--353, 2023.

\bibitem[Kumari et~al.(2023)Kumari, Zhang, Zhang, Shechtman, and Zhu]{kumari2023multi}
Nupur Kumari, Bingliang Zhang, Richard Zhang, Eli Shechtman, and Jun-Yan Zhu.
\newblock Multi-concept customization of text-to-image diffusion.
\newblock In \emph{Proceedings of the IEEE/CVF conference on computer vision and pattern recognition}, pages 1931--1941, 2023.

\bibitem[Li et~al.(2024)Li, Lin, Pathak, Li, Fei, Wu, Ling, Xia, Zhang, Neubig, et~al.]{li2024genai}
Baiqi Li, Zhiqiu Lin, Deepak Pathak, Jiayao Li, Yixin Fei, Kewen Wu, Tiffany Ling, Xide Xia, Pengchuan Zhang, Graham Neubig, et~al.
\newblock Genai-bench: Evaluating and improving compositional text-to-visual generation.
\newblock \emph{arXiv preprint arXiv:2406.13743}, 2024.

\bibitem[Liang et~al.(2024)Liang, Liu, Wright, Constable, Gu, Huang, Zhang, Feng, Huang, Wang, et~al.]{liang2024torchtitan}
Wanchao Liang, Tianyu Liu, Less Wright, Will Constable, Andrew Gu, Chien-Chin Huang, Iris Zhang, Wei Feng, Howard Huang, Junjie Wang, et~al.
\newblock Torchtitan: One-stop pytorch native solution for production ready llm pre-training.
\newblock \emph{arXiv preprint arXiv:2410.06511}, 2024.

\bibitem[Lipman et~al.(2023)Lipman, Chen, Ben-Hamu, Nickel, and Le]{lipman2023flow}
Yaron Lipman, Ricky T.~Q. Chen, Heli Ben-Hamu, Maximilian Nickel, and Matthew Le.
\newblock Flow matching for generative modeling.
\newblock In \emph{The Eleventh International Conference on Learning Representations}, 2023.
\newblock URL \url{https://openreview.net/forum?id=PqvMRDCJT9t}.

\bibitem[Liu et~al.(2025)Liu, Han, Xing, Yin, Wang, Cheng, Liao, Wang, Fu, Han, et~al.]{liu2025step1x}
Shiyu Liu, Yucheng Han, Peng Xing, Fukun Yin, Rui Wang, Wei Cheng, Jiaqi Liao, Yingming Wang, Honghao Fu, Chunrui Han, et~al.
\newblock Step1x-edit: A practical framework for general image editing.
\newblock \emph{arXiv preprint arXiv:2504.17761}, 2025.

\bibitem[Liu et~al.(2022)Liu, Gong, and Liu]{liu2022flow}
Xingchao Liu, Chengyue Gong, and Qiang Liu.
\newblock Flow straight and fast: Learning to generate and transfer data with rectified flow, 2022.

\bibitem[Lugmayr et~al.(2022)Lugmayr, Danelljan, Romero, Yu, Timofte, and Van~Gool]{lugmayr2022repaint}
Andreas Lugmayr, Martin Danelljan, Andres Romero, Fisher Yu, Radu Timofte, and Luc Van~Gool.
\newblock Repaint: Inpainting using denoising diffusion probabilistic models.
\newblock In \emph{Proceedings of the IEEE/CVF conference on computer vision and pattern recognition}, pages 11461--11471, 2022.

\bibitem[Meng et~al.(2023)Meng, Rombach, Gao, Kingma, Ermon, Ho, and Salimans]{meng2023distillation}
Chenlin Meng, Robin Rombach, Ruiqi Gao, Diederik Kingma, Stefano Ermon, Jonathan Ho, and Tim Salimans.
\newblock On distillation of guided diffusion models.
\newblock In \emph{Proceedings of the IEEE/CVF Conference on Computer Vision and Pattern Recognition}, pages 14297--14306, 2023.

\bibitem[Midjourney(2025)]{midjourney2025_home}
Midjourney.
\newblock Midjourney, 2025.
\newblock URL \url{https://www.midjourney.com/home}.

\bibitem[OpenAI(2025)]{openai2025_4o_image}
OpenAI.
\newblock Introducing 4o image generation, 2025.
\newblock URL \url{https://openai.com/index/introducing-4o-image-generation/}.

\bibitem[Pan et~al.(2023)Pan, Tewari, Leimk{\"u}hler, Liu, Meka, and Theobalt]{pan2023drag}
Xingang Pan, Ayush Tewari, Thomas Leimk{\"u}hler, Lingjie Liu, Abhimitra Meka, and Christian Theobalt.
\newblock Drag your gan: Interactive point-based manipulation on the generative image manifold.
\newblock In \emph{ACM SIGGRAPH 2023 conference proceedings}, pages 1--11, 2023.

\bibitem[Peebles and Xie(2023)]{Peebles_2023}
William Peebles and Saining Xie.
\newblock Scalable diffusion models with transformers.
\newblock In \emph{2023 IEEE/CVF International Conference on Computer Vision (ICCV)}. IEEE, 2023.
\newblock \doi{10.1109/iccv51070.2023.00387}.
\newblock URL \url{http://dx.doi.org/10.1109/ICCV51070.2023.00387}.

\bibitem[Peng et~al.(2025)Peng, Cui, Tang, Qi, Dong, Bai, Han, Ge, Zhang, and Xia]{peng2025dreambenchhumanalignedbenchmarkpersonalized}
Yuang Peng, Yuxin Cui, Haomiao Tang, Zekun Qi, Runpei Dong, Jing Bai, Chunrui Han, Zheng Ge, Xiangyu Zhang, and Shu-Tao Xia.
\newblock Dreambench++: A human-aligned benchmark for personalized image generation, 2025.
\newblock URL \url{https://arxiv.org/abs/2406.16855}.

\bibitem[Podell et~al.(2023)Podell, English, Lacey, Blattmann, Dockhorn, Müller, Penna, and Rombach]{podell2023sdxl}
Dustin Podell, Zion English, Kyle Lacey, Andreas Blattmann, Tim Dockhorn, Jonas Müller, Joe Penna, and Robin Rombach.
\newblock Sdxl: Improving latent diffusion models for high-resolution image synthesis, 2023.

\bibitem[Ramesh et~al.(2022)Ramesh, Dhariwal, Nichol, Chu, and Chen]{ramesh2022hierarchical}
Aditya Ramesh, Prafulla Dhariwal, Alex Nichol, Casey Chu, and Mark Chen.
\newblock Hierarchical text-conditional image generation with clip latents, 2022.

\bibitem[Rombach et~al.(2022)Rombach, Blattmann, Lorenz, Esser, and Ommer]{Rombach_2022}
Robin Rombach, Andreas Blattmann, Dominik Lorenz, Patrick Esser, and Bjorn Ommer.
\newblock High-resolution image synthesis with latent diffusion models.
\newblock In \emph{2022 IEEE/CVF Conference on Computer Vision and Pattern Recognition (CVPR)}. IEEE, 2022.
\newblock \doi{10.1109/cvpr52688.2022.01042}.
\newblock URL \url{http://dx.doi.org/10.1109/CVPR52688.2022.01042}.

\bibitem[Ruiz et~al.(2023)Ruiz, Li, Jampani, Pritch, Rubinstein, and Aberman]{ruiz2023dreamboothfinetuningtexttoimage}
Nataniel Ruiz, Yuanzhen Li, Varun Jampani, Yael Pritch, Michael Rubinstein, and Kfir Aberman.
\newblock Dreambooth: Fine tuning text-to-image diffusion models for subject-driven generation, 2023.
\newblock URL \url{https://arxiv.org/abs/2208.12242}.

\bibitem[Runway~AI(2025)]{runway2025_home}
Inc. Runway~AI.
\newblock Runway | tools for human imagination, 2025.
\newblock URL \url{https://runwayml.com/}.

\bibitem[Saharia et~al.(2022{\natexlab{a}})Saharia, Chan, Chang, Lee, Ho, Salimans, Fleet, and Norouzi]{saharia2022palette}
Chitwan Saharia, William Chan, Huiwen Chang, Chris Lee, Jonathan Ho, Tim Salimans, David Fleet, and Mohammad Norouzi.
\newblock Palette: Image-to-image diffusion models.
\newblock In \emph{ACM SIGGRAPH 2022 Conference Proceedings}, pages 1--10, 2022{\natexlab{a}}.

\bibitem[Saharia et~al.(2022{\natexlab{b}})Saharia, Chan, Saxena, Li, Whang, Denton, Ghasemipour, Gontijo~Lopes, Karagol~Ayan, Salimans, et~al.]{saharia2022photorealistic}
Chitwan Saharia, William Chan, Saurabh Saxena, Lala Li, Jay Whang, Emily~L Denton, Kamyar Ghasemipour, Raphael Gontijo~Lopes, Burcu Karagol~Ayan, Tim Salimans, et~al.
\newblock Photorealistic text-to-image diffusion models with deep language understanding.
\newblock \emph{Advances in neural information processing systems}, 35:\penalty0 36479--36494, 2022{\natexlab{b}}.

\bibitem[Sauer et~al.(2021)Sauer, Chitta, M{\"u}ller, and Geiger]{sauer2021projected}
Axel Sauer, Kashyap Chitta, Jens M{\"u}ller, and Andreas Geiger.
\newblock Projected gans converge faster.
\newblock \emph{Advances in Neural Information Processing Systems}, 2021.

\bibitem[Sauer et~al.(2023)Sauer, Lorenz, Blattmann, and Rombach]{sauer2023adversarial}
Axel Sauer, Dominik Lorenz, Andreas Blattmann, and Robin Rombach.
\newblock Adversarial diffusion distillation.
\newblock \emph{arXiv preprint arXiv:2311.17042}, 2023.

\bibitem[Sauer et~al.(2024)Sauer, Boesel, Dockhorn, Blattmann, Esser, and Rombach]{sauer2024fasthighresolutionimagesynthesis}
Axel Sauer, Frederic Boesel, Tim Dockhorn, Andreas Blattmann, Patrick Esser, and Robin Rombach.
\newblock Fast high-resolution image synthesis with latent adversarial diffusion distillation, 2024.
\newblock URL \url{https://arxiv.org/abs/2403.12015}.

\bibitem[Shah et~al.(2024)Shah, Bikshandi, Zhang, Thakkar, Ramani, and Dao]{shah2024flashattention}
Jay Shah, Ganesh Bikshandi, Ying Zhang, Vijay Thakkar, Pradeep Ramani, and Tri Dao.
\newblock Flashattention-3: Fast and accurate attention with asynchrony and low-precision.
\newblock \emph{Advances in Neural Information Processing Systems}, 37:\penalty0 68658--68685, 2024.

\bibitem[Sheynin et~al.(2023)Sheynin, Polyak, Singer, Kirstain, Zohar, Ashual, Parikh, and Taigman]{sheynin2023emu}
Shelly Sheynin, Adam Polyak, Uriel Singer, Yuval Kirstain, Amit Zohar, Oron Ashual, Devi Parikh, and Yaniv Taigman.
\newblock Emu edit: Precise image editing via recognition and generation tasks.
\newblock \emph{arXiv preprint arXiv:2311.10089}, 2023.

\bibitem[Simonyan and Zisserman(2014)]{simonyan2014very}
Karen Simonyan and Andrew Zisserman.
\newblock Very deep convolutional networks for large-scale image recognition.
\newblock \emph{arXiv preprint arXiv:1409.1556}, 2014.

\bibitem[Su et~al.(2024)Su, Ahmed, Lu, Pan, Bo, and Liu]{su2024roformer}
Jianlin Su, Murtadha Ahmed, Yu~Lu, Shengfeng Pan, Wen Bo, and Yunfeng Liu.
\newblock Roformer: Enhanced transformer with rotary position embedding.
\newblock \emph{Neurocomputing}, 568:\penalty0 127063, 2024.

\bibitem[Suvorov et~al.(2022)Suvorov, Logacheva, Mashikhin, Remizova, Ashukha, Silvestrov, Kong, Goka, Park, and Lempitsky]{suvorov2022resolution}
Roman Suvorov, Elizaveta Logacheva, Anton Mashikhin, Anastasia Remizova, Arsenii Ashukha, Aleksei Silvestrov, Naejin Kong, Harshith Goka, Kiwoong Park, and Victor Lempitsky.
\newblock Resolution-robust large mask inpainting with fourier convolutions.
\newblock In \emph{Proceedings of the IEEE/CVF winter conference on applications of computer vision}, pages 2149--2159, 2022.

\bibitem[Szeliski(2022)]{szeliski2022computer}
Richard Szeliski.
\newblock \emph{Computer vision: algorithms and applications}.
\newblock Springer Nature, 2022.

\bibitem[Xiao et~al.(2024)Xiao, Wang, Zhou, Yuan, Xing, Yan, Li, Wang, Huang, and Liu]{xiao2024omnigen}
Shitao Xiao, Yueze Wang, Junjie Zhou, Huaying Yuan, Xingrun Xing, Ruiran Yan, Chaofan Li, Shuting Wang, Tiejun Huang, and Zheng Liu.
\newblock Omnigen: Unified image generation.
\newblock \emph{arXiv preprint arXiv:2409.11340}, 2024.

\bibitem[Yang et~al.(2023)Yang, Gu, Zhang, Zhang, Chen, Sun, Chen, and Wen]{yang2023paint}
Binxin Yang, Shuyang Gu, Bo~Zhang, Ting Zhang, Xuejin Chen, Xiaoyan Sun, Dong Chen, and Fang Wen.
\newblock Paint by example: Exemplar-based image editing with diffusion models.
\newblock In \emph{Proceedings of the IEEE/CVF conference on computer vision and pattern recognition}, pages 18381--18391, 2023.

\bibitem[Ye et~al.(2023)Ye, Zhang, Liu, Han, and Yang]{ye2023ip}
Hu~Ye, Jun Zhang, Sibo Liu, Xiao Han, and Wei Yang.
\newblock Ip-adapter: Text compatible image prompt adapter for text-to-image diffusion models.
\newblock \emph{arXiv preprint arXiv:2308.06721}, 2023.

\bibitem[Yu et~al.(2022)Yu, Xu, Koh, Luong, Baid, Wang, Vasudevan, Ku, Yang, Ayan, Hutchinson, Han, Parekh, Li, Zhang, Baldridge, and Wu]{yu2022scalingautoregressivemodelscontentrich}
Jiahui Yu, Yuanzhong Xu, Jing~Yu Koh, Thang Luong, Gunjan Baid, Zirui Wang, Vijay Vasudevan, Alexander Ku, Yinfei Yang, Burcu~Karagol Ayan, Ben Hutchinson, Wei Han, Zarana Parekh, Xin Li, Han Zhang, Jason Baldridge, and Yonghui Wu.
\newblock Scaling autoregressive models for content-rich text-to-image generation, 2022.
\newblock URL \url{https://arxiv.org/abs/2206.10789}.

\bibitem[Zhang et~al.(2024)Zhang, Mo, Chen, Sun, and Su]{zhang2024magicbrushmanuallyannotateddataset}
Kai Zhang, Lingbo Mo, Wenhu Chen, Huan Sun, and Yu~Su.
\newblock Magicbrush: A manually annotated dataset for instruction-guided image editing, 2024.
\newblock URL \url{https://arxiv.org/abs/2306.10012}.

\bibitem[Zhang et~al.(2023)Zhang, Rao, and Agrawala]{zhang2023adding}
Lvmin Zhang, Anyi Rao, and Maneesh Agrawala.
\newblock Adding conditional control to text-to-image diffusion models.
\newblock In \emph{Proceedings of the IEEE/CVF international conference on computer vision}, pages 3836--3847, 2023.

\bibitem[Zhang et~al.(2025)Zhang, Xie, Lu, Yang, and Yang]{zhang2025context}
Zechuan Zhang, Ji~Xie, Yu~Lu, Zongxin Yang, and Yi~Yang.
\newblock In-context edit: Enabling instructional image editing with in-context generation in large scale diffusion transformer.
\newblock \emph{arXiv preprint arXiv:2504.20690}, 2025.

\end{thebibliography}

\end{document}